# Hexagonal Perovskites as Quantum Materials

Loi T. Nguyen and R.J. Cava

*Department of Chemistry, Princeton University, Princeton New Jersey 08544*

**ABSTRACT:** Hexagonal oxide perovskites, in contrast to the more familiar perovskites, allow for face-sharing of metal-oxygen octahedra or trigonal prisms within their structural frameworks. This results in dimers, trimers, tetramers, or longer fragments of chains of face-sharing octahedra in the crystal structures, and consequently in much shorter metal-metal distances and lower metal-oxygen-metal bond angles than are seen in the more familiar perovskites. The presence of the face-sharing octahedra can have a dramatic impact on magnetic properties of these compounds, and dimer-based materials, in particular, have been the subjects of many quantum-materials-directed studies in materials physics. Hexagonal oxide perovskites are of contemporary interest due to their potential for geometrical frustration of the ordering of magnetic moments or orbital occupancies at low temperatures, which is especially relevant to their significance as quantum materials. As such, several hexagonal oxide perovskites have been identified as potential candidates for hosting the quantum spin liquid state at low temperatures. In our view, hexagonal oxide perovskites are fertile ground for finding new quantum materials. This review briefly describes the solid state chemistry of many of these materials.

## CONTENTS

## 1. Introduction

Halide perovskites have recently been widely studied by chemical researchers. They can be processed under relatively mild conditions compared to oxides and, as semiconductors, can sometimes have favorable optical and charge transport properties, making them suitable for study as materials with the potential for solar energy conversion. They are not currently widely studied in the materials physics community as quantum materials. These materials are variants of the familiar perovskite $AMX_3$ formula, where the A atom or molecule (e.g. Cs or $CH_3NH_3$) is in the large cavity in the classic perovskite structure, and the $MX_6$ octahedra, made from main group M ions, share corner X ions (here Cl, Br, or I) in a three-dimensional framework. Such materials obey the familiar radius rules known for derivatives of cubic perovskites, with some potential refinements[1–9]. Many reviews of this type of perovskite can be found in the chemical literature, including in this journal[10]. They are not the subject of this review.

The classic oxide perovskites, with formulas $AMO_3$, (where A = a larger ion, typically monovalent, divalent or trivalent, and M is a smaller metal ion) are well known in the solid state chemistry and materials physics communities and have been intensively studied for many decades[11]. This is due to their interesting structural chemistry and often forefront magnetic and electronic properties. The interest in oxide perovskites as quantum materials often arises from the fact that the M ions can be members of the transition metal series, which can result in properties that can best be understood as quantum mechanical in nature.

The familiar oxide perovskites are typically made from electropositive (non-molecular) A ions (e.g. Alkali, alkaline earth or rare earth ions) that reside somewhere in a perovskite cavity where they are ideally 12 coordinated to the surrounding oxygens, and $MO_6$ octahedra sharing corners with each other to create a $MO_3$ framework structure. The *d* orbitals on the M ions hybridize in different degrees with the oxygen *p* orbitals. The quantum-like properties can arise for $d^n$ ions (where the *d* states are partly filled, and *n* is between 1 and 9) and even for $d^0$ or $d^{10}$ ions (e.g. $Ti^{4+}$ or $Pb^{4+}$) in the octahedra in the $MO_3$ framework.

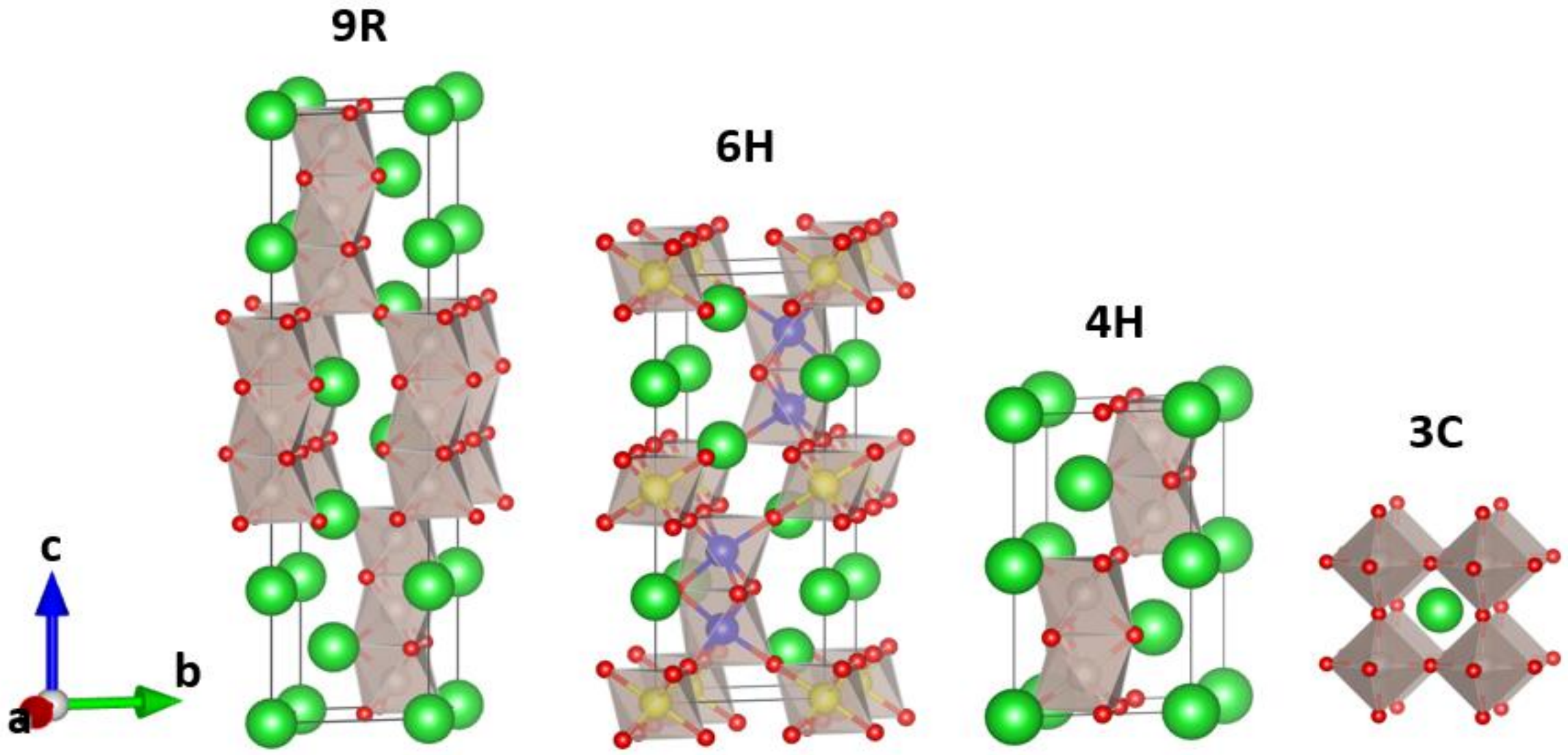

Figure 1. The different polymorphs of $BaRuO_3$. Color: Ba (green), Ru (gray octahedra) and O (red). In the 6H drawing, the two kinds of Ru – blue in the dimers and yellow in the perovskite-like layers – are shown. $BaRuO_3$ at ambient pressure is found in the 9R or 4H structure types, $SrRuO_3$ is found in a distortion of the 3C structure type, as is $BaRuO_3$ prepared at high pressures.

Structurally, the octahedra are often considered as rigid entities that tilt around their shared corners to accommodate different size A site ions and an overall cell size that is also determined by temperature or pressure. A rich set of symmetry variants is possible due to the tilting, and too many interesting combinations of atoms, physical properties and structural distortions have been studied to even list generally[12,13]. The tilting, from a chemical and quantum materials perspective, changes the degree of overlap between the oxygen *p* and metal *d* orbitals at the connections between the corner-sharing $MO_6$ octahedra in the perovskite framework, with a significant impact on the electronic and magnetic properties displayed. Even a relatively simple oxide perovskite like $SrTiO_3$ can display profound quantum mechanical properties such as superconductivity at sub-Kelvin temperatures when doped with a very small number of electrons[14–16]. Even though the tilt system in classical perovskites can sometimes yield a rhombohedral symmetry crystal structure, such materials are in general not the subjects of this review. Ruddlesden Popper and Dion Jacobsen series materials are celebrated layered oxide perovskites where different ratios and relative sizes of A to B site ions result in the break-up of the $MO_3$ perovskite network into layers of different thicknesses. Electrons and magnetic moments, when confined to layers, or largely confined to layers, can display quite unexpected quantum-derived properties, such as superconductivity[17–19]. The structural break-up of the normally three-dimensional $MO_3$ array into layers can lead to dramatically different physical properties for some of these materials, which have also been the subjects of many reviews. Double perovskites in oxides are also well known and have been the subjects of research in halide perovskites as well[6,7,20–26]. Perovskites are a well for research that never seems to run dry.

This review, however, is about a currently lesser known class of oxide perovskite quantum materials, based on hexagonal symmetry rather than cubic symmetry. These materials are also based on $AMO_3$ stoichiometries but, in contrast to the more familiar perovskites, admit for face sharing of $MO_6$ octahedra or A- or M- centered trigonal prisms in the structural framework rather than the corner-sharing geometries found in the framework of conventional perovskites. This can result in dimers, trimers, tetramers or longer fragments of chains of face-sharing octahedra in the structure. The face sharing of the octahedra within the dimers trimers tetramers and chains leads to much shorter metal-metal distances and lower metal-oxygen-metal bond angles than are seen in more familiar perovskites. It has been argued that hexagonal perovskites can best happen for selected A site ions in oxides because their ionic radii and that of oxygen are such that it is possible for them to mix in an ordered arrangement to form a closest-packed plane. $Ba^{2+}$ and $O^{2-}$ seem to be a perfect pair for engendering this kind of structural stability. A $Ba^{2+}$ ion replacing one of the oxygen ions in a close packed oxygen plane appears to be happily surrounded by the 6 near neighbor $O^{2-}$ ions that it acquires in the plane. The fact that face sharing is present has a dramatic impact on the transition metal oxygen *d* orbital oxygen *p* orbital overlap and the direct *d-d* overlap of the transition metals, and, consequently on the quantum properties.

Symmetry matters when it comes to physical properties, and the 3 or 6 fold rotational symmetries of the structural components in the hexagonal perovskites are of contemporary interest due to the potential that such geometries have to engender the geometrical frustration of the ordering of the magnetic moments or orbital occupancies of the framework ions into a single lowest energy state at low temperatures. Thus such materials are good candidates for the quantum-spin-liquid state, as described briefly here. (In quantum spin liquids, the spin state of the magnetic ions continues to fluctuate down to very low temperatures compared to the strength of the magnetic interactions as measured by the Curie-Weiss temperature.)

Further, their properties frequently challenge our current understanding of when a chemistry (i.e. an orbital based) vs. a physics (i.e., a Fermi Surface based) description is most appropriate for interpretation and prediction of the properties of a material, which is good fun[27–29] (see e.g. the review by D. Khomskii and S.V. Streltsov in this issue). Many hexagonal oxide perovskites based on face-sharing of magnetic octahedra have been studied due to their potential for hosting the quantum spin liquid state at low temperature.

As is found in the more familiar cubic-symmetry-based $AMO_3$ perovskites, structural distortions, chemical disorder, and non-$AMO_3$ stoichiometries are also found in the hexagonal perovskites. Structurally layered variants are also known. The basic crystal structures and, in some cases, the known structural complexities are the focus of this review. Although metallic, semiconducting, and dielectric hexagonal perovskites are known, the physical properties of this class of materials have so far not turned out to be of as wide interest in the materials physics community as those of their cubic-symmetry-based cousins. The properties of the materials whose chemistry and structure are the primary focus of this review are also generally summarized here when known.

This organization of this review is based on the solid state chemistry of the hexagonal perovskites, and is based primarily on their crystal structures. A reader particularly interested in the crystal structures of such materials can find alternative presentations in books on perovskites[11–13]. Our intention is to present a structural description of as many of them as we can, although some interesting materials may be omitted by accident. On the other hand, occasionally we may describe variants that are be considered by some researchers to be "not-really-hexagonal-perovskites", or better classified in a different way, and we apologize to people whose views are stricter than ours if our classification appears to include too many materials; we do not mean to exclude other classifications for such materials. In our view, hexagonal perovskites are fertile ground for finding interesting quantum materials and we hope that the descriptions presented in this review will be accessible to a wide range of scientists who may be interested in such things.

## 2. Mixtures of cubic plus hexagonal packing

Let's start with the $SrRuO_3$-$BaRuO_3$ perovskites as our initial example. $SrRuO_3$ is a slightly distorted conventional cubic perovskite due to a small tilt of the $RuO_6$ octahedra in the $RuO_3$ framework, while, at conventional pressures, the $Ba^{2+}$ ion is not a good fit for the cubic perovskite cavity. It is a better fit when inserted into a close packed plane of oxygen, however, and thus, the crystal structures of the ambient pressure $BaRuO_3$ polymorphs are hexagonal perovskites.

For A site ions of size between Sr and Ba, the resulting crystal structure is an ordered mixture of cubic perovskite plus hexagonal perovskite. There is no electropositive divalent ion with a size intermediate between Sr and Ba, but as is often found in solid state chemistry, such an ion can be simulated by mixing Sr and Ba together. Although any individual position in one unit cell is occupied by either Ba or Sr, their average radius is what impacts the long range average crystal structure observed. The (111) plane of conventional cubic perovskites has three fold rotational symmetry. When used as a building block a small section of the (111) plane is a layer of isolated $MO_6$ octahedra lying on their faces in a triangular array. The A ions are found between these layers. When the (111) layers are included in a structure, the octahedra share corners with others. When this layer is used as a structural building block in the creation of a new material and mixed with hexagonal symmetry $AMO_3$ layers that consist of face-shared octahedra, then materials with overall hexagonal symmetry are the result.

The compounds in the $SrRuO_3$-$BaRuO_3$ system allow us to introduce the simple short-hand nomenclature that is used to describe hexagonal perovskites. In this nomenclature, a number is first, which indicates the number of structural layers within a unit cell. The designation then ends with a letter, an R, an H or a C, which indicates whether the material is based on rhombohedral, hexagonal or cubic layer stacking (like in all perovskites, small distortions are possible, typically due to octahedral tilting) Using this nomenclature, classic cubic-symmetry-based $SrRuO_3$ has a 3C structure (three cubic layers per cell stacked along (111) and $BaRuO_3$ has three polymorphs, one with a 4H structure, one with a 6H structure and one with a 9R structure[30] (**Figure 1**). (In rhombohedral symmetry hexagonal perovskite crystal structures the number of layers is always a multiple of 3). At high pressures, $BaRuO_3$ is stable in a different crystal structure, transforming to a conventional 3C perovskite (**Figure 1**). This cubic $BaRuO_3$ perovskite is metallic and shows a ferromagnetic transition at Tc = 60 K, significantly lower than the Tc = 160 K of $SrRuO_3$, as shown in **Figure 2**. A related designation is when a layer of $MO_6$ octahedra is connected by corner sharing to the next layer, as is found in cubic perovskites, then that connection is designated by a lowercase c, when the octahedra are connected by sharing faces then that kind of connection has 3 or 6 fold symmetry and is designated by a lowercase h. Thus the structure of many 9-layer hexagonal perovskites may be designated as 9R or $(chh)_3$.

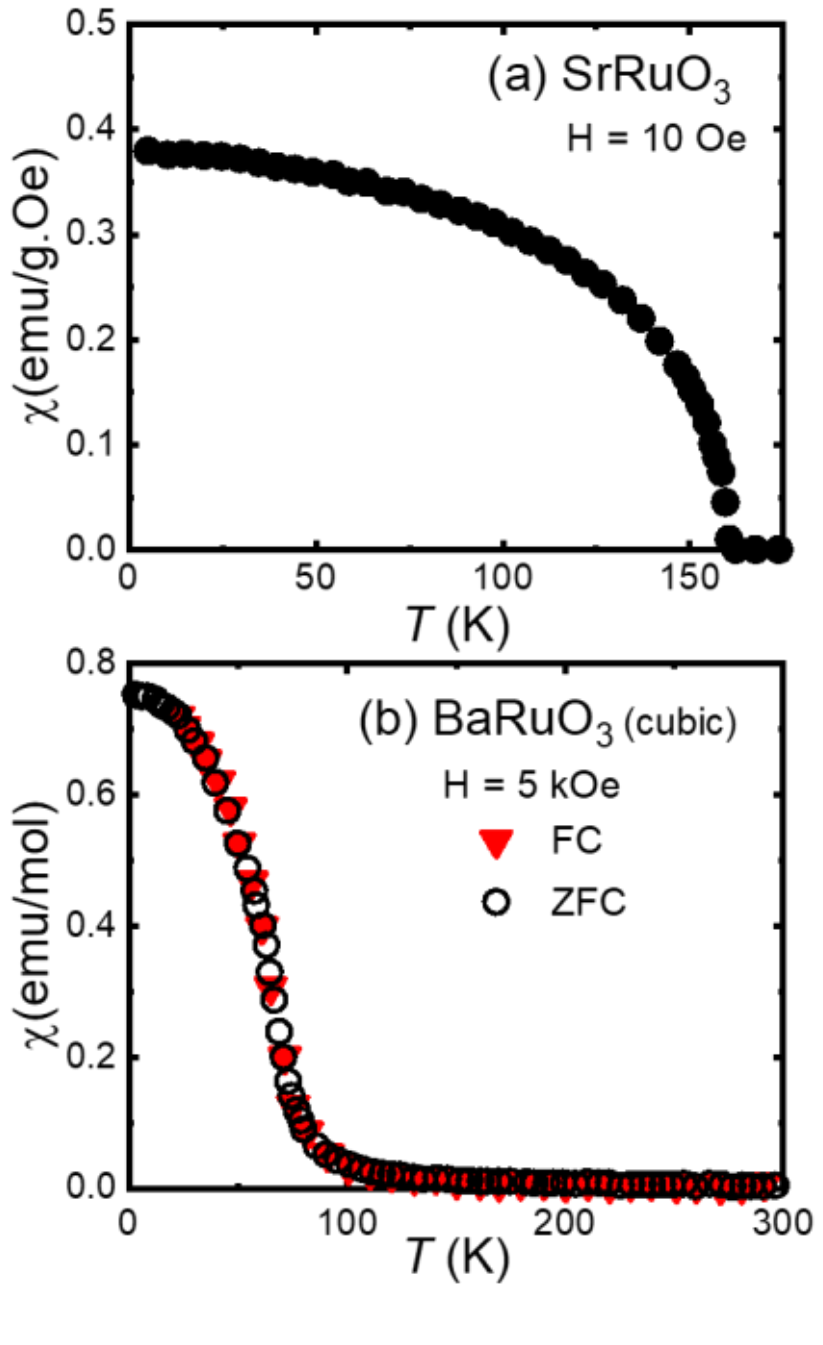

Figure 2. The ferromagnetic transitions in (a) orthorhombic $SrRuO_3$ [31] and (b) cubic $BaRuO_3$ conventional perovskites [30].

In the case of the $Ba_{1-x}Sr_xRuO_3$ solid solution, the resulting crystal structure is clearly determined by the average radius of the A site ions; an on-average-smaller A site ion yields a crystal structure with a larger fraction of cubic-like perovskite character, while a larger A site ion yields a hexagonal-like perovskite structure[32–34]. The same may be the case for other $(Ba_{1-x}Sr_x)MO_3$ perovskites where the Ba end member ($x$=0) is a hexagonal perovskite and the Sr end member ($x$=1) is a cubic perovskite. In solid solutions of the type $Ba_{1-x}Sr_xRuO_3$, when increasing the values of $x$, the nine-layer structure with stacking sequence $(chh)_3$ (9R) transforms to the four-layer structure $(ch)_2$ (4H) and finally to the classical perovskite type structure $(c)_3$ (3C) at x is about 1/6 and 1/3, respectively[33]. The 4H and 9R forms of $BaRuO_3$ have substantially different electronic properties[35], and even the $Sr_{1-x}Ca_xRuO_3$ perovskites in the more familiar cubic perovskite structure types are famously enigmatic quantum materials[36–38]. In the event that readers are not familiar with magnetism we point out that the A ions in perovskites, with the exceptions of the rare earths, are not magnetic (i.e. $Ba^{2+}$, $Sr^{2+}$ and $Ca^{2+}$) nor are the oxygen ions.

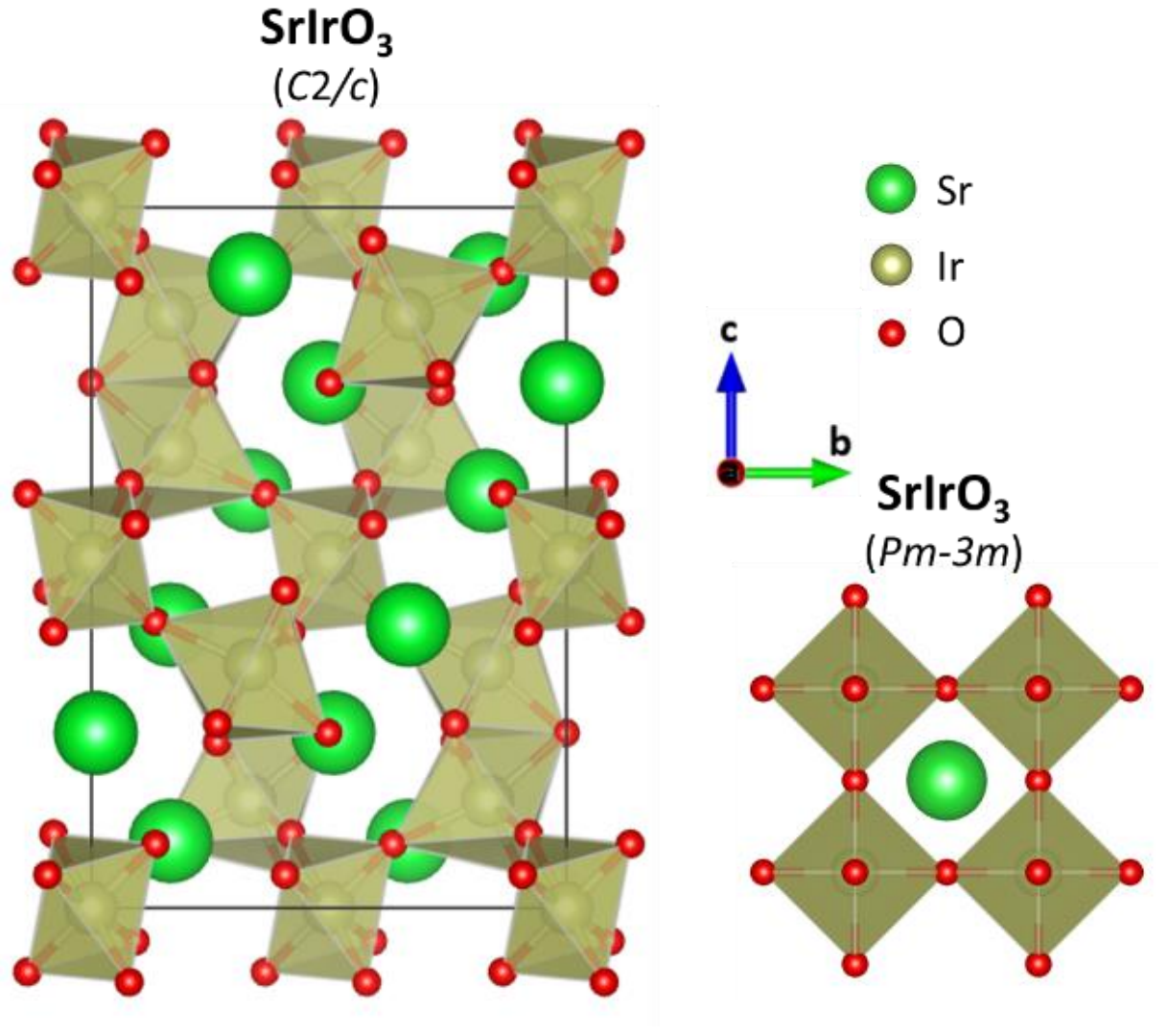

Figure 3. The crystal structures of distorted monoclinic $SrIrO_3$ (left), the ambient pressure phase, and cubic $SrIrO_3$, a stable high pressure phase (right). The former is a distorted variation of a 6H hexagonal perovskite.

Iridium oxides are of special interest in the study of quantum materials because, for these materials, spin-orbit-coupling can have a dramatic effect on the energies of the valence electrons associated with the M ions and consequently the materials' electronic and magnetic properties. The strong influence of spin-orbit-coupling can explain some of the properties of iridates such as their widely insulating behavior. $SrIrO_3$ and $BaIrO_3$ are quantum materials closely related to the ruthenate perovskites. In this series, however, $SrIrO_3$ and $BaIrO_3$ are both monoclinic $C2/c$ distortions of the 6H-$BaTiO_3$ hexagonal $(chc)_2$ structure[39,40] (former: $a$ = 5.604 Å, $b$ = 9.618 Å, $c$ = 14.17 Å, beta = 93.26°). These structural distortions have not yet been theoretically modeled to our knowledge, but they may well be related to the influence of the strong spin orbit coupling in Ir, which is widely known to be influential in determining the magnetic and electrical properties of iridium oxide based quantum materials[41–47]. The monoclinic and cubic crystal structures of $SrIrO_3$ are shown in **Figure 3**. Both $SrIrO_3$ and $BaIrO_3$ are non-magnetic electrical insulators[39,48,49] and adopt the ideal cubic $Pm$-3$m$ perovskite structure at high pressure. A substitution of Zn or Li for Ir in the 6H-$SrIrO_3$ polymorph destabilizes the face-sharing octahedra, and hence $SrIr_{1-x}Zn_xO_3$ is found to be an orthorhombic distortion of the classical cubic perovskite for $0.25 \le x \le 0.33$ and $SrIr_{1-x}Li_xO_3$ has that type of crystal structure for $x$ = 0.25 only[50]. The reason for the destabilization of the hexagonal perovskite structure for Zn and Li substitutions has not been proven theoretically.

## 3. Dimers and trimers. Relation to classic double perovskites.

In classical double perovskites, there are two different "M type" ions in an ordered array or a partially ordered array within the $MO_3$ framework. Double perovskites can be written in the form $A_2MM'O_6$, where M and M' can be different ions such as in $Sr_2FeWO_6$, a material that has a particularly high magnetoresistance[51] or even the same M metal with different oxidation states, such as is seen in $BaBiO_3$ ($Ba_2Bi^{3+}Bi^{5+}O_6$ formally)[52–57], the basis for superconductivity in several perovskite oxides[58,59]. Disordered and partially ordered M-M' double perovskites are known. "Double perovskites" are also known for ratios of M to M' that are not 1:1, for example also 2:1, where the formula is $A_2M_{4/3}M'_{2/3}O_6$ i.e. $A_2\{M[M_{1/3}M'_{2/3}]\}O_6$, with M and M' selected to yield charge neutrality. An example of this kind of double perovskite is $Ca_4Nb_2O_9$ [60,61]. Of further interest are oxide double perovskites where one of the ions does not display octahedral coordination[52,62].

Double hexagonal perovskites are also known – when there is more than one chemically distinct type of M site ion present. The materials can have ordered M-M' or disordered M-M' structures. Many ordered materials and disordered materials are known and briefly described in this review. Consider the 6H structure shown in **Figure 1** as the basis. There are two distinctly different octahedral sites for the M ions present. One (yellow in the figure) shares only corners with other octahedra, making a separate triangular plane (seen in the figure at levels 0, ½, and 1 along the crystallographic $c$ axis, and the other shares corners with one set of $(M,M')O_6$ octahedra but a face with another set (blue in the figure). When there are two octahedra in the face sharing sub-structure, the basic formula is $A_3MM'_2O_9$. Occasionally M and M' can be the same ion (as in some of the polymorphs of $BaRuO_3$) but much more frequently they are two distinctly different ions. While the M ions form triangular planar lattices of potentially isolated $MO_6$ octahedra, the M' ions form a dimer array on a triangular lattice. When either M or M' has unpaired electrons, the magnetism can make these quite interesting quantum materials due to the triangular lattice and the localized vs. delocalized character of unpaired electrons in the dimers[47]. For consistency with more complex structures, we refer to these materials simply as dimer-based hexagonal perovskites.

### 3.1. Dimer-based hexagonal perovskites

Dimer-based hexagonal oxide perovskites are by far the most frequently studied quantum materials in this family. The most common dimer-based hexagonal double perovskite family of quantum materials is found when the M'-site is occupied by Ru or Ir, making the formula $Ba_3MRu_2O_9$ or $Ba_3MIr_2O_9$. For purposes of organization in this group, which consists of many frequently studied materials, we consider the generic formula to be $Ba_3MM'_2O_9$, and we classify the materials into subgroups that differ in the total number of electrons on the $M'_2O_9$ dimer. The formal oxidation state of M can range from 1+ to 4+ - as the formal charge on the M site ion increases, to maintain neutrality, the formal oxidation state of the M' ion, and thus the total number of electrons on the $M'_2O_9$ dimer, decreases. (For formal electron counts, the electropositive Ba ion is always taken to be $Ba^{2+}$ and the oxygen ion is taken to be $O^{2-}$.) As in many quantum materials, determination of the physical properties is not related in a trivial way to the formal electron count, as how the electrons are distributed on the $M'_2O_9$ dimers depends on some important chemical characteristics of how the orbitals on M' and oxygen overlap, discussed in the review by Khomskii and Streltsov in this issue.

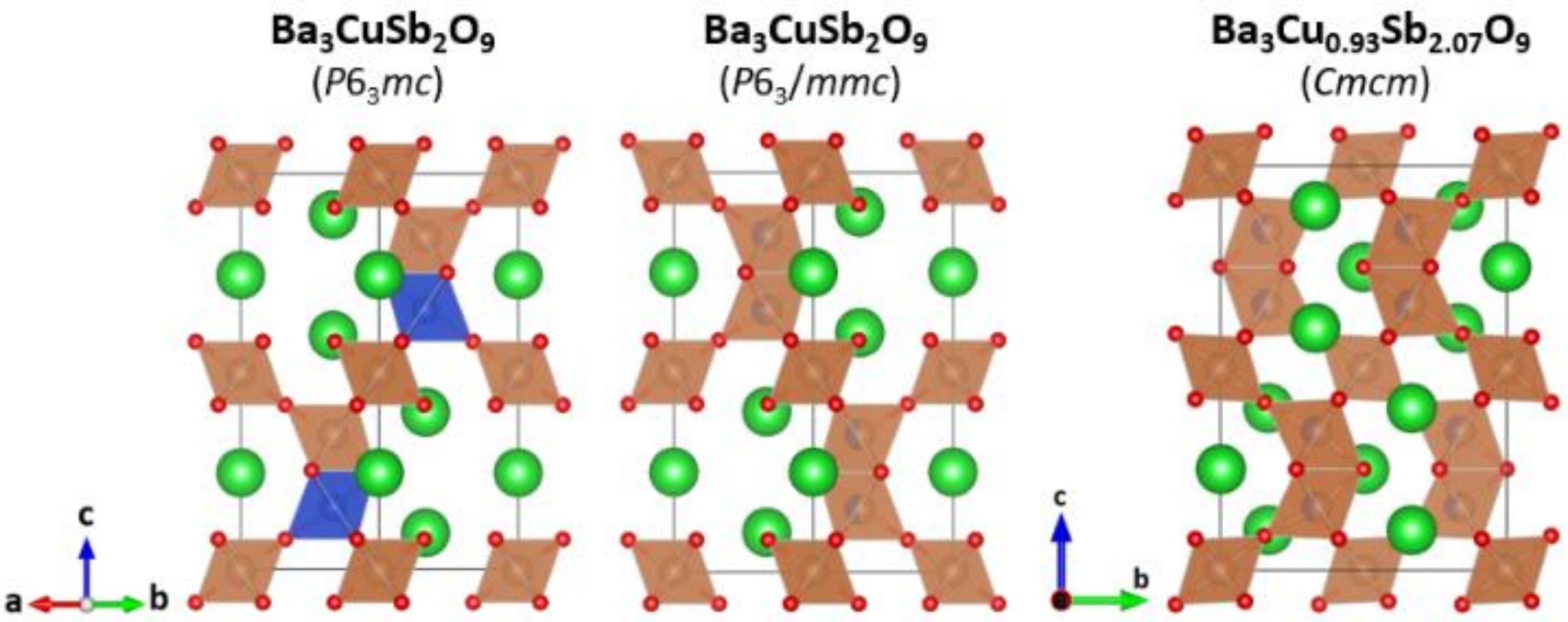

Figure 4. The reported crystal structures of "$Ba_3CuSb_2O_9$".

In $Ba_3MRu_2O_9$, when M is a non-magnetic monovalent cation (eg. Li or Na), the total number of electrons donated by the Ru to the oxygens in a $Ru_2O_9$ dimer is 11 [63–65]. (The Ru can be considered formally as a 1:1 mixture of $Ru^{5+}$ and $Ru^{6+}$, but "where's the charge, really?" is a good question that has been asked by some for decades for such highly charged transition metal ions in oxides.) Both of these compounds crystallize in the hexagonal space group $P6_3/mmc$. While the hexagonal structure of $Ba_3LiRu_2O_9$ remains stable down to 100 K, $Ba_3NaRu_2O_9$ undergoes a structural transition at 225 K to a lower symmetry space group, *Cmcm,* a simple dimensional distortion of the hexagonal perovskite. This has been attributed to a well-known phenomenon in materials physics and solid state chemistry known as charge ordering; in this case where the $Ru^{5+}$ ions are said to occupy one of the sites in the dimer and the $Ru^{6+}$ ions to occupy the other site in the dimer, in an ordered array. $Ba_3NaRu_2O_9$, which, in contrast has not been shown to undergo charge ordering transition, has a negative Curie-Weiss temperature, indicating antiferromagnetic interactions between the Ru spins, but remains paramagnetic down to 1.7 K. Its magnetic effective moment is calculated to be about 0.91μB/mol-f.u, and the conductivity activation gap is about 0.1 eV. Why the $Li^{1+}$ material behaves so differently from the $Na^{1+}$ material is a potentially interesting puzzle to address in future research. These are excellent examples of quantum materials.

When M is clearly a divalent cation in $Ba_3MRu_2O_9$, the hexagonal 6H crystal structure that we have been discussing has been reported for in the cases of $Ni^{2+}$, $Ca^{2+}$, $Mg^{2+}$, and $Zn^{2+}$. For these materials there would be 10 electrons donated to the $Ru_2O_9$ dimer by its constituent metal atoms, for a formal charge on both Ru of 5+. For the cases above when M is a transition metal (i.e. $Ni^{2+}$, which is an S=1 ion), there must be an internal equilibrium within the compound between the formal valence of the M ion and the formal valence of the Ru in the dimers. Thus dimer-based hexagonal perovskites such as $Ba_3CoRu_2O_9$ and $Ba_3FeRu_2O_9$ would be interesting if studied in this context because $Fe^{3+}$ or $Co^{3+}$ may more likely be in equilibrium with a formally $Ru^{4+}$-$Ru^{5+}$ mixture than a $Co^{2+}$ or $Fe^{2+}$ mixture would be with $Ru^{5+}$. Some of the divalent ions listed above are non-magnetic (i.e. Ca, Mg and Zn), an important characteristic for M ions that results in the magnetism, and the resulting quantum properties, arising from transition metal ions in the dimers and their orbital overlap with the oxygens.

Sb is a common 5+ ion in octahedral coordination in oxides and is an important constituent in some hexagonal perovskites. It is non-magnetic, so when it is present, the magnetism arises from the other ions present. The dimer-based hexagonal perovskite which mixes Cu and Sb as the M and M' ions is both important and unusual. An ordered 6H hexagonal perovskite crystal structure for $Ba_3CuSb_2O_9$ was reported in 1978[66]. This structure consists of ordered face-sharing of $SbO_6$ octahedra and isolated $CuO_6$ octahedra, in a normal hexagonal perovskite type structure as shown in **Figure 4**. If this is the crystal structure of the phase, then the $Cu^{2+}$ spin ½ ions are in triangular layers, a nice ordered triangular plane spin ½ configuration, the canonical arrangement for a quantum material[67–69]. Recent powder and single crystal XRD structural analysis conforms that the material is hexagonal (the senior author wonders what happened to the Jahn-Teller distortion virtually always seen for $Cu^{2+}$ in octahedral coordination), but that it is something more like an inverse spinel than an ordered 6H perovskite, with the Cu and Sb randomly occupying positions in the dimers and Sb found on the individual octahedral layers (**Figure 4**), a $Ba_3(Sb)(SbCu)O_9$ configuration in spite of the fact that the formula suggests otherwise. In this structure, the magnetic spin ½ $Cu^{2+}$ ion is disordered in the dimers with a nonmagnetic ion, $Sb^{5+}$. This material has been proposed to be a quantum spin liquid candidate with an S = ½ triangular lattice, tested by magnetic susceptibility and neutron scattering experiments down to 0.2 K. There is no magnetic ordering observed, while the Curie-Weiss temperature is calculated to be -55 K. The specific heat shows a T-linear dependence, and a relatively large Sommerfeld constant was found, 43.4 mJ/mol-$K^2$ below 1.4 K[70], although the material is insulating (which should lead to a Sommerfeld constant near 0). A study of $Ba_3CuSb_2O_9$ single crystals, which also have disordered Cu and Sb in the dimers, showed that the hexagonal symmetry and the presence of dynamical Jahn-Teller distortions remain down to 3.5 K. There is a very big investment in the "quantum materials" properties of this material by the materials physics community, a community that does not appear to be disturbed by the extensive Cu/Sb disorder present in the dimers. When the Sb/Cu ratio is slightly different from 2/1, a distorted orthorhombic structure is observed. Further, the stoichiometry and symmetry of the phase appear to be dependent on the synthesis temperature[71]. A particularly interesting set of materials of this type are the $Ba_3CuOs_2O_9$ double hexagonal perovskites, for which both the structural symmetry and the magnetic properties depend profoundly on whether the material is prepared with an ordered $M'M_2$ array or one where an inverse hexagonal perovskite of the type $Ba_3(Os)(CuOs)O_9$ is made[72].

In the same structural vein, we also have $Ba_3CoSb_2O_9$, another dimer-based quantum material that crystallizes in the hexagonal space group $P6_3/mmc$. That is, it is another classical dimer hexagonal perovskite, with the lattice parameters a = 5.857 Å and c = 14.459 Å[73–77]. This material, nominally consisting of $Co^{2+}$ and $Sb^{5+}$ in an ordered array is an ideal effective spin = ½ (that is, when the orbital and spin contributions to the moment are both considered, the effective spin of the $Co^{2+}$ turns out to be ½, Co is unique among the 3*d* transition metals in that the orbital contribution to the magnetic moment is not quenched) triangular lattice antiferromagnet. The extensive continuum of available energies seen for the spin system above the spin wave excitations by neutron scattering, is claimed to be clear evidence that $Ba_3CoSb_2O_9$ is close to the proximity of a quantum spin liquid state in a 2D triangular lattice system[73–82]. As for $Ba_3CuSb_2O_9$, there is a big investment in this material by the materials physics community.

Much more obscure, but none-the-less chemically interesting, with substantially different magnetic properties, are $Ba_3MoCr_2O_9$ and $Ba_3WCr_2O_9$[83]. These materials are fun from a chemical perspective because Cr, Mo and W are in the same column in the periodic table and yet Cr is in a nominally 3+ state, while Mo and W are formally in a 6+ ($d^0$) state. ($Cr^{3+}$ is arguably the mother of all strongly Hund's rule coupled ions, and is therefore quite stable, but when combined with Ba and O, Cr can be in more highly oxidized valence states. These materials appear to require an extremely reducing ambient for synthesis). The crystal structures are classical dimerized hexagonal perovskites like we have been describing here, and while the W variant is argued to be chemically ordered with Cr in the dimers, displaying a "dimer like" magnetic susceptibility, the Mo variant is said to be disordered, with Cr in an oxidation state greater than 3+, displaying a classical Curie-Weiss magnetic behavior. Further work on the chemistry, structures and properties of these materials would be of interest.

We now consider the group of compounds where the isolated M octahedra are occupied by 3+ ions. The only distorted hexagonal perovskite having the M formal valence = 3 in the 6H family of ruthenates is monoclinic $Ba_3ScRu_2O_9$. (Sc is an electropositive, non-magnetic, 3+ ion in oxides.) Some structures can also be stabilized in the orthorhombic space group *Cmcm*, such as $Ba_3CuRu_2O_9$ or the low temperature phase of $Ba_3CoRu_2O_9$[84–89], but the most studied 6H hexagonal perovskites are those where M is a rare earth element. Chemically, this is presumably because they are relatively easy to make, but from the point of view of magnetic properties they are interesting because the rare earth ions are found in a triangular planar lattice, and all the rare earths behave magnetically differently due to their *f* electron count. The crystal structures adopt the usual hexagonal space group $P6_3/mmc$ for all compounds in the series (M = Y, La to Lu). The magnetic ordering temperatures are summarized in **Table 1**.

In the M = $M^{3+}$ group, M = $In^{3+}$ $Ba_3InRu_2O_9$ is reported to be hexagonal. For this material, researchers claim that due to the strong hybridization of Ru and O orbitals, the $Ru_2O_9$ dimers are better explained as being molecular units, where an s = ½ moment is delocalized in the dimers. This kind of consideration - whether the face sharing units are best considered as "molecules" with distinct molecular orbitals, or simply should be thought of as the sum of the electron configurations of the transition metals that often occupy them - is one of the fundamental interesting aspects of the hexagonal perovskites. A spin glass transition was observed at 3.5 K, which can either be due to geometric frustration or disorder on the magnetic lattice (i.e. some so-far undetected structural mixing of In and Ru the hexagonal perovskite M+M' sites). In addition, the static magnetic "ground state" (Theoretically a T = 0 Kelvin state but from a practical perspective materials physicists mean the lowest temperature state that they can attain in their measurements.) in this material, studied by μSR, has been said by the researchers involved to indicate that $Ba_3InRu_2O_9$ is more complicated than a conventional spin glass[90–92].

In the case of $M^{4+}$ cations, there are several reported compounds, such as $Ba_3BiRu_2O_9$, $Ba_3TiRu_2O_9$ and $Ba_3ZrRu_2O_9$. $Ba_3BiRu_2O_9$ is particularly unusual, in that although the average formal valence of Bi is claimed to be 4+, charge disproportionation of 2Bi4+ to $Bi^{3+}$ + $Bi^{5+}$ is said to be present, similar to what is seen in $BaBiO_3$. The evidence for this exotic state (said to be a "negative U" state in materials physics) is Bi L3-edge X-ray absorption near-edge (XANES) spectroscopy. This compound also undergoes a spin gap opening at T*=176 K, meaning that above 176 K, $Ba_3BiRu_2O_9$ is a S=1 magnetic dimer system, and a spin gap state (in other words that there is a distinct energy range of forbidden spin excitations, which are usually continuously available through "magnons") emerges below 176 K. The effect on magnetoelastic transitions in $Ba_3BiRu_2O_9$ by external physical or internal chemical pressure has also been studied. It is claimed to reduce the spin gap opening temperature[93,94]. Moreover, a small amount of La-doping (~10%) on the Bi-site suppresses the spin gap and also eliminates the sharp structural and magnetic transitions that accompany it, suggesting that the spin-gap behavior is strongly associated with the unstable charge of $Bi^{4+}$ in this material[95,96]. ($Ba_3BiIr_2O_9$ also shows the opening of a spin gap, at 74 K. Magnetoelastic transitions induced by external physical or internal chemical pressure reduce the gap opening temperature[93,94]). More conventional in this group is $Ba_3TiRu_2O_9$, which shows structural disorder between Ti and Ru in both the corner-sharing octahedra and face-sharing dimers[97,98]. The system is magnetically disturbed by a substantial amount of Ti/Ru chemical disorder, with the Curie-Weiss temperature of −29.5 K and the effective magnetic moment of 1.82 μB/f.u. A bifurcation of the zero field cooled and field cooled magnetic susceptibility is observed below 4.7 K, which, along with the structural disorder, indicates that $Ba_3TiRu_2O_9$ is a compositionally-disordered spin-glass system. The material is a semiconductor with an activation energy for charge transport of approximately 0.14 eV[98].

6H hexagonal oxide perovskites are not the only $Ba3MM'_2O_9$ materials known in this family. Materials with dimers can also have rhombohedral (9R) rather than hexagonal (6H) stacking along c. $Ba_3MnRu_2O_9$ (Mn has the formal valence of $Mn^{2+}$, apparently) adopts a 9R structure while the majority of the $Ba_3MRu_2O_9$ phases form the 6H structure. When the A site ion is too small, hexagonal perovskites are not stable, and $Ca_3MnRu_2O_9$ for example adopts the standard *Pbnm* space group seen often for orthorhombically distorted conventional perovskites. In this material, Mn and Ru are randomly distributed over the M-sites. $Ba_3MnRu_2O_9$, unsurprisingly due to its magnetic disorder, shows a short range spin glass transition at 30 K while for the orthorhombically distorted conventional perovskite $Ca_3MnRu_2O_9$ a glass transition is seen at 150 K [99] also a likely indication of Mn-Ru disorder. Similarly, $Sr_3CaRu_2O_9$ and $Sr_3CaIr_2O_9$ have monoclinic structures, distortions of the conventional perovskite structure, with space group $P2_1/c$ [100,101].

Moving now to iridates, of particular interest as quantum materials due to the fact that Ir is a 5*d* ion and therefore its orbital energies are known to be influenced spin orbit coupling, the materials and crystal structures are similar to what is seen in the $Ba_3MRu_2O_9$ system. Again, the oxidation state of M in the $Ba_3MIr_2O_9$ series of dimer-based hexagonal perovskites can range from 1+ to 4+, resulting in the total number of electrons donated by Ir to the $Ir_2O_9$ dimer ranging from 11 to 8, respectively. For monovalent M ions, both $Ba_3NaIr_2O_9$ and $Ba_3LiIr_2O_9$ compounds crystallize in the hexagonal space group $P6_3/mmc$, however, only $Ba_3NaIr_2O_9$ undergoes a structural transition to a monoclinic of a hexagonal perovskite, space group $C2/c$, below 100 K. For divalent M ions, hexagonal structures are stabilized when M = Mg, Zn and Ni, and distorted monoclinic structures are observed in the cases of M = Ca and Sr. While all rare earth M elements result in structures in the hexagonal space group $P6_3/mmc$ in the $Ba_3MRu_2O_9$ ruthenates, the structural trends in the $Ba_3MIr_2O_9$ system are quite different, in that the small rare earth ions (M = Sm-Lu) result in hexagonal structures, while the larger rare earth ions (eg. La and Nd) lead to structures in the distorted monoclinic space group $C2/c$. Finally, analogous to $Ba_3BiRu_2O_9$, $Ba_3BiIr_2O_9$ also adopts a monoclinic structure. This material displays a spin gap opening at a lower temperature of 74 K, as described briefly above.

In the $Ir_2O_9$ dimers, pure $Ir^{5+}$ is expected to have a non-magnetic ground state. For $Ir^{4+}$ dimers, small the observed effective moments in a localized spin picture can be taken to indicate the presence of strong AFM interactions between the $Ir^{4+}$ ions within the dimers. For the case of the mix between $Ir^{4+}$ and $Ir^{5+}$, magnetic ordering of $Ir^{4+}$ is reported to occur at 10 K in $Ba_3ScIr_2O_9$ and $Ba_3InIr_2O_9$[102]. $Ba_3MIr_2O_9$ compounds with M = Rare Earth (RE)) adopt the usual ordered hexagonal perovskite $P6_3/mmc$ space group, except for RE= La and Nd (which have the monoclinically distorted $C2/c$ space group) as demonstrated in **Figure 5**. $Ba_3MIr_2O_9$ shows antiferromagnetic ordering at 4 K for M = Y, at 5.1 K for M = Lu and 2.0 K for M = Tb[103]. The fact that the 6H iridates are studied less frequently in the materials physics community than the 6H ruthenates are is very likely due to the difficulty in growing single crystals of easily accessible size for iridates.

The $Ba_3InIr_2O_9$ hexagonal perovskite has been proposed to be a quantum spin liquid, where the unpaired electrons are localized on a mixed valence state between $Ir^{4+}$ and $Ir^{5+}$ within the $Ir_2O_9$ dimers. The gapless ground state and persistent spin dynamics that remain active down to 20 mK, probed by magnetic susceptibility, heat capacity, neutron diffraction, NMR and μSR, are cited as evidence. As shown in **Figure 6**, no long range magnetic ordering or spin freezing down to 20mK is seen in this S = ½ dimer $Ir_2O_9$ compound[104].

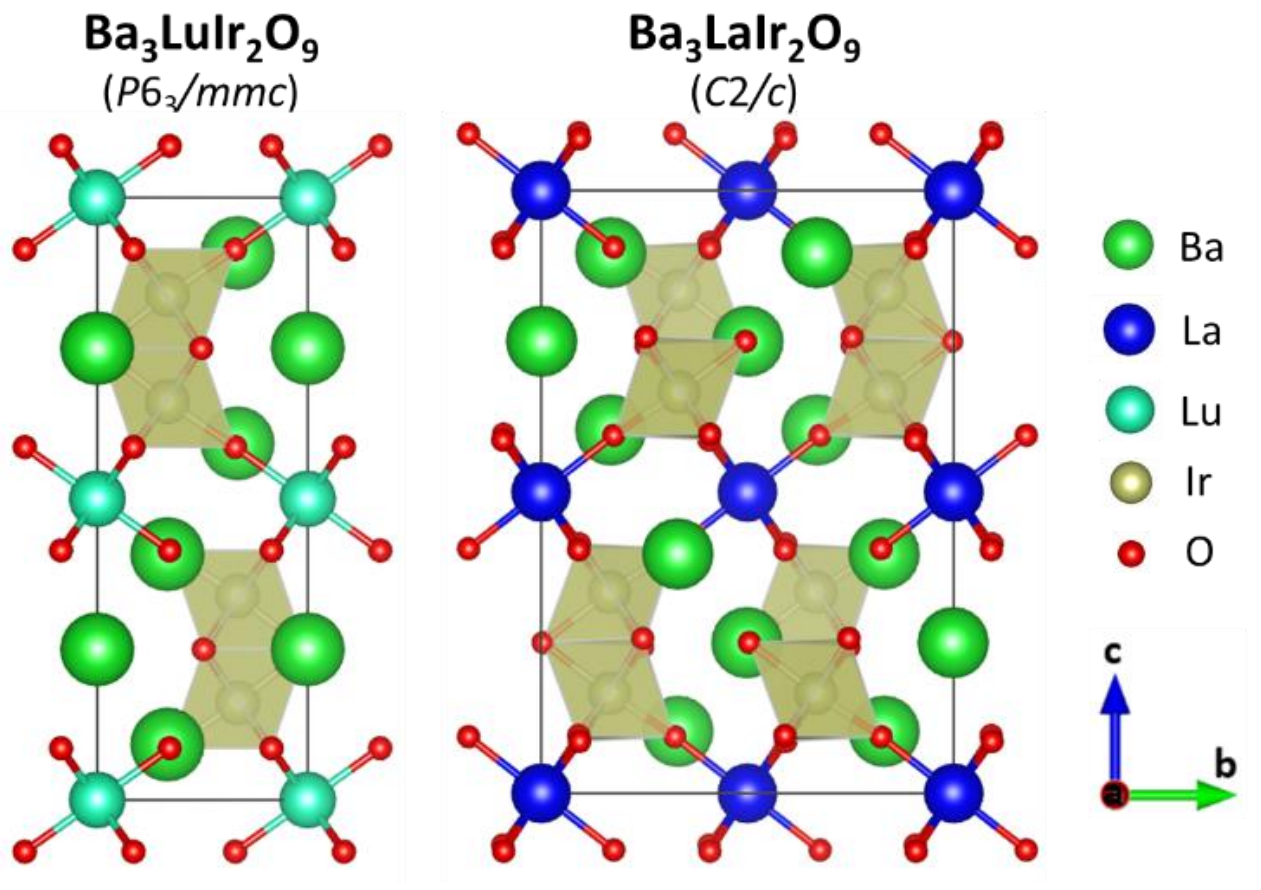

Figure 5. The crystal structures of the $Ba_3LuIr_2O_9$ and monoclinic $Ba_3LaIr_2O_9$ "hexagonal perovskites". The $Ir_2O_9$ dimers are shaded. The rare earth ions are shown by teal or blue balls.

$Ba_3IrTi_2O_9$, where the magnetism comes from $Ir^{4+}$, a spin ½ ion, reportedly in the isolated triangular layers, (the $Ti^{4+}$ is non-magnetic) crystallizes in the hexagonal space group $P6_3/mmc$, has been proposed to be an embodiment of the triangular Heisenberg-Kitaev model, and has been used to study the interplay between geometric and exchange frustration[105,106]. Materials physicists argue that $Ba_3IrTi_2O_9$ is a spin ½ Mott insulator[107] and that Kitaev exchange (Usually a weak kind of magnetic exchange that loses out to ordinary antiferromagnetic-exchange derived ordering. Materials physicists use the term Mott insulator to explain cases where the electrons are largely confined to their individual atomic orbitals, e.g. that they are localized in solids, rather freely from the perspective of the senior author of this paper, because such a classification seems to him to ignore the necessity for M-O-M orbital overlap extending through a whole network for the appearance of metallic conductivity. He would not classify many double perovskites that physicists believe to be Mott insulators, as actually being Mott insulators for example. ) destabilizes the 120̊ ordering expected for Heisenberg spins on a triangular lattice, and thus leads to the formation of a Z2-vortex quantum spin liquid state, tested by spherical neutron polarimetry[106–108]. Magnetic susceptibility displays a very strong antiferromagnetic interaction with the Curie-Weiss temperature of -130 K, but according to the researchers involved, no magnetic ordering is observed down to 23 mK. The frustration index is (= $\Theta_{CW}/T_M$ where $\Theta_{CW}$ is the Curie Weiss theta and $T_M$ is the temperature of the magnetic ordering. Materials are expected to magnetically order at a ratio between 1 and 10) therefore calculated to be larger than 5000, strongly implying the presence of a quantum spin liquid ground state[105–107,109–111]. Although there may be Ir/Ti chemical disorder in both the $MO_6$ octahedra and the $M'_2O_9$ dimers, the researchers involved claim that the material is not a spin glass due to the absence of spin freezing in magnetic susceptibility measurements, heat capacity, and µSR measurements. Further, DFT calculation based researchers report that a Dirac-type nodal line should be present along the A-L direction in *k*-space (i.e. reciprocal space), and is "symmetry-protected", meaning that it absolutely has to be there due to the material's symmetry, suggesting the possible existence of emerging quantum phenomena[108,112] (i.e. a novel spin-orbital singlet state). The senior author of this paper believes none of this by the way, because such predications in complex materials are always "within the model calculated" in his view, although he is happy to defer to the views of the large number of researchers involved in the study of this material.

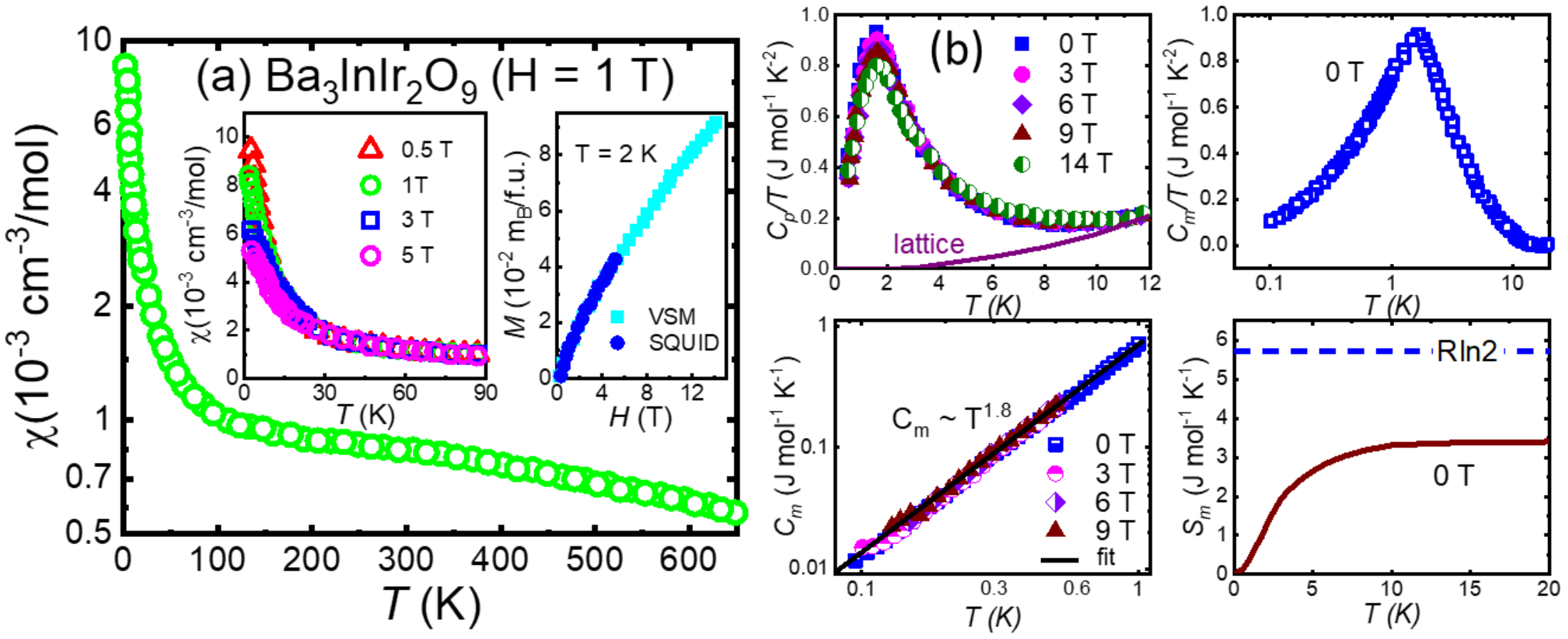

Figure 6. (a) Temperature-dependent magnetic susceptibility in the Quantum Spin Liquid candidate $Ba_3InIr_2O_9$ and (b) heat capacity measurements and magnetic entropy calculation in this $Ba_3InIr_2O_9$ material [104].

The reader can easily see from what we summarize here that the dimer-based hexagonal perovskites are a mainstay of quantum materials research. Structural disorder is sometimes detected for these materials and sometimes not, and as frequently is the case in materials physics, disorder is not universally believed to be a bad thing when it comes to quantum materials. It is only considered significant when its influence on the quantum properties can be clearly detected through physics characterization[113–115].

We now briefly consider a few complex dimer-based hexagonal perovskites. The first of these is very highly structurally disordered, but in an interesting way. The complex disordered dimer material $Ba_3Fe_{1.56}Ir_{1.44}O_9$ [116] adopts the hexagonal space group *P*3*m*1, with the lattice parameters $a$ = 5.740 Å and $c$ = 14.160 Å as shown in **Figure 7**. Interestingly, the site disorder between the Fe and Ir occurs within the dimers, where the dimerized sites are said to be unequally occupied by Fe and Ir, and in one of the triangular layers of octahedra but not the other. This kind of crazy uneven occupancy is what reduces the symmetry from the usual space group for 6H perovskites. The symmetry of this material is such that the complicated chemical disorder results in a non-centrosymmetric polar structure. The valence states of $Fe^{3+}$ and $Ir^{5+}$ were confirmed by X-ray absorption near edge spectroscopy (XANES). $Ba_3Fe_{1.56}Ir_{1.44}O_9$ is semiconducting and it has a near room temperature antiferromagnetic ordering transition of 270 K.

Table 1. The magnetic ordering temperatures (in degrees Kelvin) of the $Ba_3MRu_2O_9$ and $Ba_3MIr_2O_9$ (M = Y, La-Lu) hexagonal perovskites. [47,84,117–121] PM: paramagnetic.

| M ions | $T_C$ or $T_N$ of $Ba_3MRu_2O_9$ | $T_C$ or $T_N$ of $Ba_3MIr_2O_9$ |
|---|---|---|
| Y | 4.5 | 4 |
| La | 6.0 | 14.6 |
| Ce | PM | PM |
| Pr | PM | PM |
| Nd | 24.0 | 17.4 |
| Sm | 12.5 | 14.4 |
| Eu | 9.5 | 8.2 |
| Gd | 14.8 | 10.2 |
| Tb | 9.5 | 2.0 |
| Dy | 27.8 | 8.9 |
| Ho | 10.2 | 5.6 |
| Er | 6.0 | 6.0 |
| Tm | 8.3 | 5.5 |
| Yb | 4.5 | 5.3 |
| Lu | 9.5 | 5.1 |

The 6-layer hexagonal structure of $Ba_6Rh_{2.33}Yb_2Al_{1.67}O_{15}$ [122] is said to adopt the non-standard 6H space group *P*-6*m*2, with lattice parameters $a$ = 5.854 Å and $c$ = 14.660 Å, as displayed in **Figure 8**. Unlike the "usual" 6H hexagonal perovskites, this material has three different types of M-M' ions present. The crystal structure is slightly different from what is seen in the 6H-$Ba_3MRu_2O_9$ family, in that the face-sharing $Rh_2O_9$ dimers, which are reported to contain only Rh, while a second dimerized layer is said to be occupied by a disordered mix of Al plus Rh. The $YbO_6$ octahedra are confined to the isolated octahedra. To balance the charge, there appear to be oxygens missing in a random way in certain sites in the crystal structure. Given that some of the dimers are heavily Al-rich, the structure may be even more complex that this, with the missing oxygens associated with AlO4 tetrahedra. This is definitely a challenging structural problem to figure out.

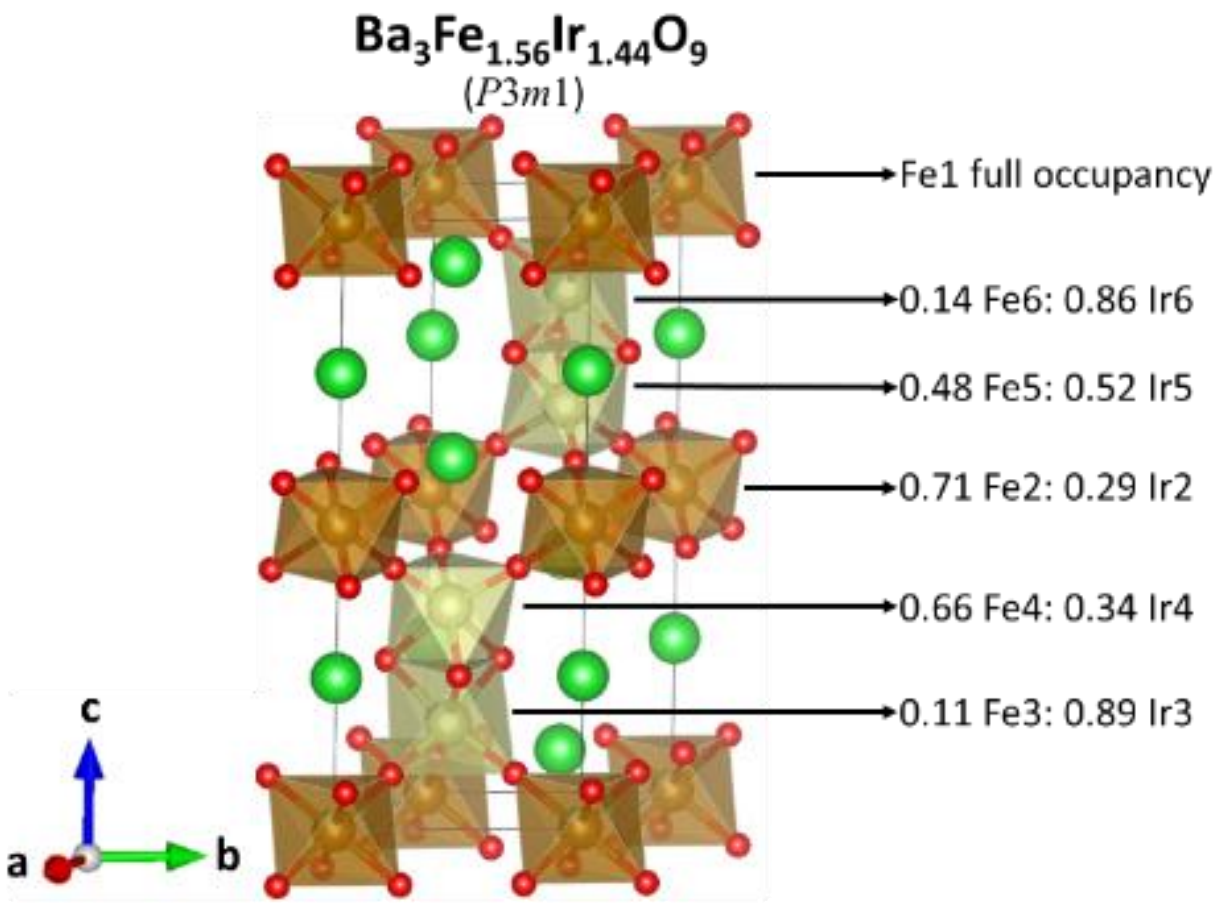

Figure 7. The crystal structure of $Ba_3Fe_{1.56}Ir_{1.44}O_9$

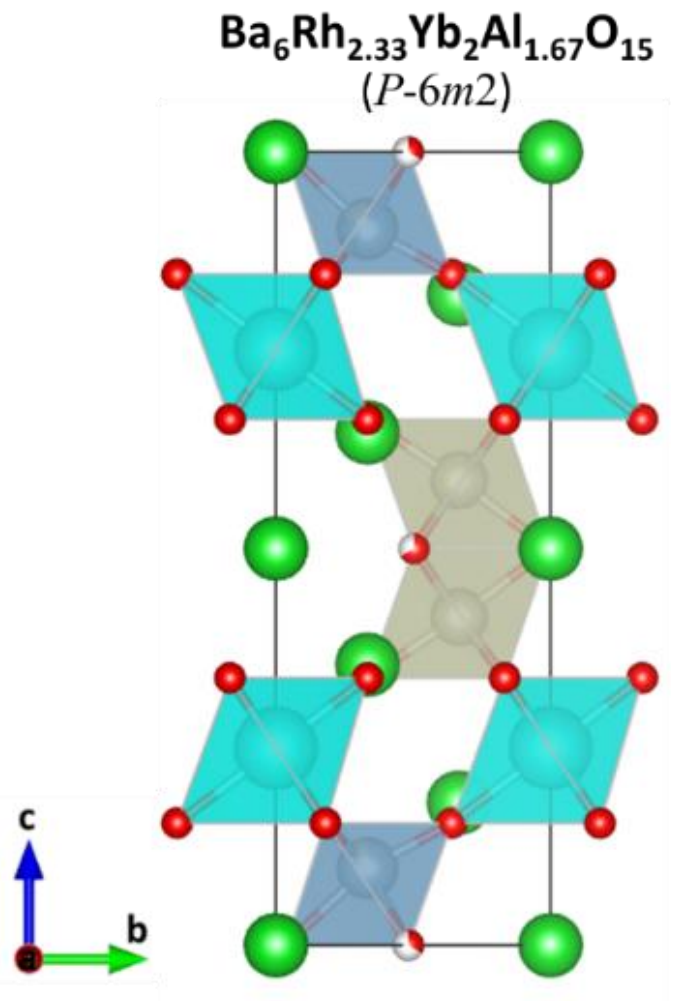

Figure 9. The crystal structure of $Ba_6Rh_{2.33}Yb_2Al_{1.67}O_x$ The $Rh_2O_9$ dimers are shown in light gray, the mixed $(Al,Rh)_2O_9$ dimers in dark gray and the $YbO_6$ octahedra shown in light blue. The oxygens are shown in Red, with partially occupied sites shown as mixtures of red and white spheres. The Ba ions are shown in green.

## 3.2. Trimer-based hexagonal perovskites

When the ratio of M to M' ions is 1:3 instead of 1:2 then variations of hexagonal double perovskites can form with formulas $A_4MM'_3O_{12}$. In these materials, there are three $M'O_6$ octahedra sharing faces with each other and corners with a triangular lattice of $MO_6$ octahedra. These trimer materials are much more rare than the hexagonal perovskite oxides based on dimers, and have been studied proportionally much less frequently. The site in the middle of the M' trimer is crystallographically distinct from that of the two outer octahedra (The central octahedron shares only faces with other octahedra while the two peripheral ones share a face on one side and corners on the other side). Thus it is occasionally noted that the central octahedron can have a different ion in it, described further below, making such materials hexagonal triple perovskites ($A_4MM'_2M''O_{12}$).

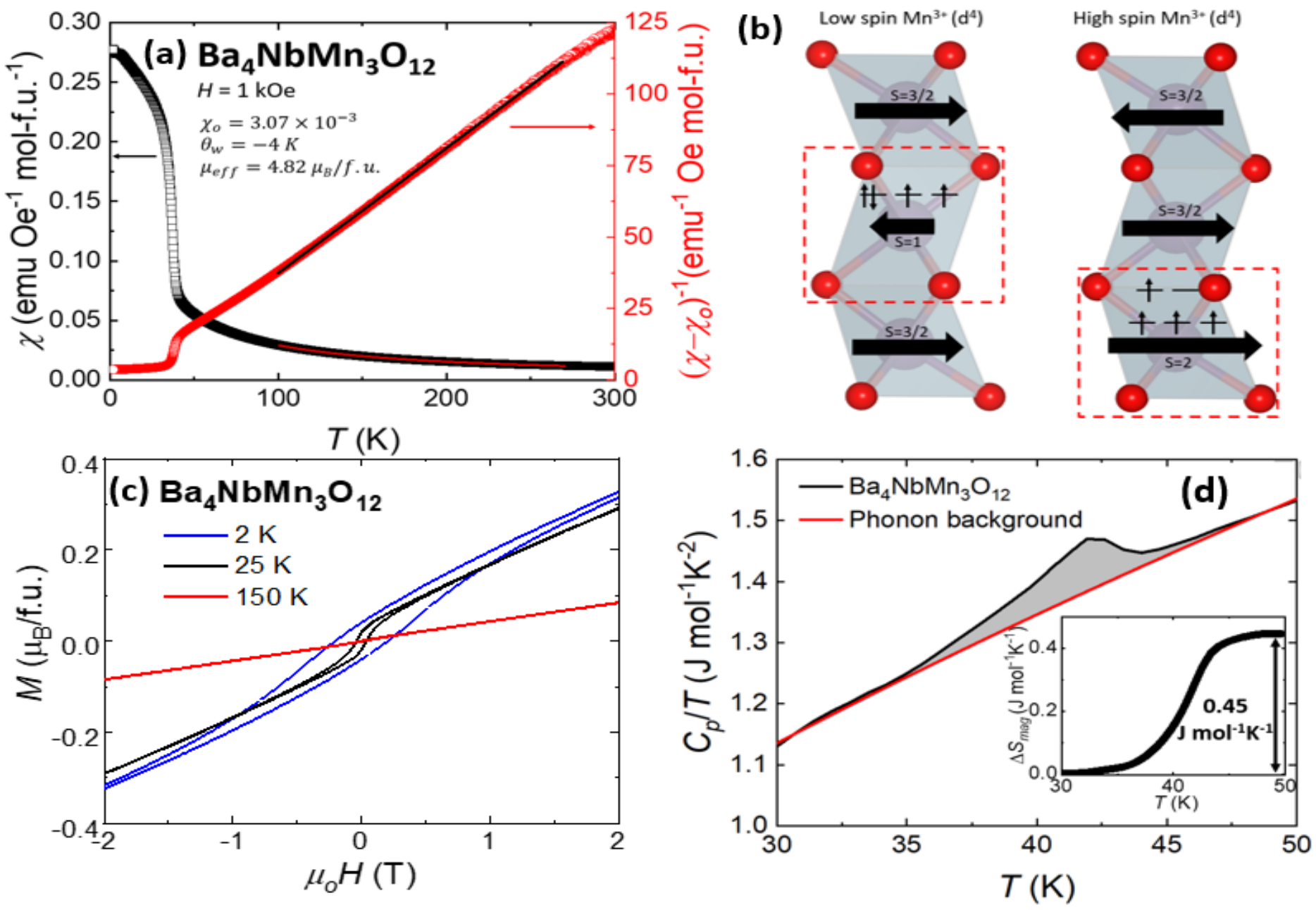

Figure 8. (a) Temperature-dependent magnetic susceptibility of $Ba_4NbMn_3O_{12}$, (b) The two magnetic models account for the observed magnetic moment of 4.82 $\mu B/Mn_3O_{12}$ per trimer, (c) Magnetic hysteresis loops above and below the ferrimagnetic transition temperature T ≈ 42 K, and (d) The heat capacity measurement that shows the partial entropy released at the ordering temperature at T ≈ 42 K [126].

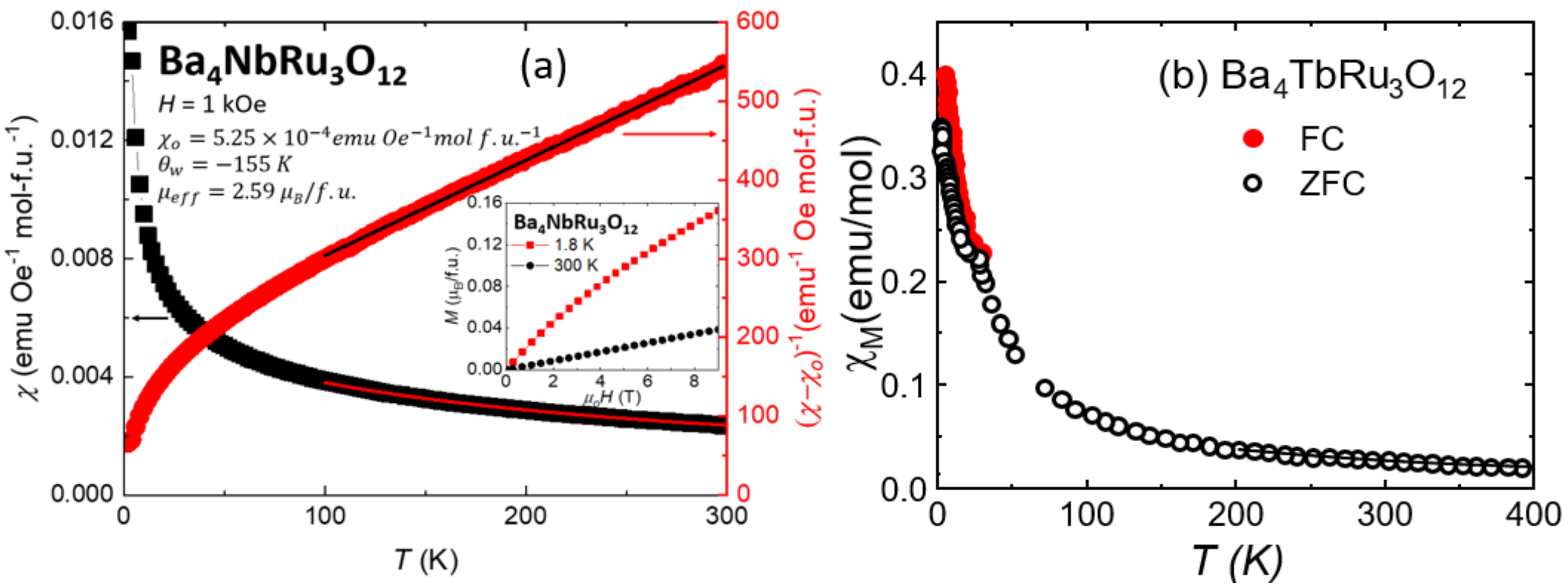

Figure 10. Temperature-dependent magnetic susceptibility of (a) $Ba_4NbRu_3O_{12}$ [127] and (b) $Ba_4TbRu_3O_{12}$[129].

From a property perspective, the trimer-based materials can be interesting magnetically, as the electrons in the trimer are either localized on the ions, as they are when M' is a 3*d* element, or confused about whether they are localized or delocalized within the trimer, as has been proposed when M' is a 4*d* or a 5*d* element[47]. This is particularly of interest to spectroscopists, who have most frequently studied this issue in dimers[123] and theorists (see the review by S.V. Streltsov and D. Khomskii in this journal). The "where are the electrons" question in these trimer perovskites can make them interesting as quantum materials.

In this section, we focus our attention mostly on the three trimer systems $Ba_4NbM_3O_{12}$ (M = Mn, Ru, Ir), which have been investigated in the context of whether the unpaired electrons are localized or delocalized within the $M_3O_{12}$ trimer clusters and whether spin orbit coupling can play an important role in the magnetism of face-sharing hexagonal perovskites. There are only three $Mn_3O_{12}$ trimer compounds reported in the literature in this group so far, $Ba_4MMn_3O_{12}$ (M = Nb, Ce and Pr), all of which crystallize in the hexagonal space group *R*-3*m*. The latter two materials were only structurally reported without any magnetic analysis. This can be due to the multi-valent states of Ce and Pr, causing complicated magnetic properties, or that the phases are not pure enough for the magnetic study. Ce is 4+ (with no *d* electrons an no *f* electrons) in the $BaCeO_3$ conventional perovskite, so there may be no interference of rare earth magnetism in that case, but Pr is a notorious nightmare to deal with in the more familiar oxide perovskites[121,124,125]. Thus, we will discuss in detail the magnetic properties of $Ba_4NbMn_3O_{12}$ instead, where the Nb is 5+ and non-magnetic[126]. There are two $Mn^{3+}$ and one $Mn^{4+}$ present. As shown in **Figure 9a**, $Ba_4NbMn_3O_{12}$ shows an effective moment of 4.82 μB/$Mn_3O_{12}$ trimer and competing antiferromagnetic and ferromagnetic interactions between Mn ions are inferred from the Curie-Weiss temperature of −4 K and the ferrimagnetic ordering transition temperature of approximately 42 K. With the effective moment of 4.82 μB/$Mn_3O_{12}$, two potential magnetic ordering schemes for this material are displayed in **Figure 9b**. Neutron diffraction is necessary in order to solve the magnetic structure and determine the actual magnetic spin configuration. From this magnetic picture, we can see that the magnetism of the $Mn_3O_{12}$ trimer can be well explained by a localized electron model, meaning that the unpaired electrons reside locally at each ion site, instead of being delocalized within the $Mn_3O_{12}$ trimer unit. The magnetic hysteresis loops above and below the transition temperature are shown in **Figure 9c**. Finally, the heat capacity measurement in **Figure 9d** demonstrates the presence of an entropy loss at the around 42 K, however, only a small amount of the total magnetic entropy expected for an S = 2 Heisenberg system is recovered. So, it is fair to ask, where is the missing entropy?

Moving to a 4*d* transition metal based material, the $Ba_4MRu_3O_{12}$ system for example, $Ba_4NbRu_3O_{12}$ and $Ba_4TaRu_3O_{12}$ are reported to adopt the hexagonal space group *R*-3*m*. The crystal structures consist of a trimer-based $Ru_3O_{12}$ cluster. The exchange interaction between $Ru^{3+}$ and $Ru^{4+}$ is claimed to be the origin of the magnetic ground state[127,128]. A few other compounds in this trimer family have been structurally reported, but the magnetic properties are not yet investigated. A comprehensive study of $Ba_4LnRu_3O_{12}$ - type materials has been reported, however. In the case of non-magnetic rare earth ions (Ln = Y, La, Lu), the magnetism comes purely from the magnetic $Ru_3O_{12}$ trimers. But for other rare earth ions, the observed magnetic moments must arise from both the magnetic rare earth ions and the $Ru_3O_{12}$ trimers. The magnetic ordering temperatures of the $Ba_4LnRu_3O_{12}$ series are summarized in **Table 2**. If the trimers could be made fully nonmagnetic, then these compounds would have nicely separated triangular layers of rare earth ions, and be good candidates for low temperature quantum magnets.

The magnetic properties of isolated $Ru_3O_{12}$ trimers in $Ba_4NbRu_3O_{12}$ have been compared with the combined magnetic moments of the magnetic rare earth $Tb^{4+}$ and the $Ru_3O_{12}$ trimer in $Ba_4TbRu_3O_{12}$. The temperature-dependent magnetic susceptibilities of both compounds are shown in **Figure 10**. The Curie-Weiss fitting of the data from 200-400 K in $Ba_4TbRu_3O_{12}$ results in a Curie-Weiss temperature of -57 K and an effective moment of 8.78 μB/f.u. This large magnetic moment has been attributed to the magnetic sum of $Tb^{4+}$ (7.94 μB) plus three individual $Ru^{4+}$ (S = 1, μeff = 2.83 μB), which should be about 9.33 μB/mol-f.u. The observed moment is argued to be within experimental error of the expected moments from localized spins. From neutron diffraction, the magnetic structure of $Ba_4TbRu_3O_{12}$ was solved. It is displayed in **Figure 11**.

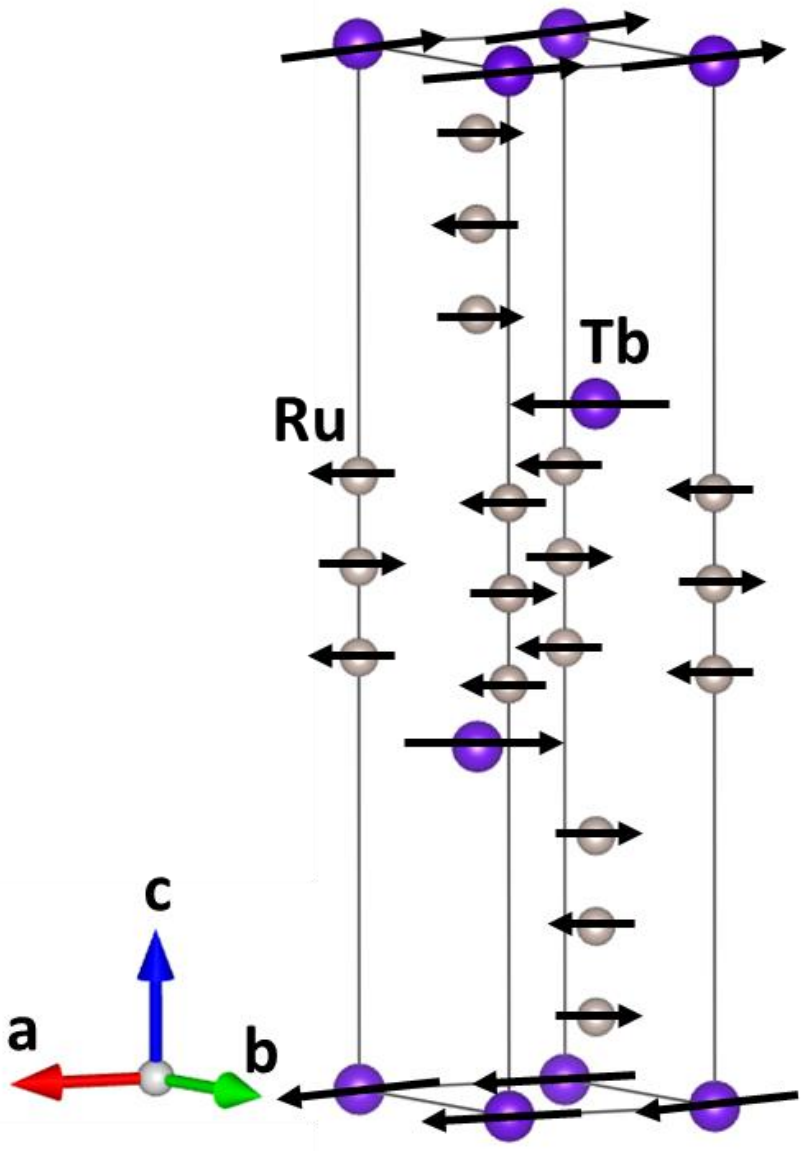

Figure 11. The magnetic structure of $Ba_4TbRu_3O_{12}$ at 2.5 K, solved from neutron diffraction[129]. Arrows show the directions of the ordered magnetic spins at low temperatures.

Table 2. The summary of the magnetic ordering temperatures in $Ba_4MRu_3O_{12}$ and $Ba_4MIr_3O_{12}$ [129–131] (M = Y, La-Lu) systems. DM: diamagnetic, CW: Curie-Weiss, VV: Van-Vleck.

| M ions | $T_C$ or $T_N$ of $Ba_4MRu_3O_{12}$ | $T_C$ or $T_N$ of $Ba_4MIr_3O_{12}$ |
|---|---|---|
| La | 6.0 | DM |
| $Ce^{4+}$ | CW | 10.5 |
| $Pr^{4+}$ | 2.4 | 35 |
| Nd | 11.5 | CW |
| Sm | 3.5 | VV |
| Eu | 4.0 | VV |
| Gd | 2.5 | CW |
| $Tb^{4+}$ | 24 | 16 |
| Dy | 30 | CW |
| Ho | 8.5 | CW |
| Er | 8.0 | CW |
| Tm | 8.5 | CW |
| Yb | 13 | CW |
| Lu | 8.0 | DM |

In the case of $Ba_4NbRu_3O_{12}$, since $Nb^{5+}$ is non-magnetic, the only magnetic moments come from the isolated $Ru_3O_{12}$ trimer. Analysis of the magnetic susceptibility leads to a Curie-Weiss temperature of -155 K and an effective moment of 2.59 μB/mol-f.u. This observed moment is between the value of spin ½ (1.73 μB) and spin 1 (2.83 μB) per trimer, and cannot therefore be due to free-ion-like localized magnetic moments. The Curie-Weiss temperature is relatively large and negative, indicating the presence of strong antiferromagnetic interactions between the Ru ions. A freezing temperature to a glassy state was observed at around 4 K. The corresponding frustration index is calculated to be larger than 30, implying that $Ba_4NbRu_3O_{12}$ is a strongly frustrated magnet. A recent neutron study employing both elastic and inelastic neutron scattering (in preparation by T. Halloran et. al) shows that only 1/3 of the spins freeze below 4 K, with the remaining spins still fluctuating. Hence, $Ba_4NbRu_3O_{12}$ is proposed to be a spin liquid or even a quantum spin liquid candidate. As shown in **Figure 12**, both the trimer energy level diagram, 1, and the molecular orbital diagram, 2, result in the overall spin ½ in each $Ru_3O_{12}$ trimer.

Spin-orbit coupling is the interaction of a non-zero electron spin with its orbital motion. This is a relativistic effect, and is proportional to the fourth power of the atomic number Z[132,133]. Thus, it has the strongest effect in 5*d* transition metals (eg. Ir) compared to 3*d* ions (e.g. Mn) and 4*d* ions (e.g. Ru). $Ba_4BiIr_3O_{12}$, which crystalizes in the distorted monoclinic space group $C2/c$, has $Ir_3O_{12}$ trimers bridged by corner-sharing $BiO_6$ octahedra[95,134]. It contains an unusual $Bi^{4+}$ formal oxidation state, but actually has charge disproportionation of the $Bi^{4+}$ into $Bi^{3+}$ and $Bi^{5+}$. This material shows an anomaly at T* = 215 K. This transition is not seen at the spin gap opening temperature, as is seen in $Ba_3BiRu_2O_9$ or $Ba_3BiIr_2O_9$. The Curie-Weiss temperature is found to be -15 K and the effective moment is 0.22 μB/Ir. The relatively low effective moment comes from strong spin orbit coupling for the 5*d* Ir transition metal. Antiferromagnetic interactions within the $Ir_3O_{12}$ trimer are claimed to lead to the formation of an S = ½ doublet. The change in the magnetic state at T* is claimed to result in a structural distortion to balance the amount of energy for the formation of the S = ½ doublet. The Ir-Ir distance within the trimers is relatively short, 2.6 A, implying the possibility of direct metal-metal bonding[40,96,135,136].

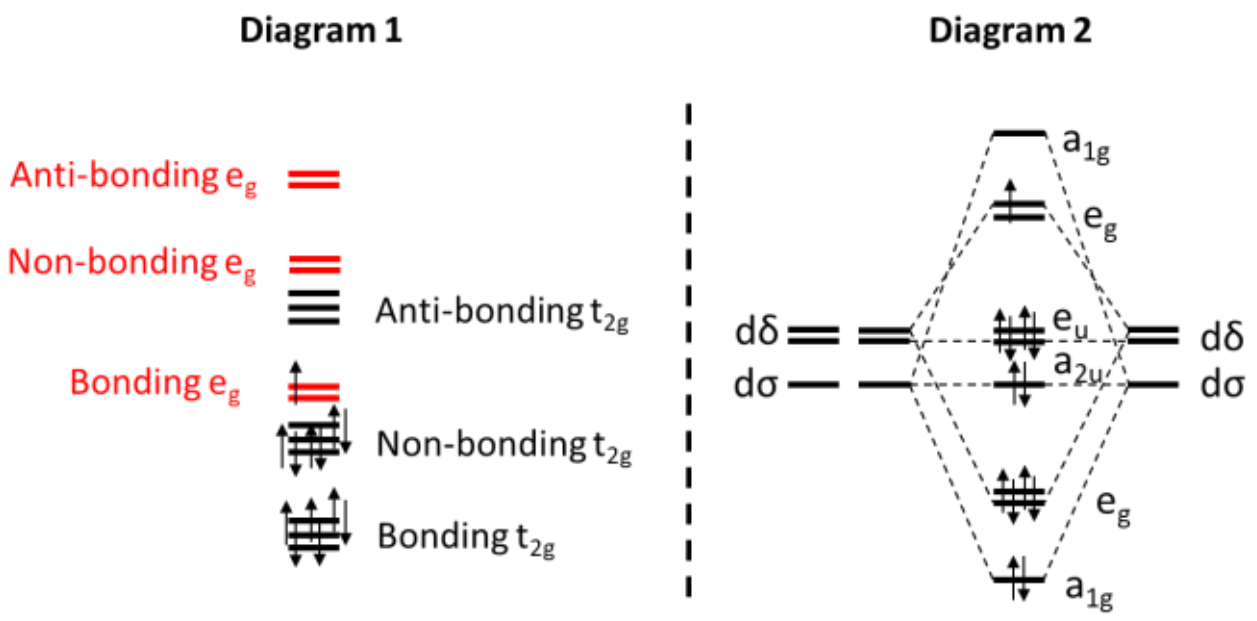

Figure 12. 1 The possible energy level diagram of a trimer[137] and 2 the alternative molecular orbital diagram [47]. The electron counting shown is taken from the example of $Ba_4NbRu_3O_{12}$, which has a total of 13 electrons in each $Ru_3O_{12}$ trimer. Diagrams are redrawn from the references.

While $Ba_4RERu_3O_{12}$ ($RE^{3+}$) compounds show magnetic ordering at low temperatures, $Ba_4REIr_3O_{12}$ compounds frequently remain paramagnetic down to 1.8 K. The exception is the case where the RE is $Ce^{4+}$. $Ba_4CeIr_3O_{12}$ magnetically orders at 10.5 K, while $Ba_4CeRu_3O_{12}$ is paramagnetic down to 1.8 K[131,138,139]. The ordering temperature and effective magnetic moments of the two trimer-based series, $Ba_4RERu_3O_{12}$ and $Ba_4REIr_3O_{12}$ ($RE^{3+}$), are summarized in **Table 2**.

### 3.3. A structure based on both trimers and dimers

There is one example that we can find of a material that has both trimers and dimers in its hexagonal oxide perovskite structure. This material is 10H- $Ba_5Fe_4NiO_{13.5}$, synthesized by the decomposition of nitrate salts, which crystallizes in the hexagonal space group $P6_3/mmc$ with lattice parameters $a$ = 5.771 Å and $c$ = 24.581 Å. The crystal structure consists of the stacking of clusters of three face-sharing octahedra (trimers) corner-sharing with clusters of two face-sharing octahedra (dimers). Structural refinement shows that the Ni ions (in gray in **Figure 13**) preferentially occupy the central octahedron of the trimer and Fe ions (in brown in **Figure 13**) occupy the other sites. Fitting of the temperature-dependent magnetic susceptibility results in a Curie-Weiss temperature of -255 K and an effective moment of 5.45 μB per formula unit. It is not known how the magnetic moments are distributed. The ordering of the Ni and Fe in different octahedral sites implies that new phases with this kind of unusual stacking might be stabilized for other transition metal combinations as well[140]. We note that if the oxygen content of this material is really 13.5 per formula unit then it is quite oxygen deficient. The missing oxygen, which is on the order of 10% of the total, is either missing randomly in the structure, or the coordination in one or more of the layers of ions is not actually octahedral, which is definitely a possibility for $Fe^{3+}$. More on this concept can be found in subsequent sections of this summary.

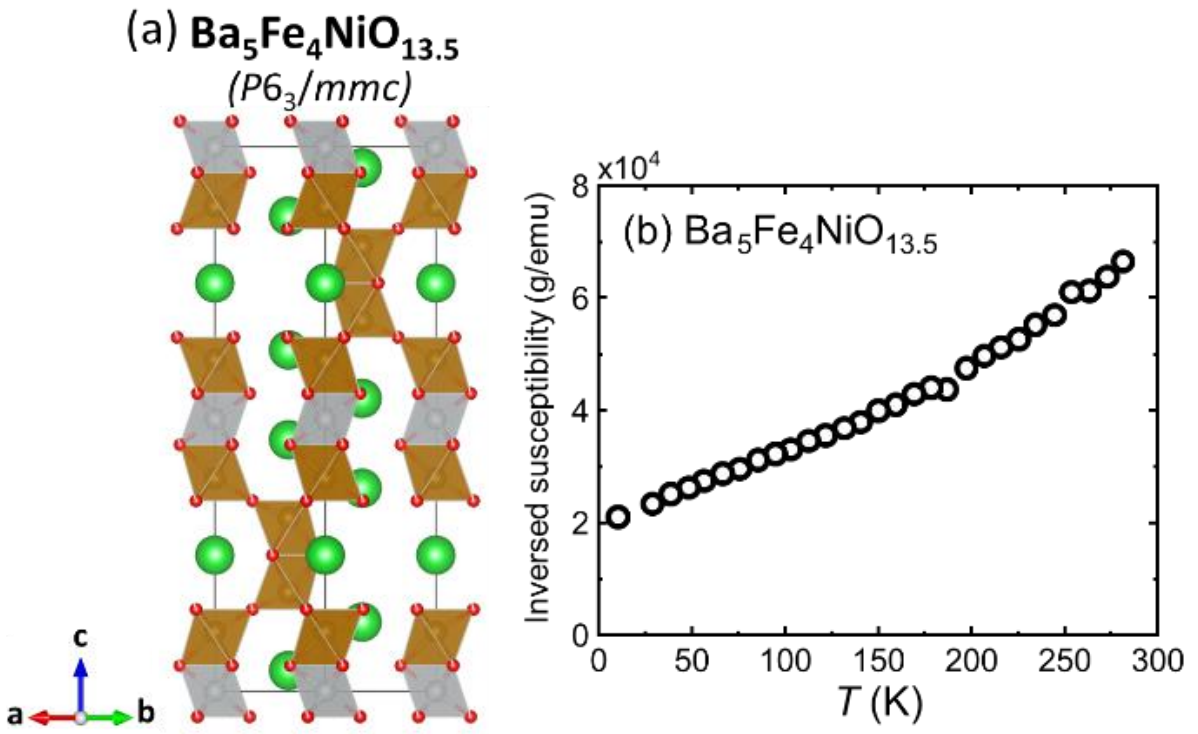

Figure 13. (a) The crystal structure of 10H- $Ba_5Fe_4NiO_{13.5}$, (b) Inverse magnetic susceptibility of $Ba_5Fe_4NiO_{13.5}$ under the applied magnetic field of 1 kOe [140].

### 3.4. 12R structures based on trimers

An example of this kind of structure is $Ba_4Fe_3NiO_{12}$. In this material the transition metals are ordered and there does not appear to be an issue with the oxygen content. While the two end members $BaNiO_3$ and $BaFeO_3$ crystallize in the 2H and 6H hexagonal perovskite structures, a composition in between, $Ba_4Fe_3NiO_{12}$, adopts the 12R hexagonal structure, space group $R$-3$m$ with lattice parameters $a$ = 5.66564(7) Å and $c$ = 27.8416(3) Å)[141]. Similar to other trimers discussed above, the crystal structure of $Ba_4Fe_3NiO_{12}$ consists of three face-sharing octahedra (the trimers) separated by a layer of single $FeO_6$ octahedra, as shown in **Figure 14a**. A combination of synchrotron and neutron diffraction data refinements shows that the $Fe^{3+}$ ions (in brown) occupy the layer of corner-sharing octahedra while $Fe^{4+}$ (again in brown) and $Ni^{4+}$ (in gray) occupy the face-sharing octahedra. The Ni occupy the octahedron in the middle of the trimer. A small amount of oxygen vacancies (~5%) were found at the intersections of the corner-sharing octahedra only. The magnetic susceptibility is claimed to show a weak ferro/ferrimagnetic transition that couples $Fe^{3+}$ and $Fe^{4+}$ at around 200 K, confirmed by magnetic hysteresis loops, as shown in **Figures 14b-c**.

Similar to $Ba_4Fe_3NiO_{12}$, $Ba_4Ni_2Ir_2O_{12}$ also crystallizes in a hexagonal 12R perovskite structure, space group $R$-3$m$, with the lattice parameters a = 5.7309(2) Å and $c$ = 28.6318(12) Å, as shown in **Figure 15a**. The crystal structure consists of face-sharing $Ir_2NiO_{12}$ trimers and layers of single $NiO_6$ octahedra. This material is claimed to have the unusual mix of $Ni^{2+}$ and $Ni^{4+}$ in its structure, according to XANES measurements[142] (**Figure 15b**). Fitting of the magnetic susceptibility data yields a Curie-Weiss temperature of -33 K and the effective moment of 4.40 μB/f.u, with data as shown in **Figure 15c**. This observed magnetic moment is consistent with the oxidation states of Ni seen in the XANES measurements, namely that the $Ni^{2+}$ ions occupy the sites within the layer of single $NiO_6$ octahedra while the $Ir_2NiO_{12}$ trimers consist of an alternating face-sharing $Ir^{5+}O_6$-$Ni^{4+}O_6$-$Ir^{5+}O_6$ arrangement. Thus, the magnetism has been attributed to the contributions of both the $Ir_2NiO_{12}$ trimer and the two unpaired electrons from the $Ni^{2+}$ ions.

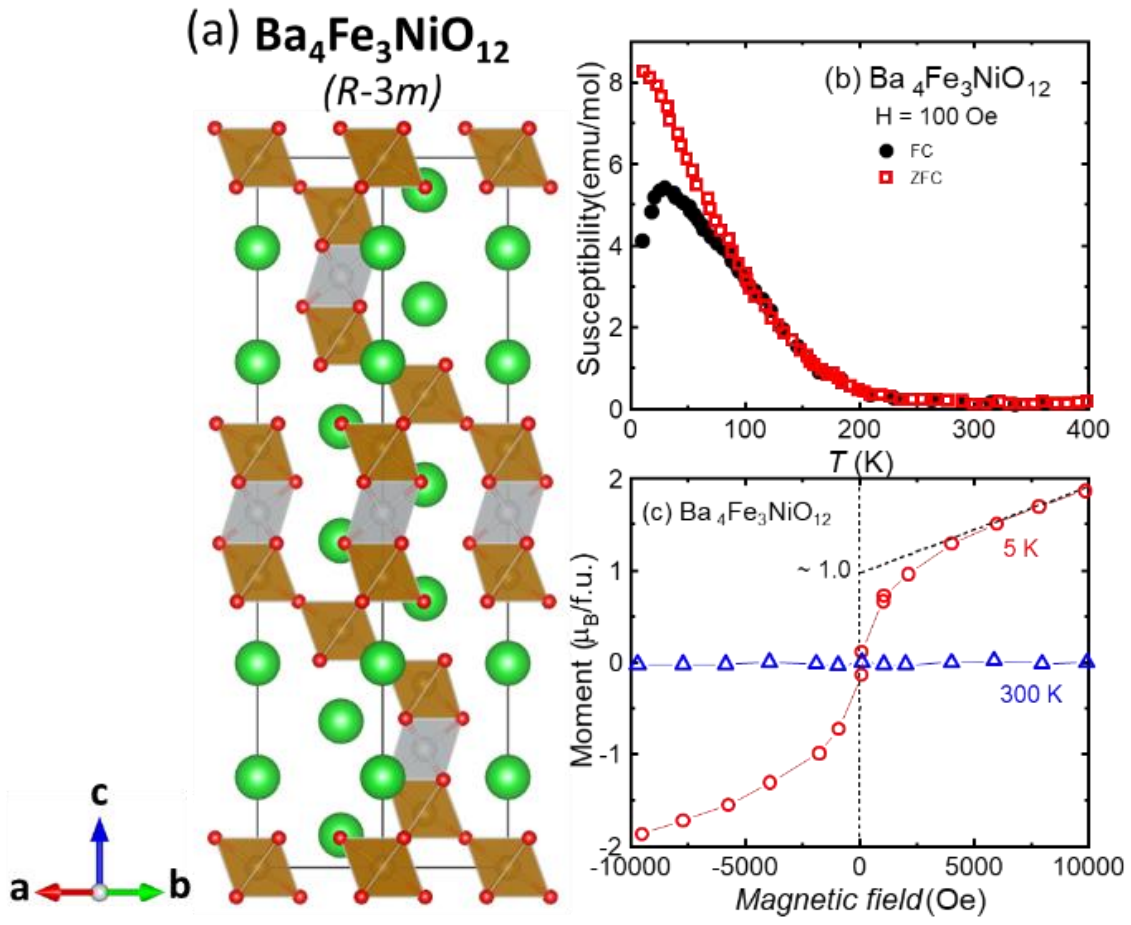

Figure 14. (a) The crystal structure of 12R- $Ba_4Fe_3NiO_{12}$, (b) Temperature-dependent FC/ZFC magnetic susceptibility of $Ba_4Fe_3NiO_{12}$ under the applied magnetic field of 100 Oe, and (c) The field-dependent magnetic hysteresis loops of $Ba_4Fe_3NiO_{12}$ at 300 K and 5 K [141].

A polycrystalline sample of $Ba_4NbIr_3O_{12}$ has been synthesized by the solid state method. The material also has a 12R-type structure. It crystalizes in the rhombohedral space group $R$-3$m$, with lattice parameters $a$ = 5.7827(2) Å and $c$ = 28.7725(9) Å. The structure of $Ba_4NbIr_3O_{12}$ is based on a triangular planar geometry of $Ir_3O_{12}$ trimers. The material displays a very low effective moment of 0.80 μB/f.u, reported to be due to the strong spin-orbit coupling effect of the 5$d$ Ir-ions. The Curie-Weiss temperature is -13 K, taken to indicate antiferromagnetic interactions between Ir ions. However, there is no magnetic ordering down to 0.35 K, confirmed by heat capacity measurements. The heat capacity of $Ba_4NbIr_3O_{12}$ shows a linear upturn in Cp/T below 2 K under applied magnetic fields of 0 and 1 T. The upturn is suppressed under 2 and 3 T, as previously observed in the reported spin liquid material $H_3LiIr_2O_6$[143]. The heat capacity data for $Ba_4NbIr_3O_{12}$ below 2 K fits the power law $Cp(T) = kT^{\alpha}$, where k = 0.1 and $\alpha \approx 3/4$. The power being smaller than 1 is taken to suggest a possible spin liquid state in this material[143]. The crystal growth of this material with slight off-stoichiometry and disorder has been reported[144,145]. As shown in **Figure 16**, $Ba_4Nb_{0.8}Ir_{3.2}O_{12}$ experiences site disorder between Nb and Ir in the isolated octahedra only. A recently reported crystal growth method produces large 3mm size crystals. Magnetic measurements on these single crystals has been found to be consistent with a possible spin liquid state, and is also consistent with the report on the stoichiometric polycrystalline samples. With the availability of large crystals, the potential quantum spin liquid state in this material can be studied by thermal conductivity, μSR, inelastic neutron scattering and heat capacity down to 50 mK.

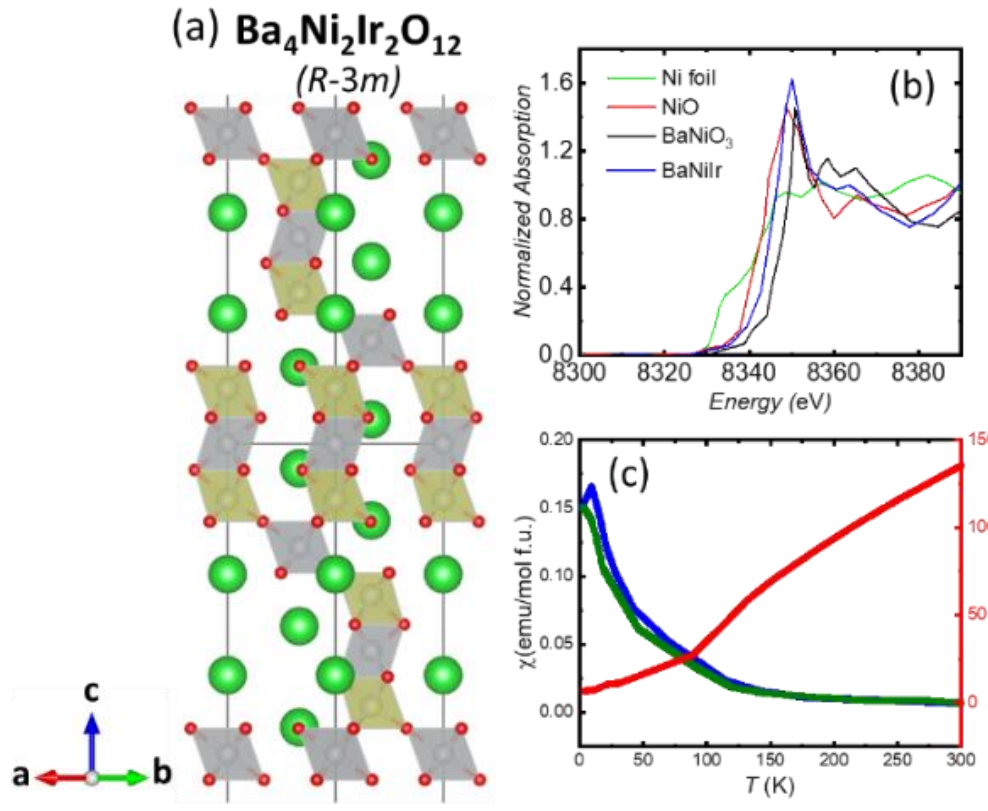

Figure 15. (a) The crystal structure of 12R- $Ba_4Ni_2Ir_2O_{12}$, (b) XANES characterization of $Ba_4Ni_2Ir_2O_{12}$ and other standards, Ni, NiO, $BaNiO_3$ employed to confirm the coexistence of $Ni^{2+}$ and $Ni^{4+}$in this material. (c) Temperature-dependent FC/ZFC magnetic susceptibility of $Ba_4Fe_3NiO_{12}$ under the applied magnetic field of 1 T [142].

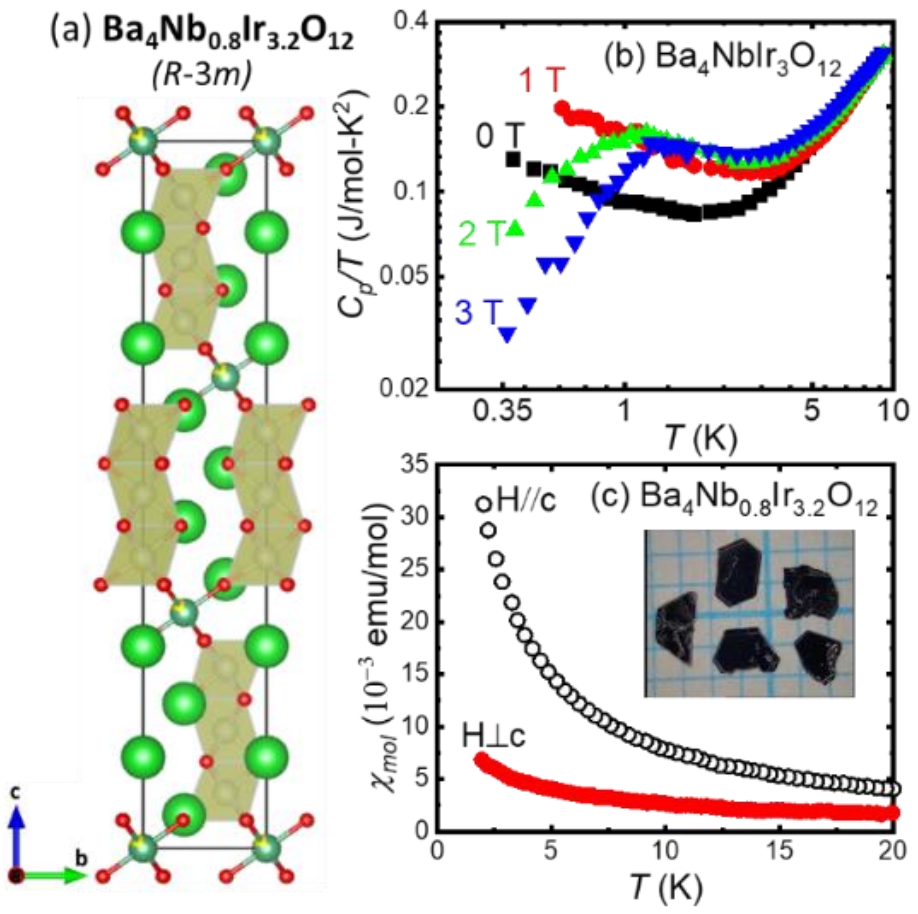

Figure 16. (a) The rhombohedral crystal structure of $Ba_4Nb_{0.8}Ir_{3.2}O_{12}$ [144]. Ba are light green and oxygens are red. The brown trimers contain only Ir and the isolated octahedra contain disordered mixtures of Nb and Ir. (b) Heat capacity data in $Ba_6Y_2Rh_2Ti_2O_{17}$ under different magnetic fields. (c) Temperature-dependent magnetic susceptibility of $Ba_4Nb0._8Mn_{3.2}O_{12}$ crystals. The inset shows the large size crystals up to 3mm.

## 3.5. $A_{4-n}A'_nMM'_2O_{12}$: Missing an octahedron in a trimer

This is a derivative of the $Ba_4MRu_3O_{12}$ family. The crystal structure is very similar to the 12L-$Ba_4NbRu_3O_{12}$ hexagonal perovskite, except that there is a missing $M'O_6$ octahedron in the middle of the trimers. The crystal structure of $Sr_4CoRe_2O_{12}$, an example of a material in this $A_{4-n}A'_nMM'_2O_{12}$ family,[146,147] is shown in **Figure 17a**. (*n* can range from 0 to 2, with the different A site ions chosen to accommodate the charge on the M,M' sublattice.) The inverse DC magnetic susceptibility and the AC susceptibility for $Sr_4CoRe_2O_{12}$ are displayed in **Figures 17b-c**. The magnetic ordering temperature, Curie-Weiss temperature and effective moment of a few compounds in this series are known; they are summarized in **Table 3**. Two general observations can be made about such phases: (1) the missing octahedron results in a material that has only corner shared octahedra and thus can also be considered as a layered oxide perovskite based on (111) cubic rather than (100) cubic layers, and (2) if the magnetism is really acting as described in the publication, then the $Co^{2+}$ present is undergoing a spin state transition on cooling and in addition the magnetic ordering is not frustrated ($T_M$ and $\Theta_{CW}$ are nearly equal).

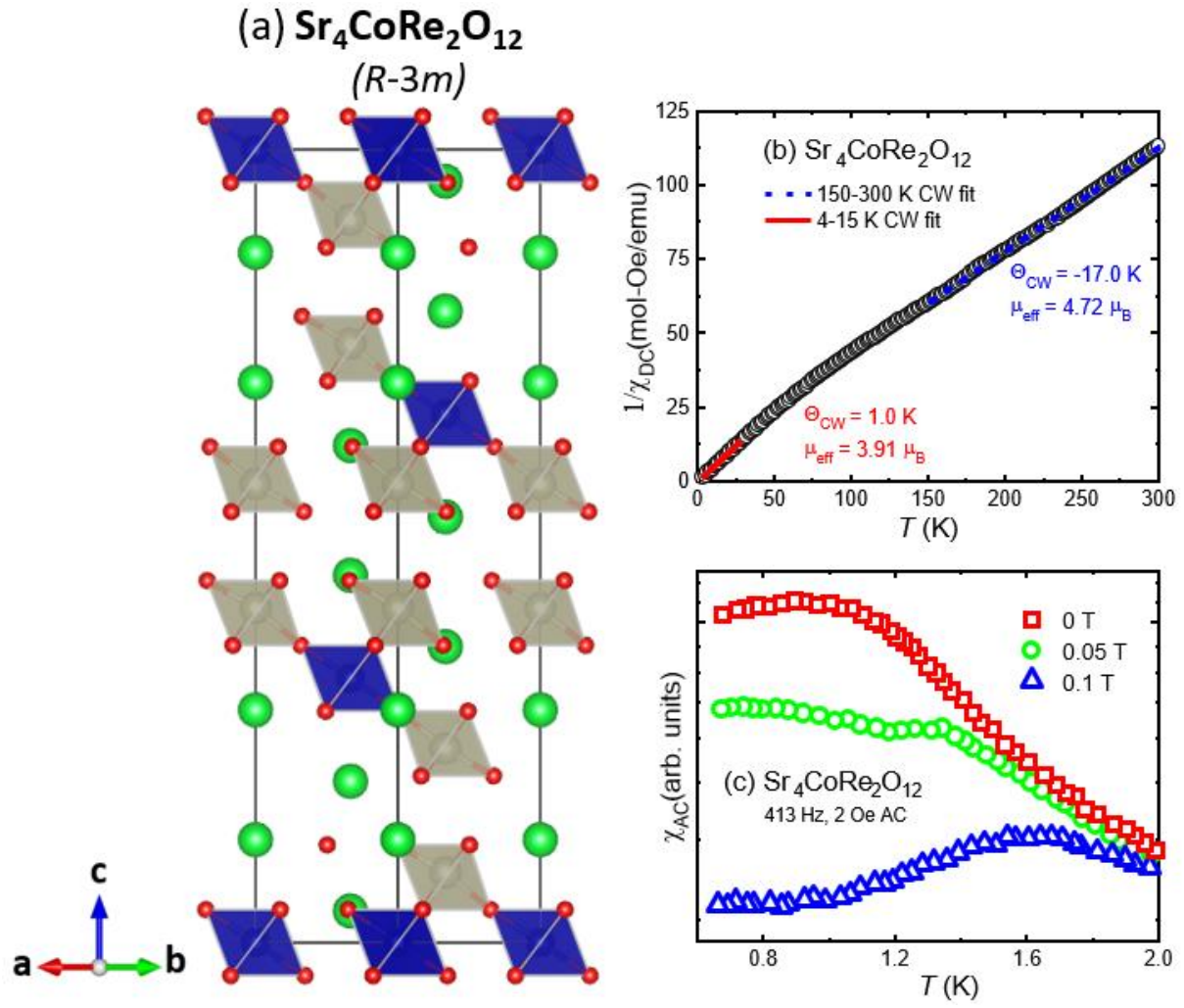

Figure 17. (a) The crystal structure of $Sr_4CoRe_2O_{12}$, (b) Inverse DC magnetic susceptibility of $Sr_4CoRe_2O_{12}$ under the applied field of 5 kOe, and (c) AC susceptibility of $Sr_4CoRe_2O_{12}$ under the 2 Oe-excitation field[146]

Table 3. The magnetic ordering temperatures, Curie-Weiss temperature and effective magnetic moment per formula unit for some examples of materials in the $A_{4-n}A'_nMM'_2O_{12}$ series.

| Compounds | $T_M$ (K) | $\Theta_{CW}$ (K) | $\mu_{eff}$ (μB) |
|---|---|---|---|
| $Sr_4CoRe_2O_{12}$ | 1.0 | 1.0 | 3.91 |
| $Sr_2La_2CoW_2O_{12}$ | 1.26 | 1.5 | 3.95 |
| $Ba_2La_2CoW_2O_{12}$ | 1.28 | 1.6 | 4.01 |
| $Ba_3LaCoReWO_{12}$ | 0.83 | 0.78 | 3.89 |
| $Ba_4CoRe_2O_{12}$ | 0.87 | 0.61 | 3.87 |
| $Ba_2La_2NiW_2O_{12}$ | 6.2 | 25.5 | 3.19 |
| $Ba_2La_2MnW_2O_{12}$ | 1.7 | -10.7 | 5.73 |

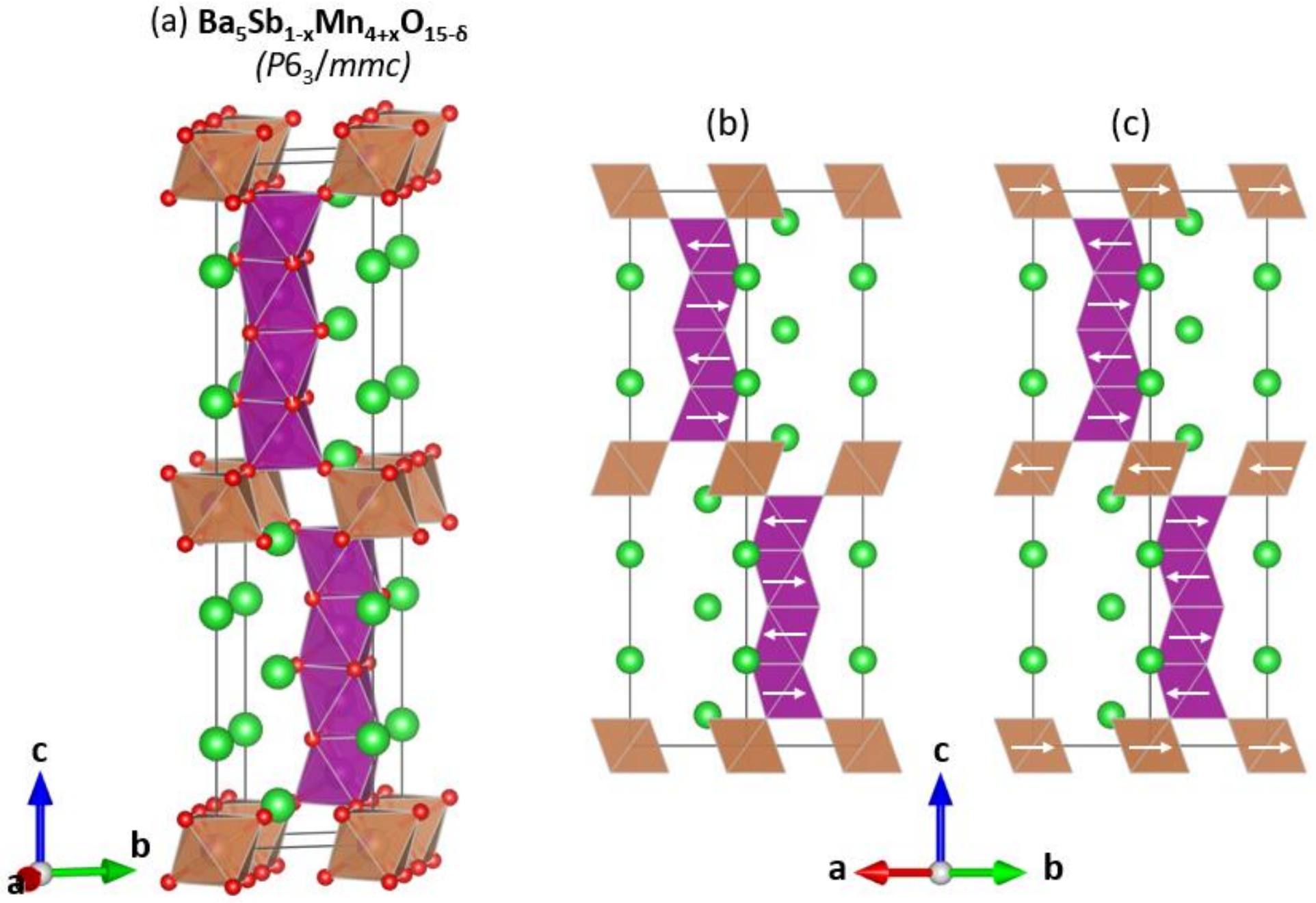

Figure 18. (a) The crystal structure of 10H- $Ba_5Sb_{1-x}Mn_{4+x}O_{15-\delta}$, (b) and (c) Two proposed magnetic structures for $Ba_5Sb_{1-x}Mn_{4+x}O_{15-\delta}$ [148].

### 3.6. Tetramers

Very few tetramer-based materials are known; we describe one here and another in a later section where we consider the presence of tetrahedral in the crystal structures. The return of $Sb^{5+}$ to our systems allows for the formation of $Ba_5Sb_{1-x}Mn_{4+x}O_{15-\delta}$, which is a 10 layer, 10H-hexagonal perovskite based on tetramers. It crystallizes in the space group $P6_3/mmc$ with lattice parameters $a$ = 5.7095(1) Å, and $c$ = 23.4866(3) Å. It has different stacking from $Ba_5Fe_4NiO_{13.5}$. The crystal structure of $Ba_5Sb_{1-x}Mn_{4+x}O_{15-\delta}$ consists of four face-sharing $MnO_6$ octahedra (tetramers) and a layer of corner-sharing disordered $(Sb/Mn)O_6$ octahedra, as shown in **Figure 18**. The magnetic properties of $Ba_5Sb_{1-x}Mn_{4+x}O_{15-\delta}$ ($0.24 \le x \le 0.36$) solid solutions show two distinct magnetic transitions. For example, when x = 0.24, the upper transition at 129 K corresponds to the formation of ferro/ferrimagnetic interactions within the tetramers, and the lower transition at 13 K is the freezing temperature of a glassy state, this latter transition evidenced by the bifurcation of FC/ZFC magnetic susceptibility at low applied magnetic field. This material has an antiferromagnetic Curie-Weiss temperature of -133 K and an effective moment of 3.55 μB/f.u[148].

## 4. Materials with $MO_4$ tetrahedra in the framework

The crystal structure of $Ba_5AlIr_2O_{11}$ was first reported in 1989, but its quantum magnetism has only recently been investigated. $Ba_5AlIr_2O_{11}$ crystallizes in the orthorhombic space group *Pnma*, with the lattice parameters $a$ = 18.8360 Å, $b$ = 5.7887 Å and $c$ = 11.1030 Å, it is a distorted version of a hexagonal perovskite. The crystal structure consists of two face-sharing $IrO_6$ octahedra ($Ir_2O_9$ dimers), corner-sharing with $AlO_4$ tetrahedra. However, within the $Ir_2O_9$ dimers, there are two inequivalent Ir sites, as shown in **Figure 19a**, one that could contain $Ir^{4+}$ ($5d^5$) and the other one containing $Ir^{5+}$ ($5d^4$). In contrast to this local charge picture, the formation of $Ir_2O_9$ dimer orbitals in this material has been detected by Resonant Inelastic x-ray scattering (RIXS). Sharp excitations of electrons within the dimer orbitals were observed by this method, and analyzed by using Density Functional Theory (DFT) and theoretical simulations. The spin-orbit coupling present for the Ir-5$d$ transition metal is said to play an important role in the competing interaction between intra-atomic exchange and covalent bonding in this material. The researchers involved argue that Ir-Ir metal-metal bonds are present, causing a substantial reduction of the magnetic moment. This material undergoes a structural phase transition at 210 K, which is supported by temperature-dependent electrical resistivity measurements (**Figure 19b**). The material is *very* highly resistive, so it is basically an insulator. A clear magnetic transition is at 4.5 K, observed in the magnetic susceptibility, where a clear field dependence to the susceptibility even above the ordering temperature is seen. (**Figure 19c**) [149–153].

The 10L-$Ba_5In_2Al_2ZrO_{13}$ hexagonal perovskite is another example of a material with $MO_4$ tetrahedra in the structural framework. The crystal structure is shown in **Figure 20**. It adopts the space group $P6_3/mmc$ with lattice parameters $a$ = 5.8707(7) Å and $c$ = 24.445(3) Å. In this case the hexagonal symmetry is maintained. A special feature in the crystal structure of this material is where $AlO_4$ tetrahedra share corners with $InO_6$ octahedra, separated by a triangular layer of $ZrO_6$ octahedra. This remarkable crystal structure is the result of the intergrowth along the $c$-axis of $Ba_2InAlO_5$ and $BaZrO_3$ blocks.[154,155] For this material, ALL ions are in their usually highest formal oxidation states and no unpaired $d$ electrons are present, it is thus non-magnetic and therefore no quantum properties are expected.

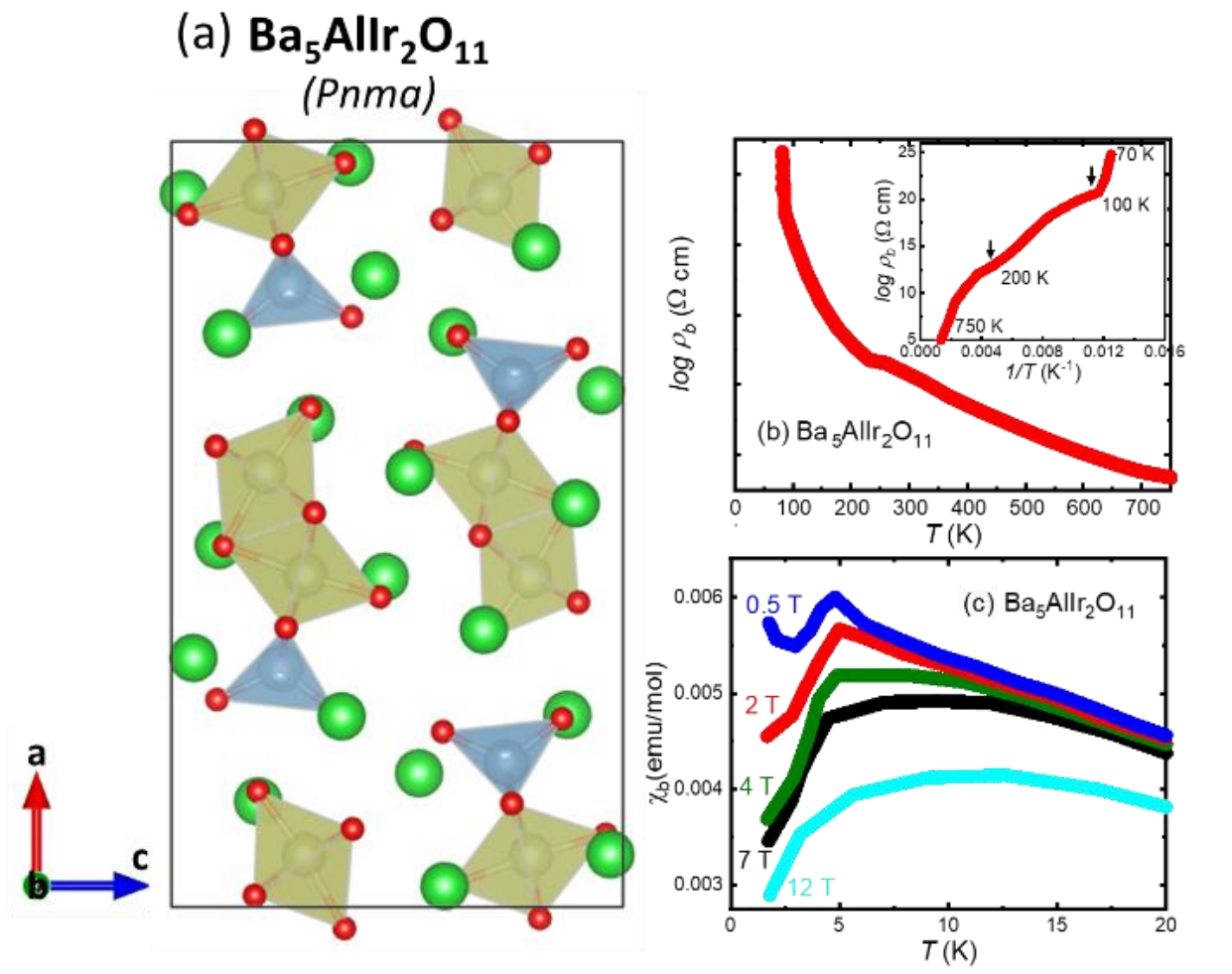

Figure 19. (a) The crystal structure of $Ba_5AlIr_2O_{11}$, with the $Ir_2O_9$ dimers and $AlO_4$ tetrahedra (blue polyhedra); (b) Resistivity measurement to show a subtle structural transition at 210 K, and (c) Temperature dependent magnetic susceptibility showing the magnetic transition at 4.5 K. The symmetry of the crystal structure is low compared to standard hexagonal perovskites [149].

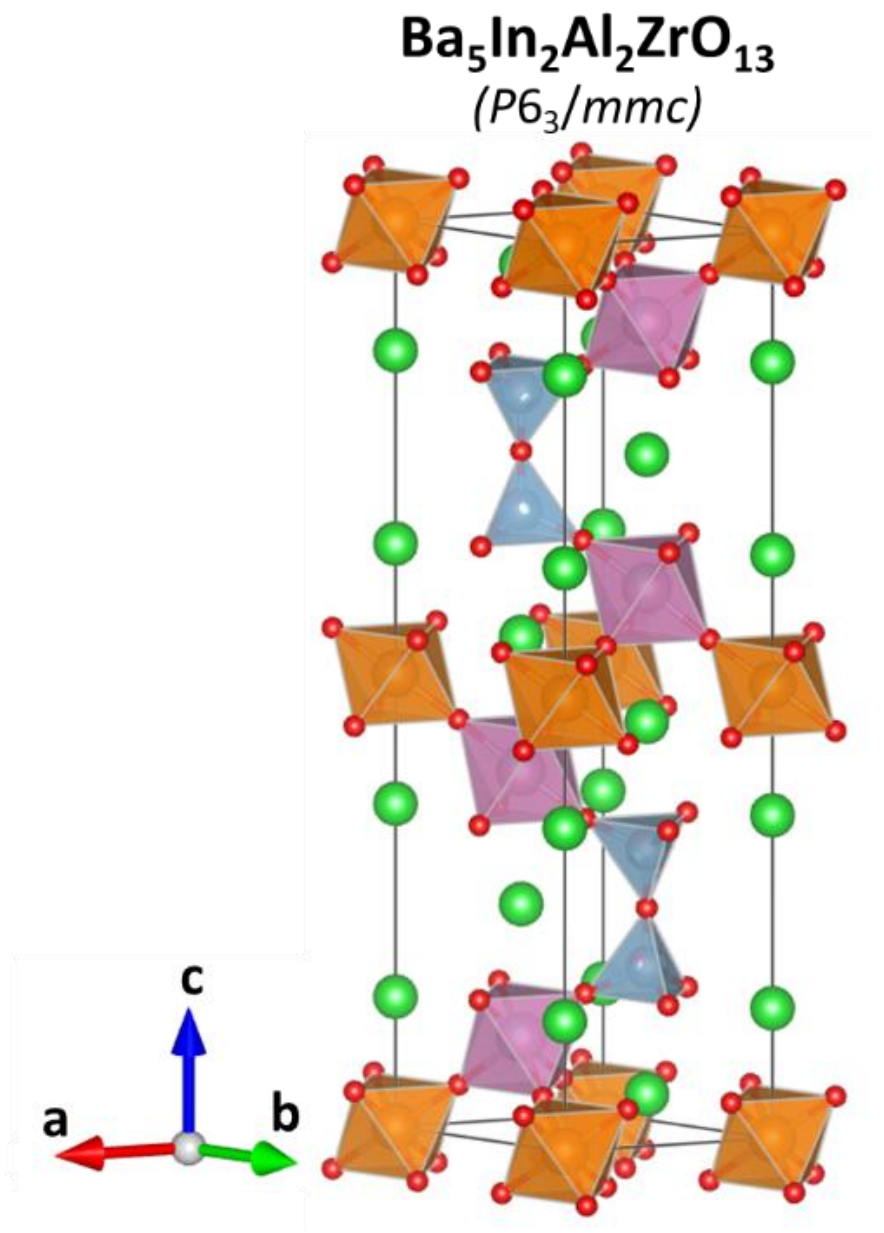

Figure 20. The crystal structure of $Ba_5In_2Al_2ZrO_{13}$. Ba ions are green, oxygens are red, $Al^{3+}O_4$ tetrahedra are shown in blue, $In^{3+}$ octahedra as purple and $Zr^{4+}$ octahedra as orange.

In another structure type consisting of octahedra plus tetrahedra of M ions, 12H-$Ba_6Na_2Ru_2V_2O_{17}$ crystalizes in a hexagonal structure, space group $P6_3/mmc$, with the lattice parameters $a$ = 5.8506(1) Å and $c$ = 29.6241(4) Å. This crystal structure contains a $Ru_2O_9$ dimer, which is corner-sharing with layers of $NaO_6$ octahedra, separated by double sheets of $VO_4$ tetrahedra (**Figure 21**). This nominally complex $Ba_6Na_2M_2X_2O_{17}$ structure type can be stabilized for many possible cation combinations, for example for M = Ru, Nb, Ta and Sb and X = V, Cr, Mn, P, and V [156]. The magnetic properties of many of these materials remain unreported.

Recently, two additional compounds with this crystal structure have reported, with formulas $Ba_6R_2Ti_4O_{17}$ (R = Nd and Y)[157]. In these materials, the triangular layers of $NaO_6$ octahedra are replaced by octahedra of rare earth ions (Nd or Y). Furthermore, the Ti cations occupy both the $TiO_4$ tetrahedra and the face-sharing $Ti_2O_9$ dimers. This is one of the very rare oxides where Ti cations adopt both tetrahedral and octahedral sites in the same structure.

In this complex yet somehow flexible structure, we successfully synthesized the new compound $Ba_6Y_2Rh_2Ti_2O_{17}$. In this material, the only magnetic ion is $Rh^{4+}$ ($4d^5$) in the $Rh_2O_9$ dimers, which helps simplify a potentially complex magnetic picture (**Figure 22a).** Preliminary magnetic analysis shows that the material displays a small effective magnetic moment, which must arise from the Rh ions present, and a negative Curie-Weiss temperature. The transport band gap and optical band gaps are very similar, at 0.16(1) eV, and thus $Ba_6Y_2Rh_2Ti_2O_{17}$ is a small band gap semiconductor. As shown in **Figure 22**, a large upturn in the heat capacity at temperatures below 1 K, suppressed by magnetic fields larger than 2 Tesla, is observed. A very large Sommerfeld-like T-linear term in the specific heat (γ=166 mJ/mol f.u-$K^2$), as is often observed for Heavy Fermion metals, is seen although the material is highly insulating at low temperatures[158]. These results suggest the possibility that a spin liquid state exists at low temperatures in $Ba_6Y_2Rh_2Ti_2O_{17}$.

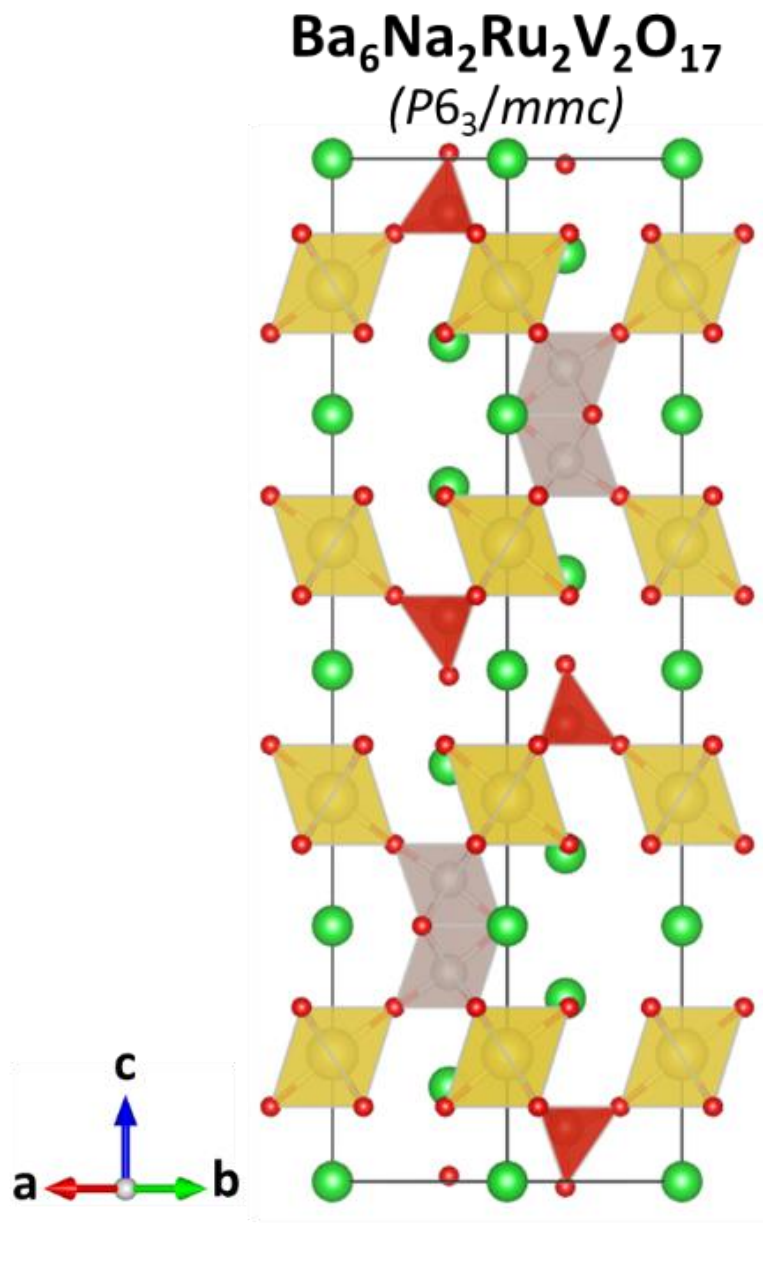

Figure 21. The crystal structure of $Ba_6Na_2Ru_2V_2O_{17}$. Ba = green spheres, oxygen = red spheres, $AlO_4$ tetrahedra depicted in red, $Ru_2O_9$ dimers in brown and individual layers of $NaO_6$ octahedra in yellow.

For another example of a material with mixed tetrahedra and octahedra, 21R-$Ba_7Mn_5Cr_2O_{20}$ crystalizes in a rhombohedral structure, space group $R$-3$m$, with the lattice parameters $a$ = 5.7401(1) Å and $c$ = 50.597(1) Å. The c axis is large for this structure, which consists of three shifted structural blocks. In **Figure 23a**, the crystal structure is shown. It consists of an ordered array of $CrO_4$ tetrahedra (when Cr is formally highly charged, then its radius is small), $MnO_6$ octahedra and face-sharing $Mn_3O_{12}$ trimers. Below 50 K, the material undergoes 2D magnetic ordering, attributed to ferromagnetic super-exchange between the "isolated" $MnO_6$ octahedra and the $Mn_3O_{12}$ trimers[159–161]. Under moderate applied magnetic fields, the material converts to 3D magnetic ordering. Magnetic susceptibility measurements reveal a Curie-Weiss temperature of 38 K and an effective moment of 9.69 μB/f.u., as shown in **Figure 23b**. The presence of ferromagnetism at 5 K is confirmed by the presence of a magnetic hysteresis loop, seen in **Figure 23c**. However, the magnetic behavior in this material is quite complex since it contains different magnetic ions with different oxidation states, such as $Cr^{5+}$, $Mn^{2+}$ and $Mn^{4+}$.

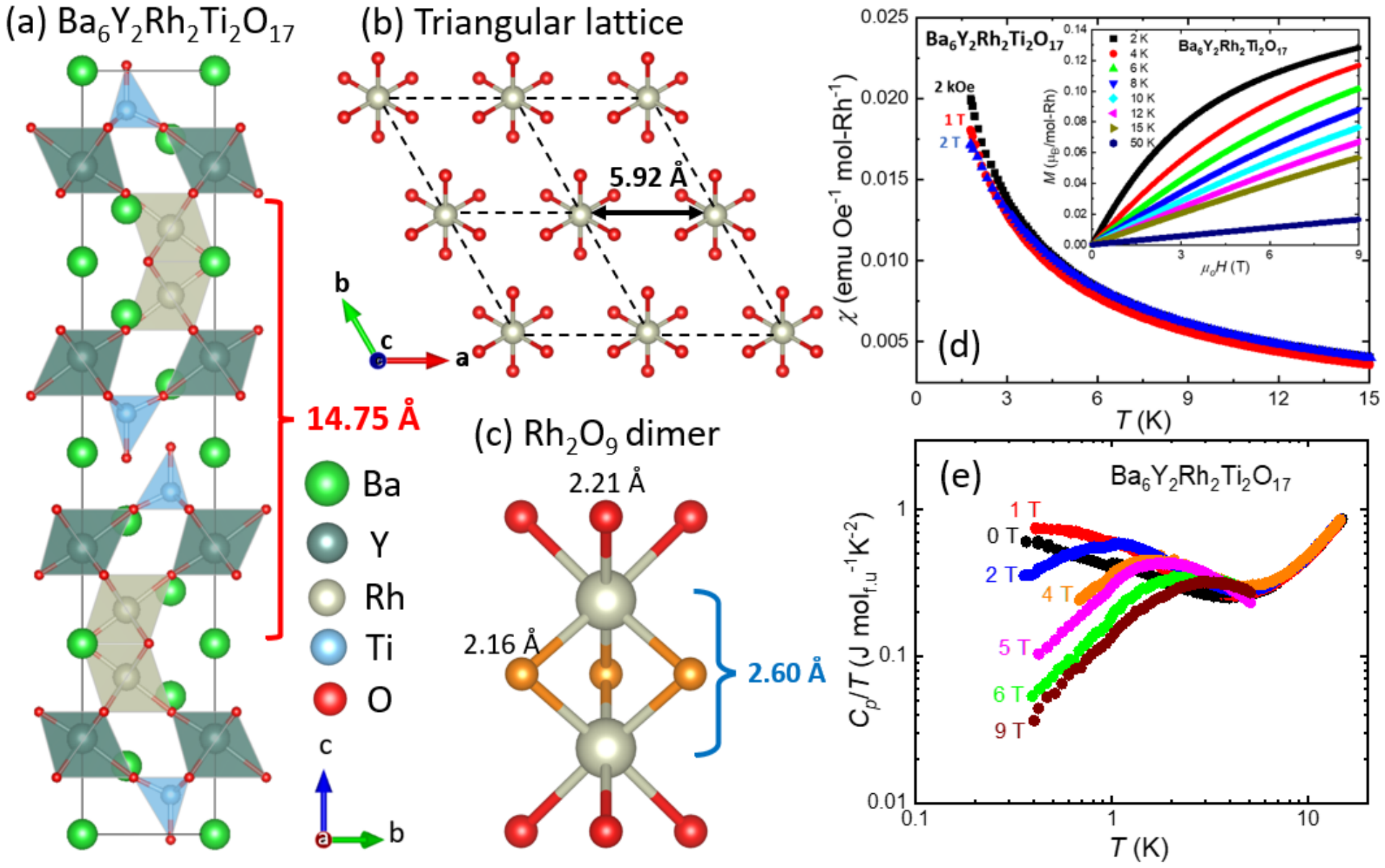

Figure 22. (a) The crystal structure of $Ba_6Y_2Rh_2Ti_2O_{17}$, (b)Triangular layer of $Rh^{4+}(d^5)$ along the c-axis, (c) The Rh-O, and Rh-Rh distances within the $Rh_2O_9$ dimer, (d) Temperature dependent magnetic susceptibility of $Ba_6Y_2Rh_2Ti_2O_{17}$ under different applied magnetic fields, and (e) Heat capacity data in $Ba_6Y_2Rh_2Ti_2O_{17}$ under different magnetic fields[158].

Finally, we end this section of this summary by describing one more multi-layered structure in the hexagonal perovskite family is 16L-$Ba_4Ca_{1-x}Mn_{3+x}O_{12-\delta}$ (**Figure 24a**). This material crystallizes in the hexagonal space group *P*-6*m*2 with the lattice parameters *a* = 5.8003(3) Å and *c* = 38.958(1) Å, with the large c axis arising because there are 8 layers in each of two building blocks. . It is one of the very first hexagonal perovskites known that contains Mn in tetrahedral sites, with Mn charge disproportionation. The crystal structure consists of Mn-tetramers (a cluster of four face-sharing $MnO_6$ octahedra), corner-sharing with $MnO_6$ octahedra and $MnO_4$ tetrahedra. Infrared spectroscopy study of $Ba_4Ca_{1-x}Mn_{3+x}O_{12-\delta}$, comparing the results with $KMnO_4$ and $BaMnO_4$ standards, has been said to confirm the possible coexistence of $Mn^{6+}$ and $Mn^{7+}$ ions in the structure, as shown **Figure 24b**. Temperature-dependent magnetic susceptibility measurements do not show any magnetic ordering down to 10 K, and high temperature inverse magnetic susceptibility fitting results in a Curie-Weiss temperature of -288 K and an effective moment of 6.7 μB/f.u (**Figure 24c**). The large negative Curie-Weiss temperature indicates the presence of strong antiferromagnetic interactions; however, the determination of Mn oxidation states and the magnetic configuration requires careful XANES and neutron diffraction study[162].

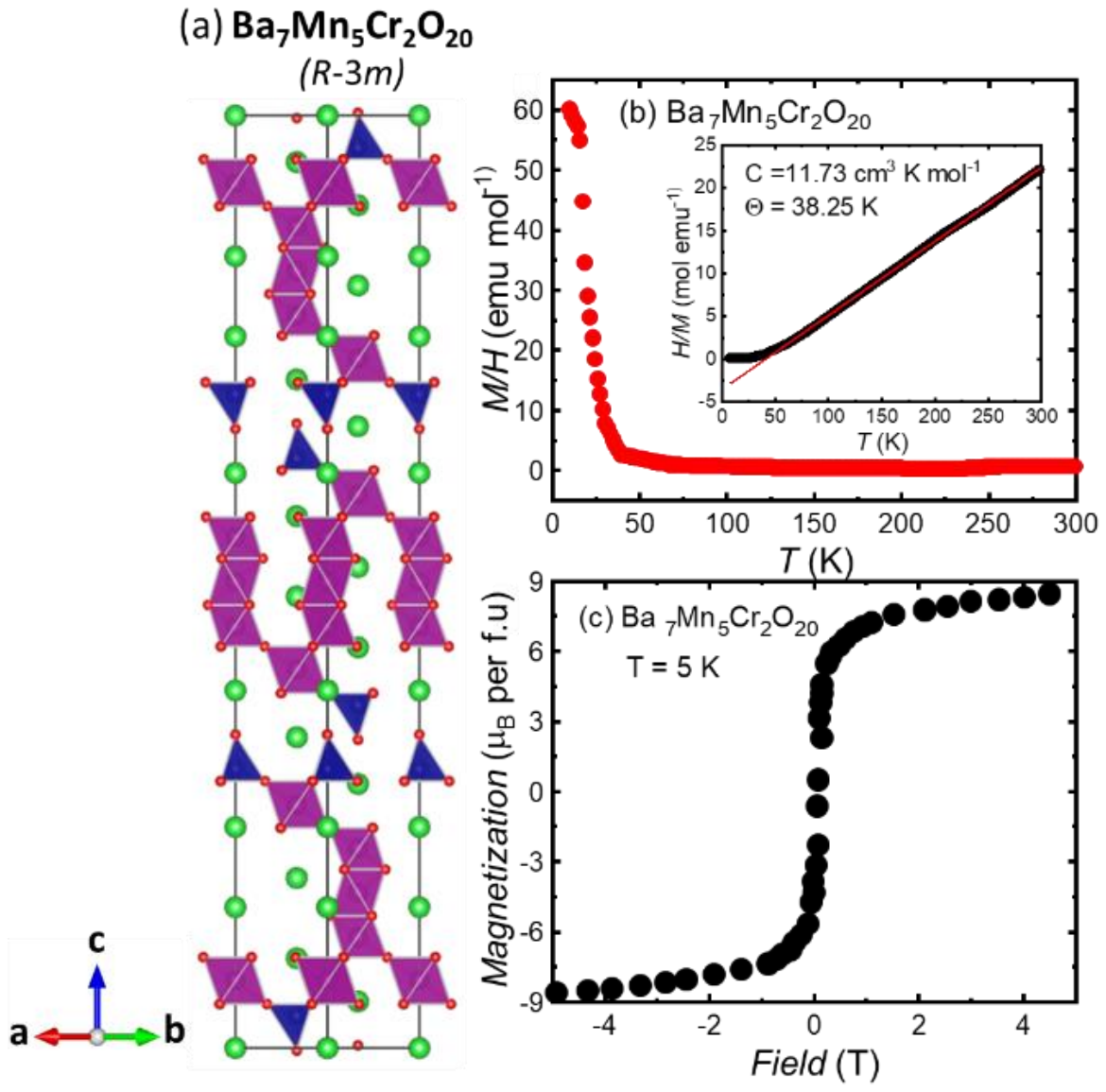

Figure 23. (a) The crystal structure of 21R-$Ba_7Mn_5Cr_2O_{20}$, (b) Temperature dependent FC/ZFC magnetic susceptibility of $Ba_7Mn_5Cr_2O_{20}$ and (c) Magnetic hysteresis loop at 5 K in $Ba_7Mn_5Cr_2O_{20}$[159].

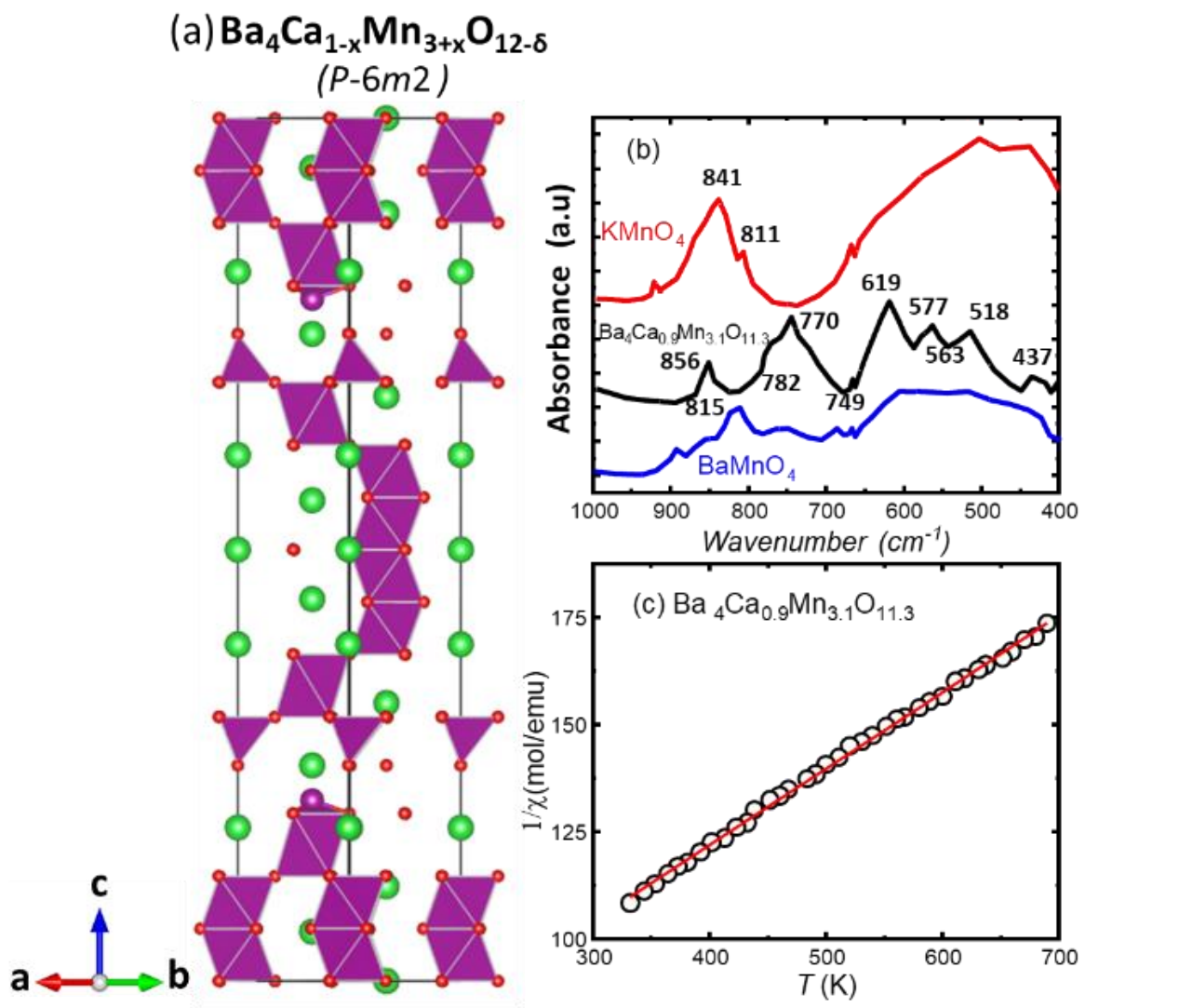

Figure 24. (a) The crystal structure of 16L-$Ba_4Ca_{1-x}Mn_{3+x}O_{12-\delta}$, (b) FTIR spectrum of $Ba_4Ca_{1-x}Mn_{3+x}O_{12-\delta}$ and the standards, $KMnO_4$ and $BaMnO_4$. (c) The inverse magnetic susceptibility of $Ba_4Ca_{1-x}Mn_{3+x}O_{12-\delta}$ under the applied magnetic field of 3 kOe [162].

## 5. Oxygen deficiency: vacancies both random and ordered

Nominally simple hexagonal oxide perovskites that actually have oxygen deficiency can form complex structures when the vacancies are ordered, especially for multi-valent transition 3*d* metal cations such as Mn, Fe and Co. Even randomly distributed vacancies can lead to a wide variety of structures, presumably to accommodate the different oxidation states if the M site metals. $BaMnO_{3-x}$, for example, can adopt different types of structures depending on the fraction of oxygen vacancies present. The published structures have randomly distributed oxygen vacancies, although structural probes that are sensitive to the local structures of the ions may eventually reveal that the structures are more complex. A similar phenomenon is also observed in $BaFeO_{3-x}$ and $BaCoO_{3-x}$ systems. In some cases, for example for the Cobalt compound $BaCoO_{2.6}$, the oxygens appear to be largely missing in an ordered fashion, resulting in the presence of $CoO_4$ tetrahedra. This is also seen for some materials in the $BaFeO_{3-x}$ system.

Different polytypes can be stabilized by the ionic size of the central atom and its valence state. Without the oxygen vacancy, $BaCoO_3$, with a typical hexagonal perovskite formula, displays infinite linear chains of face-sharing $CoO_6$ in the ideal hexagonal space group $P6_3/mmc$, with the lattice parameters $a$ = 5.645(3) Å and $c$ = 4.752(3) Å[163]. When the oxygen content is less than 3 in a formula unit, a variety of octahedral stackings occur, and the crystal structures adopt different space groups as shown in **Figure 25** [164–169]. In the 12H-$BaCoO_{2.6}$ structure, the tetramers consisting of four face-sharing $CoO_6$ octahedra are bridged by two layers of $CoO_4$ tetrahedra. The presence of tetrahedral Co in the structure may come from the small ionic size of Co ions in the tetrahedra compared to those in the octahedra. Curiously, it is reported that structures based on hexagonal perovskites convert to an ideal cubic perovskite for the very oxygen deficient material $BaCoO_{2.23}$. The most reduced form of this structure type results in another hexagonal compound, $BaCoO_2$, a material where all the Co is in the formal 2+ state. However, the crystal structure of $BaCoO_2$ consists of corner-sharing $CoO_4$ tetrahedra instead of the $CoO_6$ octahedra commonly found in the perovskite structures. Cobalt-oxide based perovskites have been intensively studied for many decades, by many different methods, and the issue of whether and how high-spin, intermediate-spin and low-spin in $Co^{3+}$ ions exist in the "simple" $LaCoO_3$ perovskite as a function of temperature is still controversial[170–174]. Similarly the spin state of $Co^{2+}$ in oxides can also be enigmatic.

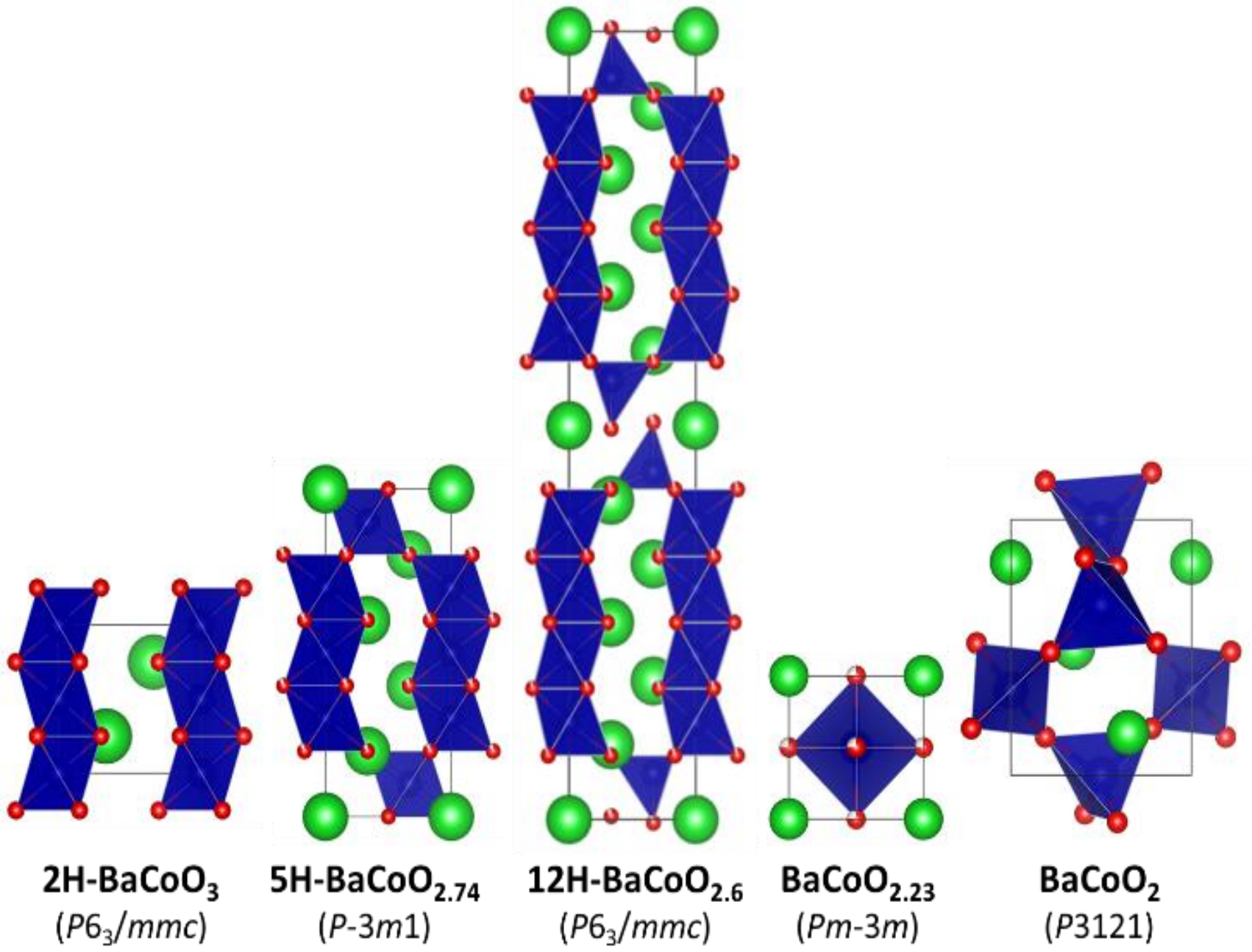

Figure 25. Different types of $BaCoO_{3-x}$ polymorphs, containing $CoO_6$ octahedra and $CoO_4$ tetrahedra. The materials with oxygen contents between those of $BaCoO_3$ and $BaCoO_2$ as a rule display randomly distributed oxygen vacancies, even in the case where the oxygen efficiency compared to the ideal $BaCoO_3$ formula is largely accommodated by $CoO_4$ octahedra.

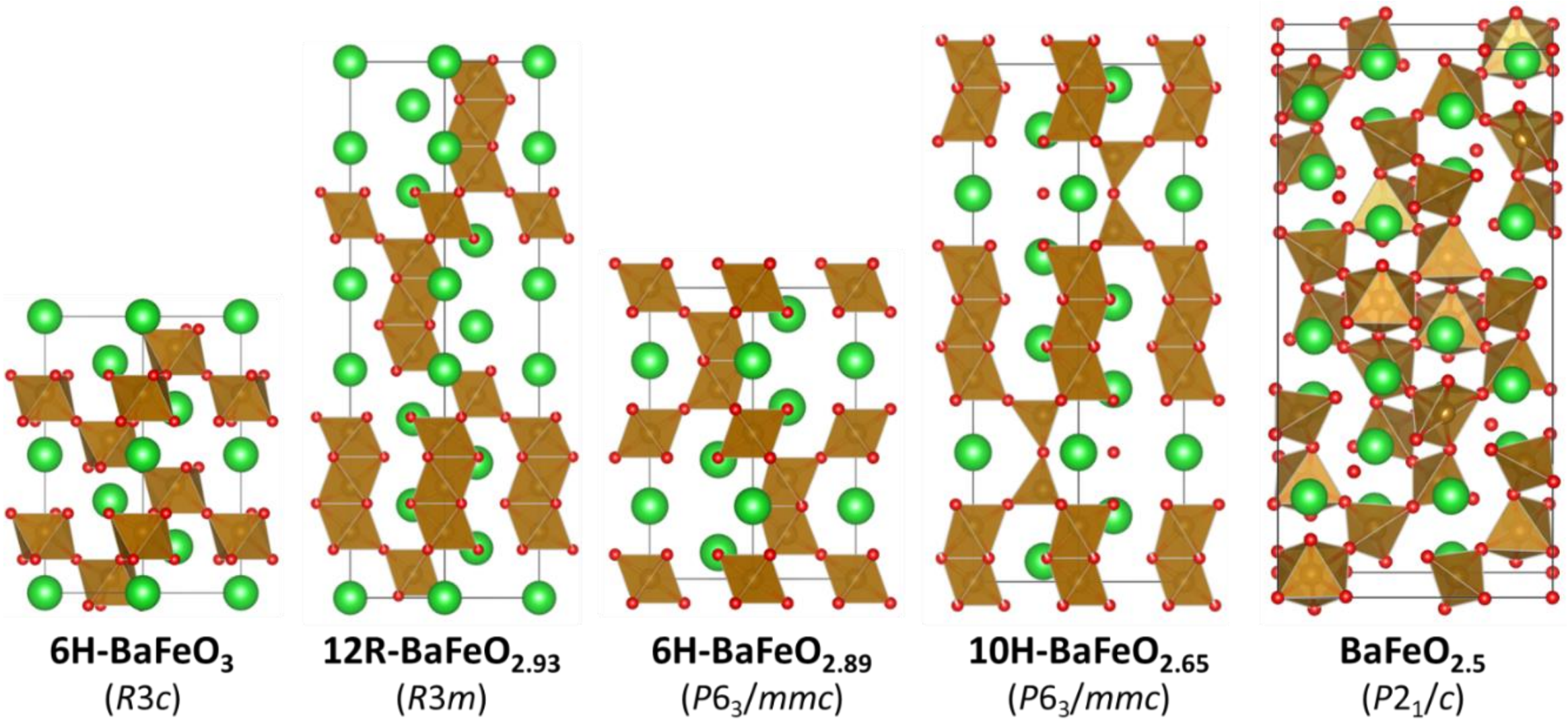

Figure 26. Different types of $BaFeO_{3-x}$ structures.

Similarly, $BaFeO_{3-x}$ also adopts different structures at different oxygen contents. This family of materials displayed in **Figure 26** . Although similar globally to the Co based materials, some interesting differences are reported. In 10H-$BaFeO_{2.65}$, the appearance of $FeO_4$ tetrahedra is similar to the case of 12H-$BaCoO_{2.6}$. Moreover, a further reduced form of $BaFeO_{3-x}$ leads to the transformation to the monoclinic structure $BaFeO_{2.5}$, space group $P2_1/c$. In $BaFeO_{2.5}$, the coordination of Fe ions ranges from 4 to 6, and the crystal structure is very highly distorted. A unique feature of this structure is that within an individual layer, both octahedra and tetrahedra are found.

For the $BaMnO_{3-x}$ system, there is a systematic trend in the number of face-sharing $MnO_6$ octahedra in the chains. The system goes from an infinite chain of $MnO_6$ octahedra in 2H-$BaMnO_3$ to a dimer consisting of two face-sharing $MnO_6$ octahedra in 4H-$BaMnO_{2.59}$ as seen in **Figure 27**[175–178]. The crystal structure and the oxygen vacancies in $BaMnO_{3-x}$ are closely related to interesting properties in manganates, such as metal-insulator transitions and colossal magnetoresistance[179–182]. Although there are a larger number of structures observed in this manganite system over a smaller range of oxygen deficiency than is seen in the Co and Fe – based oxygen deficient double perovskite, in a sense their structures appear to be simpler, and thus potentially easier to understand. To the senior author of this manuscript the oxygen deficient hexagonal perovskites for the transition metals Co, Fe and Mn literally scream for analysis through local structural methods. It does not seem to be something for impatient people to tackle.

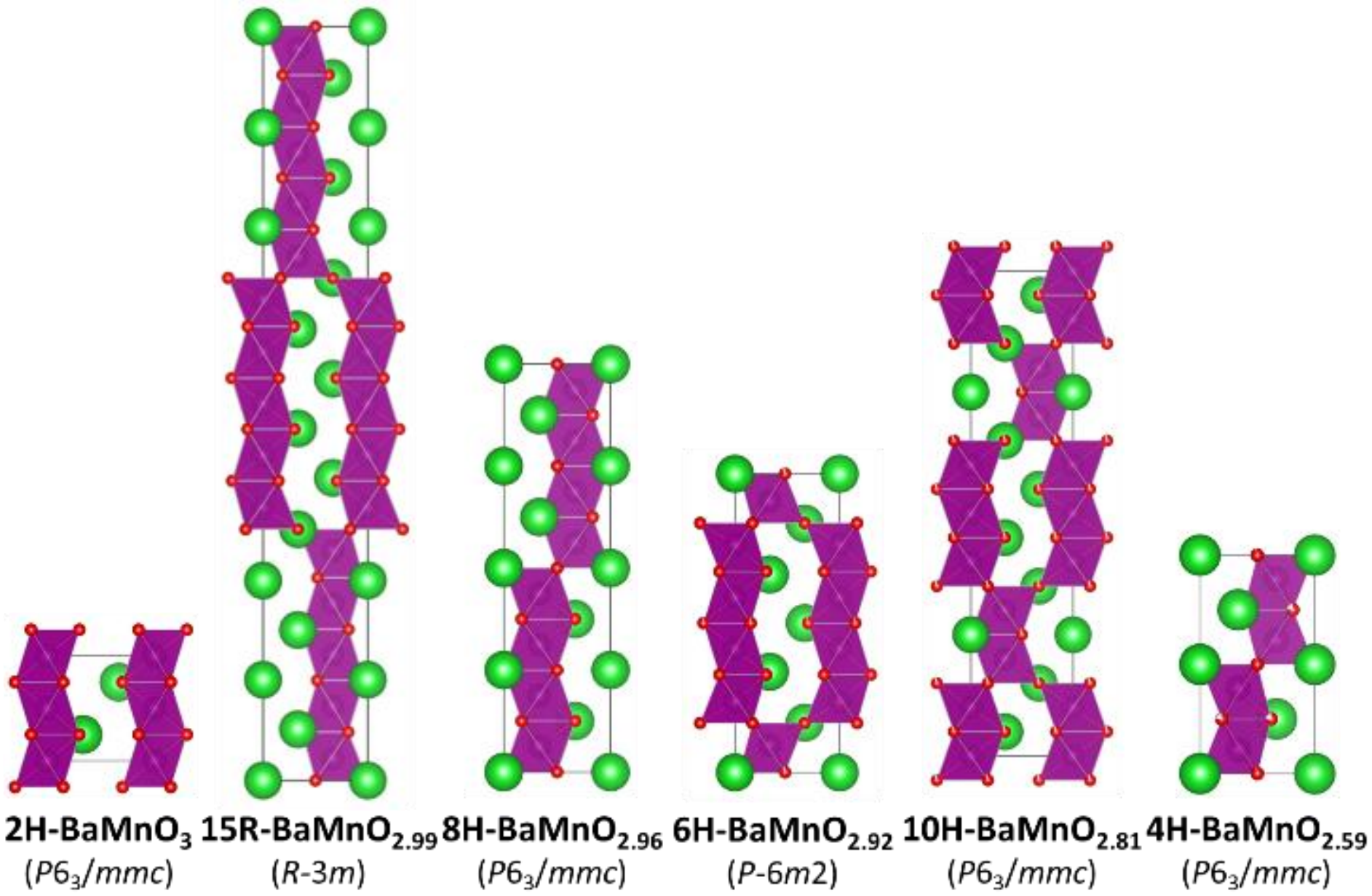

Figure 27. Different types of $BaMnO_{3-x}$ structures.

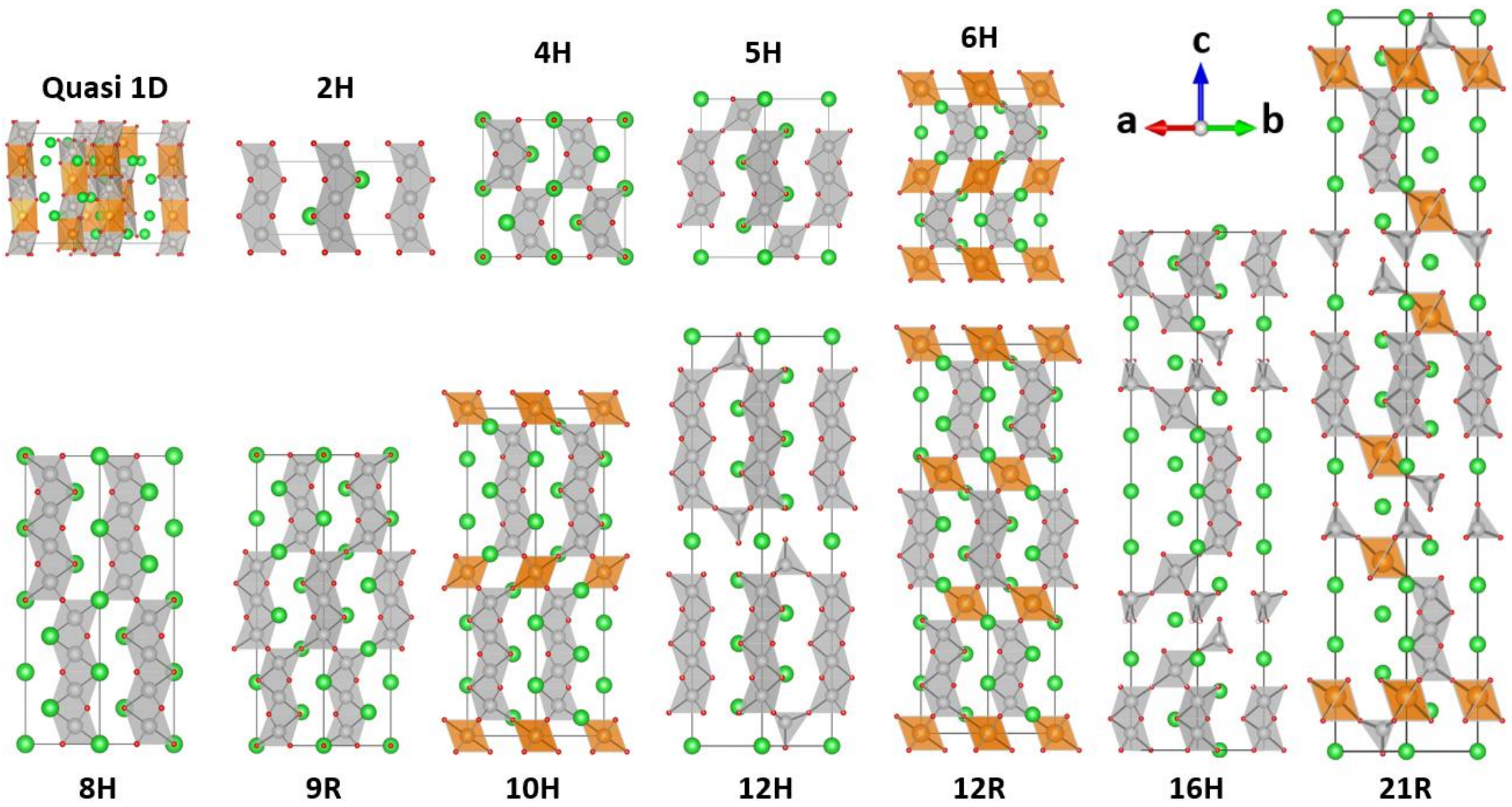

Figure 28. Schematic representation of some of the crystal structures of materials with complex stacking in hexagonal perovskites.

## 6. Polymorphism

We briefly specifically point out that hexagonal perovskites, even at a fixed $AMO_3$ formula, are not free from the kind of polymorphism that is seen in many other kinds of materials systems. Although this has been described previously in the case of $BaRuO_3$, it is seen in other hexagonal perovskites as well. A nice example is shown in **Figure 28,** which shows two different polymorphs of $BaFeO_3$. The first of these is a 3R 111-type rhombohedral distortion of a classical cubic perovskite, while the second is a 6H hexagonal perovskite with $Fe_2O_9$ dimers and layers of isolated $FeO_6$ octahedra. In the other way of looking at these materials, the first is $(ccc)_2$ type and the second is a $(cch)_2$ type. Pressure is a typical way of generating different polytypes, and he same is the case for hexagonal perovskites.

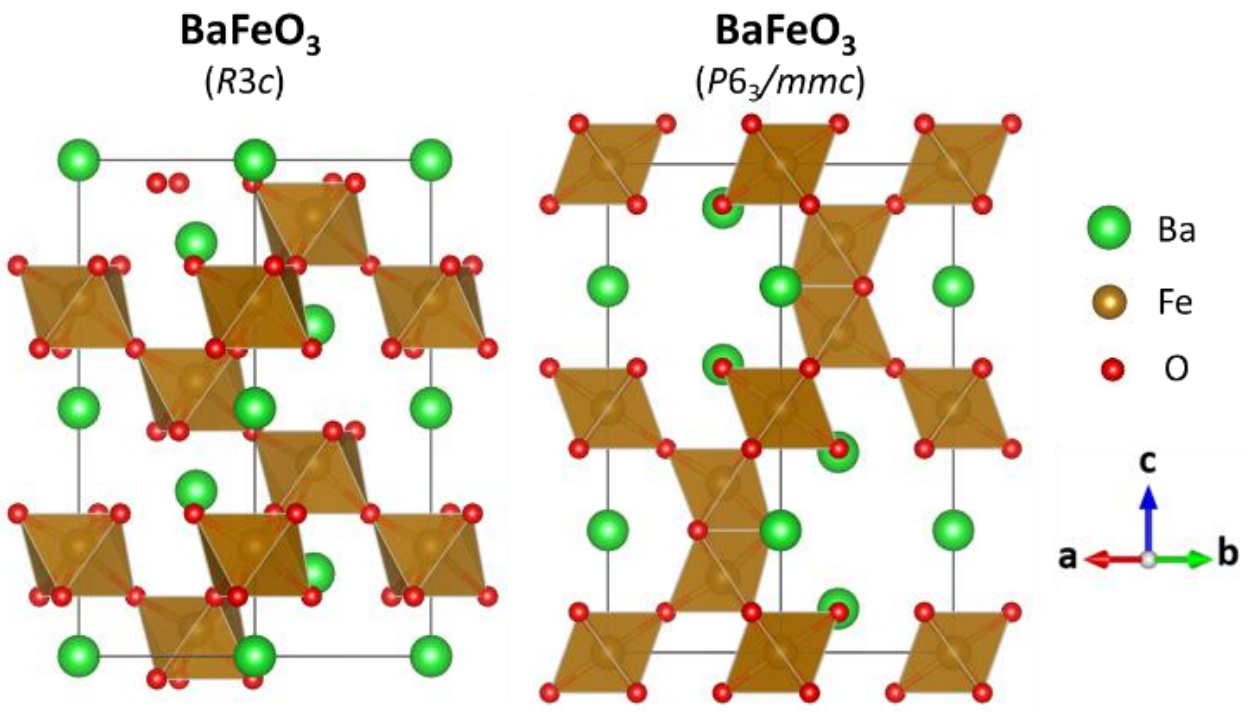

Figure 29. Two different polymorphs of $BaFeO_3$.

## 7. Summary of some of the complex variations in stacking

**Figure 29** displays, in the form of a summary image, of some of the complex stacking sequences possible in hexagonal perovskites. In most structures, the sequences of face-sharing octahedra share corners with each other (4H, 8H and 9R) or are bridged by layers of single octahedra (5H, 6H, 10H, 12R and 21R). Moreover, layers of tetrahedra in between the face-sharing octahedral blocks are found in 12H, 16H and 21R structures. Although both form 12-layer structures, the 12H structure includes tetrahedra while the 12R structure only contains octahedra. The 21R structure is an interesting and complicated structure type that has both tetrahedral and octahedral layers in between trimer building blocks. There are many materials with chains of face sharing octahedra or octahedra and triangular prisms. They are described briefly in a later section of this review.

## 8. Hexagonal Perovskites with Ba-Cl or Ba-Br layers in addition to Ba-O layers

Oxyhalide perovskites have sometimes been found as new phases in the crystal growth of oxide perovskites in a halide flux. In these materials, in addition to the Ba-O layers, Ba-X (X = F, Cl, Br, I) layers can also be stabilized between the face-sharing magnetic building blocks. Oxychlorides are the most frequently reported of this group. As shown in **Figure 30**, the crystal structure of $Ba_6Ru_2PtO_{12}Cl_2$ contains Ba-Cl layers sandwiched between $(Ru_2Pt)O_{12}$ trimers. With both suitable electronegativities and ionic radii, oxychloride and oxybromide perovskites are commonly reported. The space groups and lattice parameters of some selected materials in this family are summarized in

**Table** 4. Their magnetic properties have not yet been thoroughly studied, although they are likely to be as interesting as those of the purely oxide hexagonal perovskites. Their potential applications as photocatalysts and cathode materials have, however, recently been investigated[183–188]. When materials physicists figure out that some of them can be grown as nice single crystals, then more detailed studies can be expected.

Table 4. Summary of the space groups and lattice parameters of some of the hexagonal perovskites with Ba-Cl layers in addition to Ba-O layers.

| Compounds | Space group | Lattice parameters | Reference |
|---|---|---|---|
| $Ba_2Co_4O_7Cl$ | *R*-3*m* | *a* = 5.716 Å<br>*c* = 45.010 Å | 189 |
| $Ba_6Ru_2PtO_{12}Cl_2$ | *P*-3*m*1 | a = 5.805 Å<br>*c* =15.006 Å | 190 |
| $Ba_6Ru_3O_{12}Cl_2$ | *P*-3*m*1 | *a* = 5.815Å<br>*c* = 14.915 Å | 191 |
| $Ba_6Co_6O_{15.5}Cl$ | *P*-6*m*2 | *a* = 5.6738 Å<br>*c* = 14.5237 Å | 192 |
| $Ba_7Ru_4O_{15}Cl_2$<br>(and Br analog) | *R*-3*m* | *a* = 5.7785 Å<br>*c* = 51.6730 Å | 193 |
| $Ba_5Ru_2O_9Cl_2$<br>(and Br analog) | Pnma | *a* = 15.310 Å<br>*b* = 5.945 Å<br>*c* = 14.197Å | 194 |
| $Ba_5Co_5O_{13}Cl$ | P6₃/mmc | *a* = 5.669 Å<br>*c* = 24.304 Å | 195 |
| $Ba_5(MnO_4)_3Cl$ | *P*6₃/*m* | *a* = 10.469 Å<br>*c* = 7.760 Å | 196 |
| $KBa_2V_2O_7Cl$ | P6₃/mmc | *a* = 5.7688 Å<br>*c* = 15.0762 Å | 197 |
| $BaZn_2(SeO_3)_2Cl_2$ | *R*-3*m* | *a* = 5.5156 Å<br>*c* = 24.5840 Å | 198 |
| $Ba_8Co_2Mn_6O_{22}Cl$ | *P*-6*m*2 | *a* = 5.7207 Å<br>*c* = 19.4099 Å | 199 |
| $Ba_8Ru_{3.33}Ta_{1.67}O_{18}Cl_2$ | *R*-3*m* | *a* = 5.947 Å<br>*c* = 59.760 Å | 200 |
| $Ba_{10}Fe_8Pt_2Cl_2O_{25}$ | P6₃/mmc | *a* = 5.8034 Å<br>*c* = 24.9970 Å | 201 |

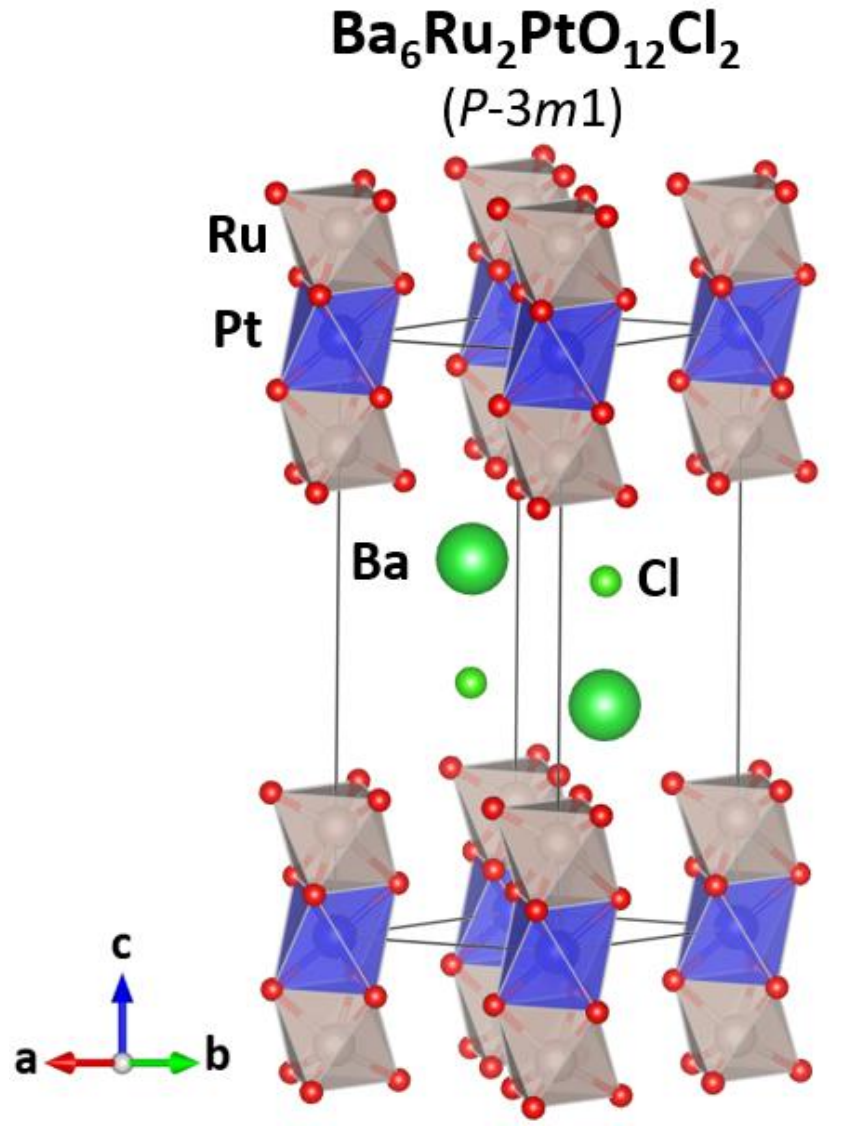

Figure 30. The crystal structure of $Ba_6(Ru_2PtO_{12})Cl_2$ [190].

# 9. Chains

One of the very nice structural aspects of hexagonal perovskites is that chain fragments of different lengths can be formed from face centered transition metal octahedra. These chains can be interrupted by face sharing with triangular prisms, which have some leeway along their prismatic lengths to shrink or stretch to accommodate ions of different preferred M-O bond lengths. This leads to some very interesting compounds. In a hexagonal perovskite structure type, materials of the type $Sr_4MO_6$ (M = Pt, Rh, and Ir), for example, crystallize in a rhombohedral structure, space group *R*-3*c*, as shown in **Figure 31**. Along the *c*-axis, the magnetic $MO_6$ octahedra are linked by non-magnetic $SrO_6$ trigonal prisms to form 1D-chains, where the magnetic octahedra and non-magnetic triangular prisms alternate. In some of these materials, the $Sr_4MO_6$ [202,203] formula ($Sr_4MO_6 = Sr_3(SrM)O_6$), because it results in a non- 1:1 A site to Chain site ratio, can lead to incommensurability of the chain periodicity and the A site periodicity.

From the quantum materials perspective, a few materials of this type have been studied. The analysis of magnetic susceptibility in $Sr_4RhO_6$[204] results in an effective moment of 2.03 μB/f.u and a Curie-Weiss temperature of -10 K. The authors of that study argue that $Sr_4RhO_6$ is the first clear example of a magnetically ordered $Rh^{4+}$ compound. The antiferromagnetic ordering temperatures of $Sr_4RhO_6$ and the related $Sr_4IrO_6$ chain compounds are 7 K and 12 K, respectively, which has been interpreted as reflecting a greater degree of covalency in the Ir5*d*-O than the Rh *4d*-O system[204,205]. Although in this review we have encountered several compounds in which "Rh4+" displays a magnetic moment, that is by far not the usual case in oxides, which, when based on the 4*d* transition metal Rh, are generally non-magnetic.

In this $K_4CdCl_6$-type structure, a series of 2H-related perovskites $(A_{3-x}Na_x)NaBO_6$ (A = La, Pr, Nd; B = Rh, Pt) has also been reported, made through synthesis by carbonate and hydroxide flux techniques. These materials are different in that the ions on the A site are mixed to yield charge neutrality. All of the materials crystallize in the rhombohedral space group *R*-3*c*. The temperature dependent magnetic susceptibility of a selected material in this family, $(Nd_2Na)NaPtO_6$ shows no magnetic ordering down to 2 K. The observed effective moment is found to be 5.06 μB/f.u, which agrees well with the expected moment of non-interacting $Nd^{3+}$ ions, 5.12 μB/f.u.[206]. In fact many rare earth oxides do not magnetically order above 2K, and thus with $Pt^{4+}$ and $Na^+$ being non-magnetic ions, these hexagonal perovskite materials can be considered as rare-earth-based paramagnets. Further study may or may not reveal magnetic ordering of the rare earth moments in these materials at lower temperatures.

In the same $A_{3n+3m}A'_nB_{3m+n}O_{9m+6n}$ family, the compound $Sr_6Rh_5O_{15}$[207] with m = 1 and n = 1, grown by a molten $K_2CO_3$ flux technique, has been reported. It crystallizes in the hexagonal space group *R*32, with the lattice parameters a = 9.6517(5) Å and c = 13.0480(5) Å. The crystal structure consists of 1D-chains parallel to the *c*-axis of four face-sharing $RhO_6$ octahedra a single $RhO_6$ trigonal prism. This material differs from the ones we have described so far because the atom in the triangular prism is the same as the atom in the octahedra. Magnetic susceptibility measurements shows an antiferromagnetic transition at 11 K, and thus that this material is another example of a magnetic Rh-based hexagonal oxide perovskite. The material is reported to have a large magnetic anisotropy, with the magnetic susceptibility when the applied field is parallel to the crystal's *c*-axis being many orders of magnitude larger than the case when the applied field is perpendicular to the crystal's *c*-axis.

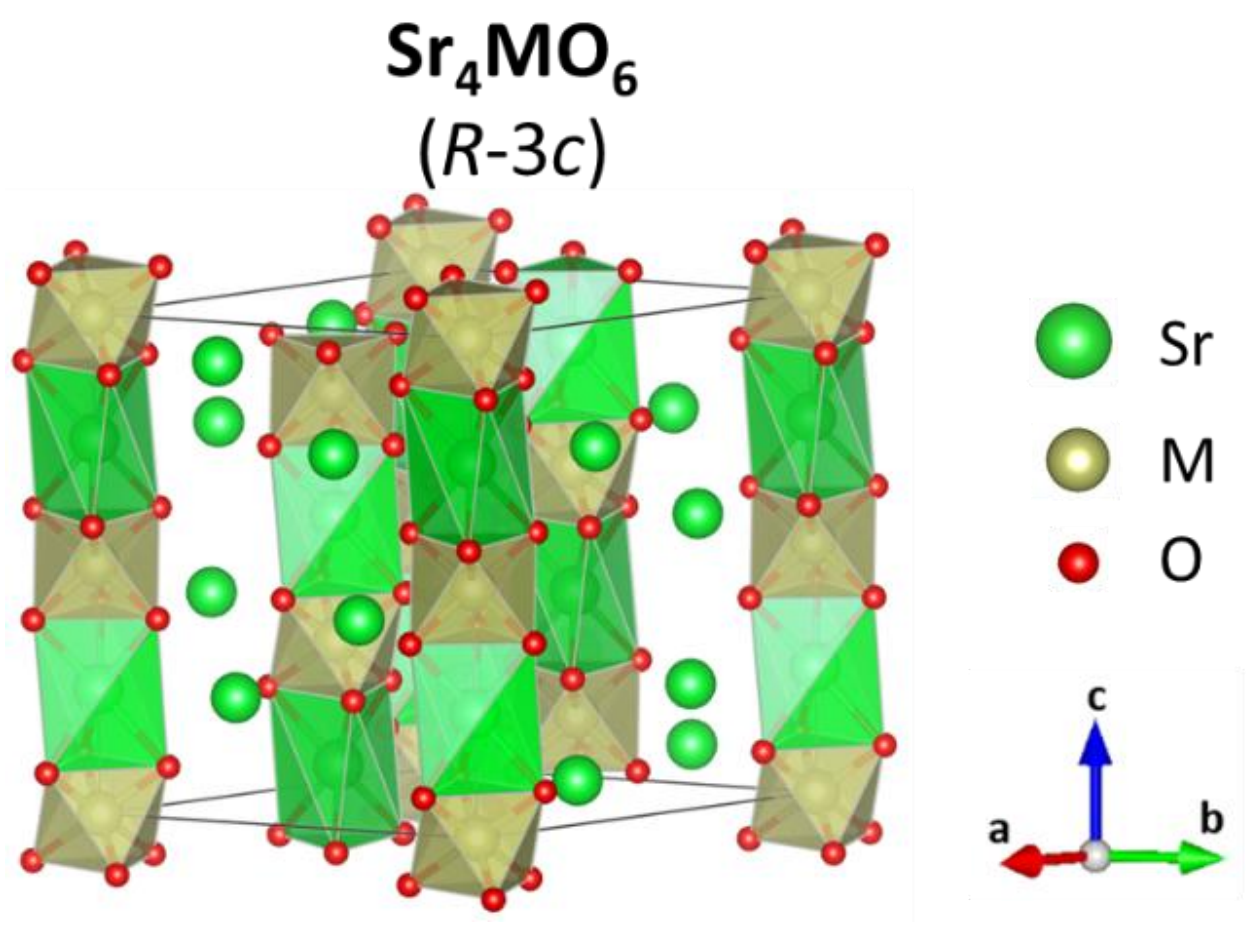

Figure 31. The crystal structure of $Sr_4MO_6$ (M = Pt, Rh, Ir).

Another example of a chain-structure variation is found for $Sr_4CuIr_2O_9$, whose crystal structure consists of $Ir_2O_9$ dimers face-sharing with $CuO_6$ trigonal prisms, to form 1 D chains along the *c*-axis (**Figure 32**). An alternative view is that the continuous chains are made of $IrO_6$ octahedra interrupted after only two share faces by sharing faces with a Cu triangular prism or a vacant prism. The crystal structure of this material, which has Cu/Ir ordering in the M site chains, is very sensitive to the conditions of synthesis: prolonged heating of a commensurately ordered trigonal material leads to the transformation to an incommensurate structure, which can be described as a mixture of two substructures with the same lattice parameter *a*, but with different lattice parameters $c_1$ and $c_2$ ($c \sim 2c_1 \sim 3c_2$)[208]. The incommensurability, as is seen in this type of compound and for other materials in the hexagonal perovskite family, comes from the difference in periodicities of the A-site and M-site chains, from which one can infer that the atoms in the chains are more strongly interacting along their lengths than with other chains. In contrast, in the $Ba_6CuIr_4O_{15}$ crystal structure, a disordered distribution of Cu and Ir, rather than an ordered distribution, is seen in both the trigonal prismatic and the octahedral sites. The incommensurate structure of this material can be considered as a combination of trigonal and rhombohedral substructures, similarly to the incommensurate hexagonal perovskites $SrMn_{1-x}Co_xO_{3-y}$[209,210].

Along the same line of study is a material where Cu and Ir are mixed in triangular prisms but there are three octahedra before the chains are interrupted by a triangular prism. The crystal structure of $Ba_5CuIr_3O_{12}$ was determined by a combination of both X-ray and neutron diffraction in 1999 [211]. As shown in **Figure 33**, its structure consists of three face-sharing $IrO_6$ octahedra forming a quasi 1D chain along the *c*-axis. DFT calculations for this material show that the localized j = ½ picture cannot fully explain the electronic and magnetic properties of $Ba_5CuIr_3O_{12}$, attributed by the authors of that study to strong covalency among the atoms that is not captured by the calculations. In other words, the very short Ir-Ir distance present within the $Ir_3O_{12}$ trimers results in strong uncaptured metal-metal interactions that may lead to the formation of molecular orbitals[212]. Recently, a spin liquid ground state in $Ba_5CuIr_3O_{12}$ has been tested by thermodynamic and high magnetic field measurements. Although weak antiferromagnetic interactions were seen in the magnetic susceptibility and heat capacity measurements, the magnetization did not saturate, even up to a 59 T applied magnetic field. This implies the existence of a random single state, which can fully explain the experimental data[213].

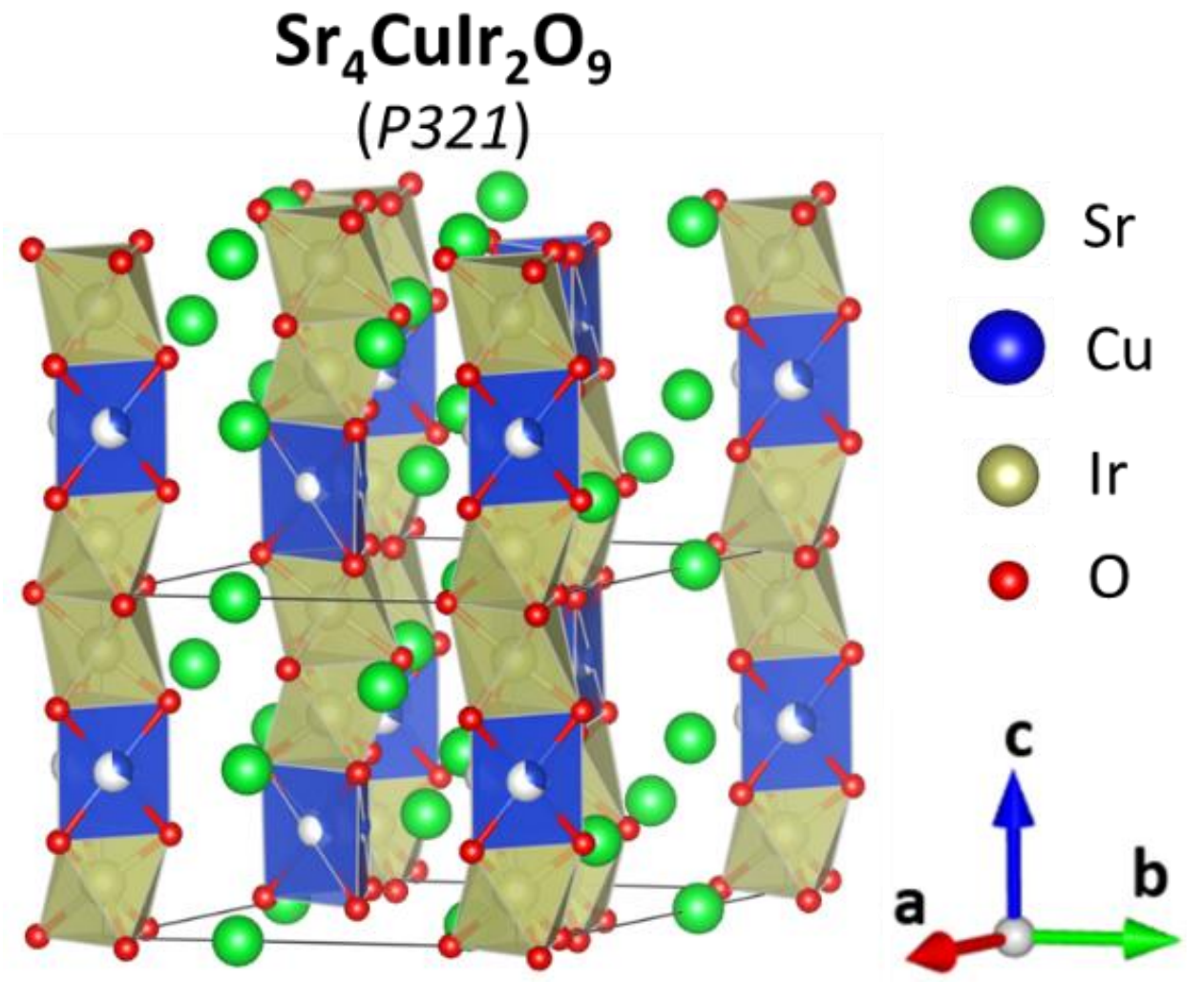

Figure 32. The crystal structure of $Sr_4CuIr_2O_9$.

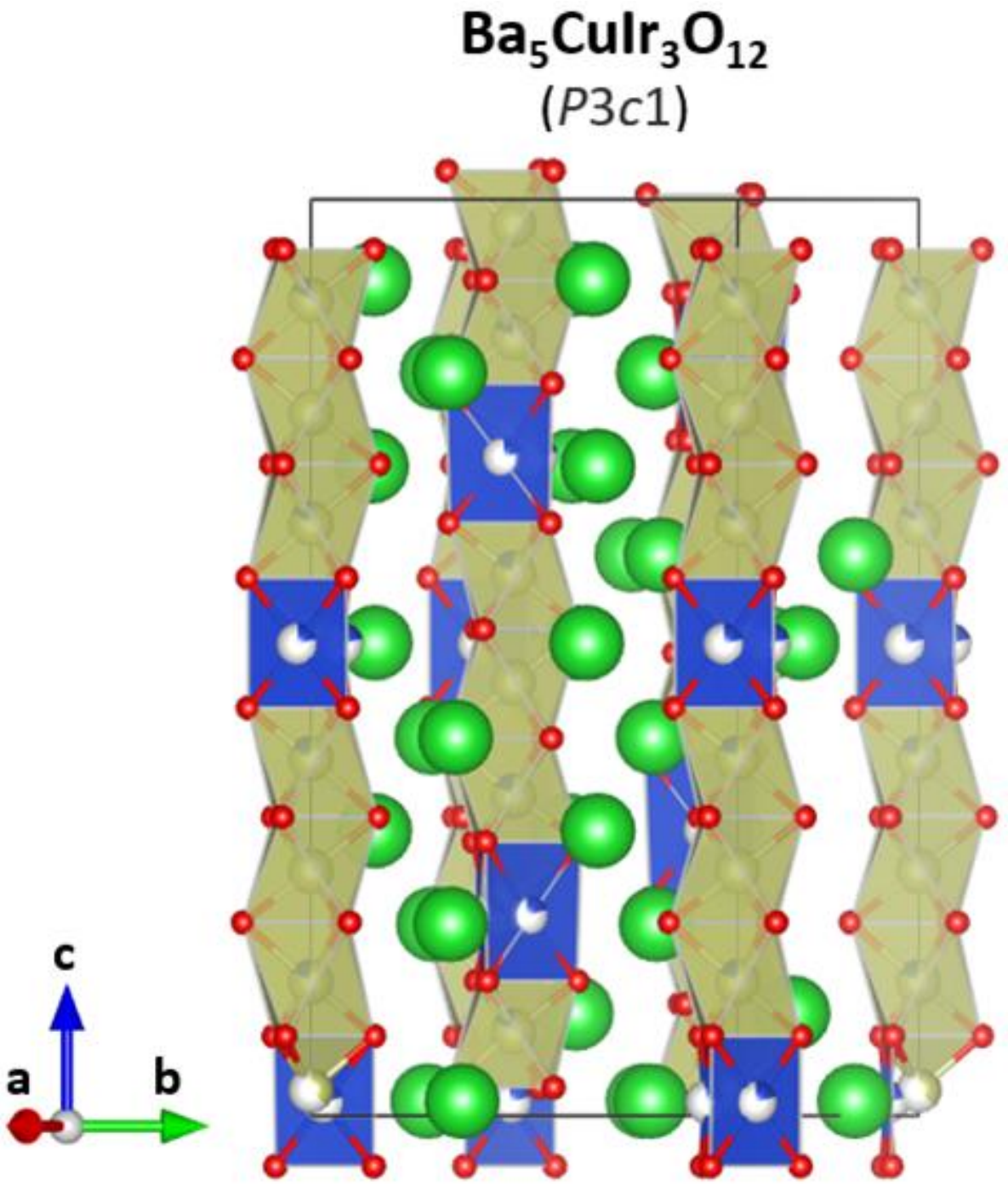

Figure 33. The crystal structure of $Ba_5CuIr_3O_{12}$.

In contrast to the other materials described here, the crystal structure of $Ca_2IrO_4$[214] displays a pure 1D chain of edge-sharing (not face-sharing) $IrO_6$ octahedra, but nonetheless has hexagonal symmetry, as shown in **Figure 34**. DFT calculations have been performed for this material and they argue that the material should be an antiferromagnetic insulator with a band gap of 0.64 eV and an effective moment of 0.68 μB/f.u. The band gap has been attributed to the long range antiferromagnetic interactions of $Ir^{4+}$ spins. Moreover, attributed to the inherent, strong spin orbit coupling present for the 5*d* Ir transition metal, large magnetic anisotropy has been observed in this system. The crystal structure of a more traditional hexagonal perovskite based on chains, 2H $BaCoO_3$, is shown in the figure for comparison. The Co-Co bond length in 2H-$BaCoO_3$ is even shorter than that in the metal; the distances being 2.38 Å and 2.50 Å along the chain, respectively[170].

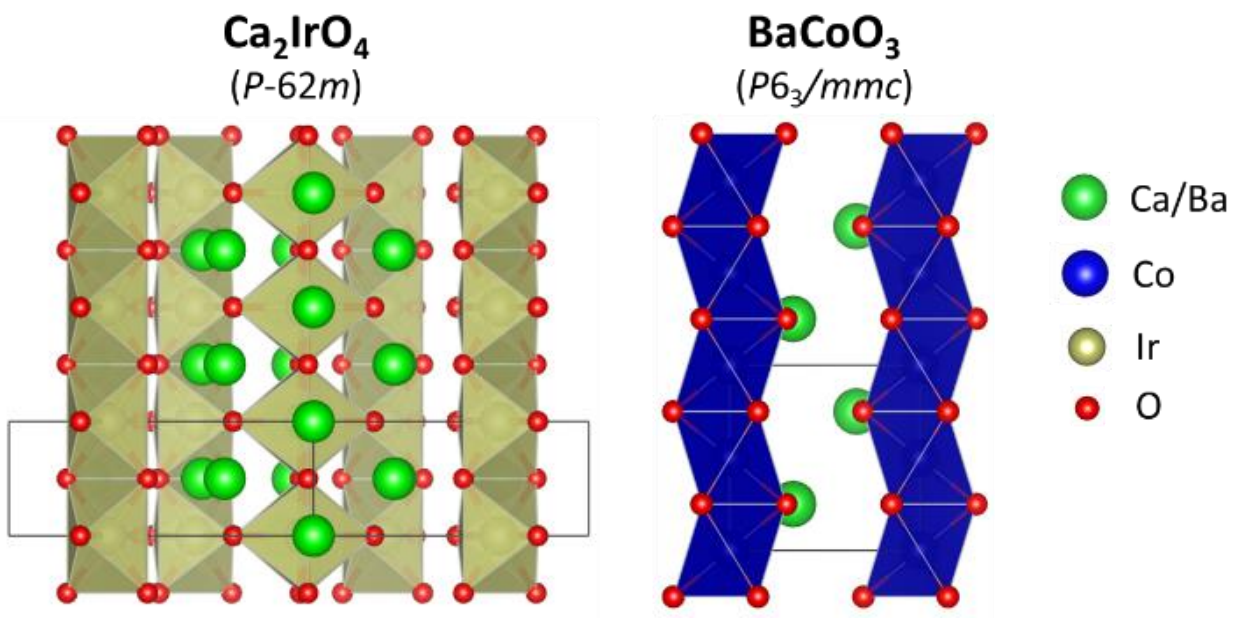

Figure 34. The crystal structure of $Ca_2IrO_4$ and 2H-$BaCoO_3$.

Sticking with continuous chains of octahedra for the moment, a continuous 1D chain of face-sharing $IrO_6$ octahedra has been observed recently reported material, $K_3Ir_2O_6$ [215]. This material is an electrical insulator, but DFT calculations show no band gap near the Fermi level. This has been interpreted to indicate that $K_3Ir_2O_6$ is a Mott insulator. More interestingly, "flat bands" (that is, electronic bands that show no dispersion of electron energy versus electron wave vector) along the Γ-F direction in the Brillouin zone in the electronic band structure suggest that the electrons are localized along certain momentum vectors in the unit call. This material may also prove to be of interest from the quantum materials perspective on further study.

A few more crystal structures of materials with 1D chains are outlined in **Figure 35**. They are obviously complex structures, because the chain fragments are quite long and have different lengths. $RhO_6$ octahedra seem to be a fertile constituent of such systems. $Ba_9Rh_8O_{24}$[216] crystallizes in the hexagonal space group $R$-3$c$ with the lattice parameters $a$ = 10.0899(4) Å, and $c$ = 41.462(2) Å. A chain of seven face-sharing $RhO_6$ octahedra is sandwiched by $RhO_6$ trigonal prisms in this material. The directional magnetic susceptibility shows divergence below 100 K, implying the onset of the magnetic anisotropy in this material. Its high temperature magnetic susceptibility does not follow the Curie-Weiss law, and it is too complex to allow the assignment of the oxidation states of Rh ions along the chain. The next two materials, $Ba_{11}Rh_{10}O_{30}$ and $Ba_{32}Rh_{29}O_{87}$ [217], both crystallize in the hexagonal space group $R$-3$m$. $Ba_{11}Rh_{10}O_{30}$ contains chains of nine face-sharing $RhO_6$ octahedra while $Ba_{32}Rh_{29}O_{87}$ has eight repeated face-sharing $RhO_6$ octahedra followed by a $RhO_6$ trigonal prism. The longest chain of face-sharing octahedra is seen in 12R-$Ba_{12}Rh_{9.25}Ir_{1.75}O_{33}$[218], for which Ir is mixed with the Rh in a disordered fashion The chain consists of ten face-sharing (Rh/Ir)$O_6$ octahedra, connected to single (Rh/Ir)$O_6$ trigonal prism. This material adopts a hexagonal structure, space group $R$32, with lattice parameters $a$ = 10.0492(2) Å and $c$ = 28.386(3) Å. Magnetic susceptibility measurements yield a Curie-Weiss temperature of -1 K, and a very low effective moment of 1.08 μB/f.u. This very low observed effective moment has been attributed to a strong spin orbit coupling effect in both the Rh and Ir ions and strong metal-metal interactions along the chains. The continuous sequence of face-sharing octahedra displays very short metal-metal bond lengths (2.5-2.6 Å) in this material, which indicates the presence of direct metal-metal bonding along the 1D chain. This apparent bonding would not lead to a preference of one type of magnetic model over any other. If these compounds could be made in quantities sufficient for detailed study, or as large enough single crystals, then they would certainly challenge theorist's understanding of quantum materials

## 10. Some unusual derivatives of hexagonal perovskites

But the crystal structures can sometimes be even more complex than these. $Pb_3Rh_7O_{15}$ [219] for example crystallizes in a hexagonal structure, space group $P6_3/mmc$ with lattice parameters $a$ = 10.3537(2) Å and $c$ = 13.2837(5) Å. The crystal structure of $Pb_3Rh_7O_{15}$ is shown in **Figure 36a**. It is a variant of a "dimer" type hexagonal perovskite. The electrical resistivity is low and isotropic, about 1mΩcm at room temperature. As seen in **Figures 36b-c**, there exists a clear transition at 185 K in both magnetic susceptibility and heat capacity measurements. This comes from a structural phase transition at 185 K that has been attributed to the charge modulation of $Rh^{3+}$ and $Rh^{4+}$. The substitution of Bi for Pb leads to the disappearance of the structural transition at 185 K, which the authors argue confirms that the transition in $Pb_3Rh_7O_{15}$ is due to $Rh^{3+}$-$Rh^{4+}$ charge modulation.

The crystal structure of a recently described rhodate $Bi_{1.4}CuORh_5O_{10}$ [220] is shown in **Figure 37**. This material crystallizes in a monoclinic space group, $C2/m$, with the lattice parameters a = 17.310(3) Å, b = 3.0775(9) Å, c = 16.009(3) Å and β = 95.903(15)°, with a more severe distortion of a hexagonal perovskite structure then the Pb rhodate described above. The crystal structure contains undulating layers of both edge-sharing and corner-sharing $RhO_6$ octahedra, with Cu ions occupying square planar coordinate sites. This type of octahedral framework with channels is commonly known in rhodates, such as rutile $RhO_2$, hollandite $Ba_{1.72}Rh_8O_{16}$, and todorokite $Bi_6Rh_{12}O_{29}$.

As shown in **Figure 38**, $Sr_7Mn_4O_{15}$ [221] is a distorted dimer-based perovskite. Studied by neutron diffraction, this material crystallizes in the monoclinic space group, $P2_1/c$, with the lattice parameters $a$ = 6.81825(9) Å, $b$ = 9.6228(1) Å, $c$ = 10.3801(1) Å, and β = 91.8771(9)° . The crystal structure consists of dimers of face-sharing $MnO_6$ octahedra that are then corner shared with other dimers, but the crystal structure is highly distorted, with an unusual ratio of seven $Sr^{2+}$ ions to four Mn ions, which in the end remarkably yields a uniform formal charge state of $Mn^{4+}$, a 3$d^3$ electron configuration to accompany the highly distorted crystal structure that accommodates the 7:4 metal ion ratio. Magnetic susceptibility measurements show the presence of three different magnetically ordered states in this material as a function of temperature at ambient temperature and below. Above 150 K, intradimer ordering is observed. 2D-magnetic clusters are formed in the temperature range of 150-75 K, and finally, full 3D antiferromagnetic ordering among the $Mn^{4+}$ spins is established below 75 K. The magnetic structure of $Sr_7Mn_4O_{15}$, determined by using neutron diffraction, was interpreted as showing the presence of those three different magnetic phases. Mn seems to be highest among royalty in its ability to couple lattice and electronic and magnetic states in oxides.

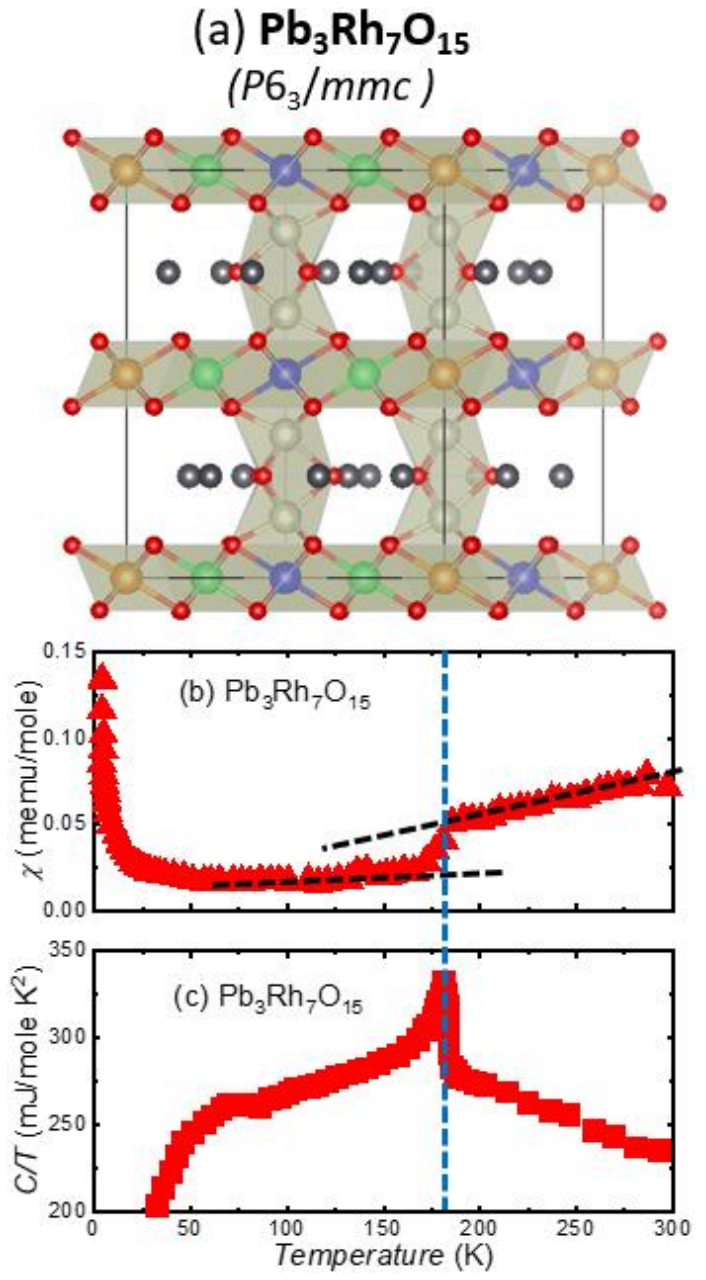

Figure 36. (a) The crystal structure of $Pb_3Rh_7O_{15}$, (b) Temperature dependent magnetic susceptibility and (c) Heat capacity of $Pb_3Rh_7O_{15}$ [219].

The ternary Ba-Ru-O and Ba-Ir-O hexagonal perovskite chemical systems display many unusual derivatives, largely based on different Ba:M ratios, with interesting quantum properties. First of all, $Ba_5Ru_2O_{10}$ [222] crystallizes in the hexagonal space group P6$_3$/mmc, and its crystal structure shown in **Figure 39** consists of two face-sharing $RuO_6$ octahedra forming $Ru_2O_9$ dimers. The Ru-Ru distance in the dimer is 2.071 Å, and the Ru in this material has a formal charge of 5+. The magnetic susceptibility of $Ba_5Ru_2O_{10}$ has a large maximum at around 280 K, which has been ascribed to the antiferromagnetic interactions of the Ru moments within the $Ru_2O_9$ dimers, and the coupling exchange is calculated to be -116 K. With the same synthesis conditions in an $O_2$ rich environment, $Ba_5Ru_2O_{11}$ is formed instead. The only difference in the crystal structures between $Ba_5Ru_2O_{10}$ and $Ba_5Ru_2O_{11}$ is that peroxide groups $O_2^{2-}$ are stabilized in the $Ba_2O_2$ layers in $Ba_5Ru_2O_{11}$ while only isolated oxygen ions are found in the unit cell in $Ba_5Ru_2O_{10}$. The high stability of the peroxide $O_2^{2-}$ group has been said to limit reversible electrochemical reactions of the $O_2^{2-}/O^{2-}$ redox couples used in oxygen evolution reaction catalysts and Li-ion battery materials[223]. In the mind of the senior author of this manuscript, the fact that peroxide ions are actually observed in the crystal structure of $Ba_5Ru_2O_{11}$ really raises the question of how to think about the distribution of electrons in materials like these that have high rations of electropositive ions (i.e. $Ba^{2+}$) to highly oxidized transition elements (i.e. Ru); he realizes of course that people who do DFT calculations will say that they know where the charges are, but for his generation at least, proposed charge distributions are best determined experimentally not through DFT calculations. A famous case of this is found in cuprate superconductors such as $La_{2-x}Sr_xCuO_4$. As Sr is substituted for La, do the electrons come out of Cu orbitals, making the actual number of electrons that can be associated with the Cu go down, or is the charge actually removed from oxygen *p* states? Much of the observed phenomenology is not sensitive to this. $Ba_5Ru_2O_{10}$ by the way appears to be a simple example of relatively isolated $Ru_2O_9$ dimers on a layered hexagonal lattice.

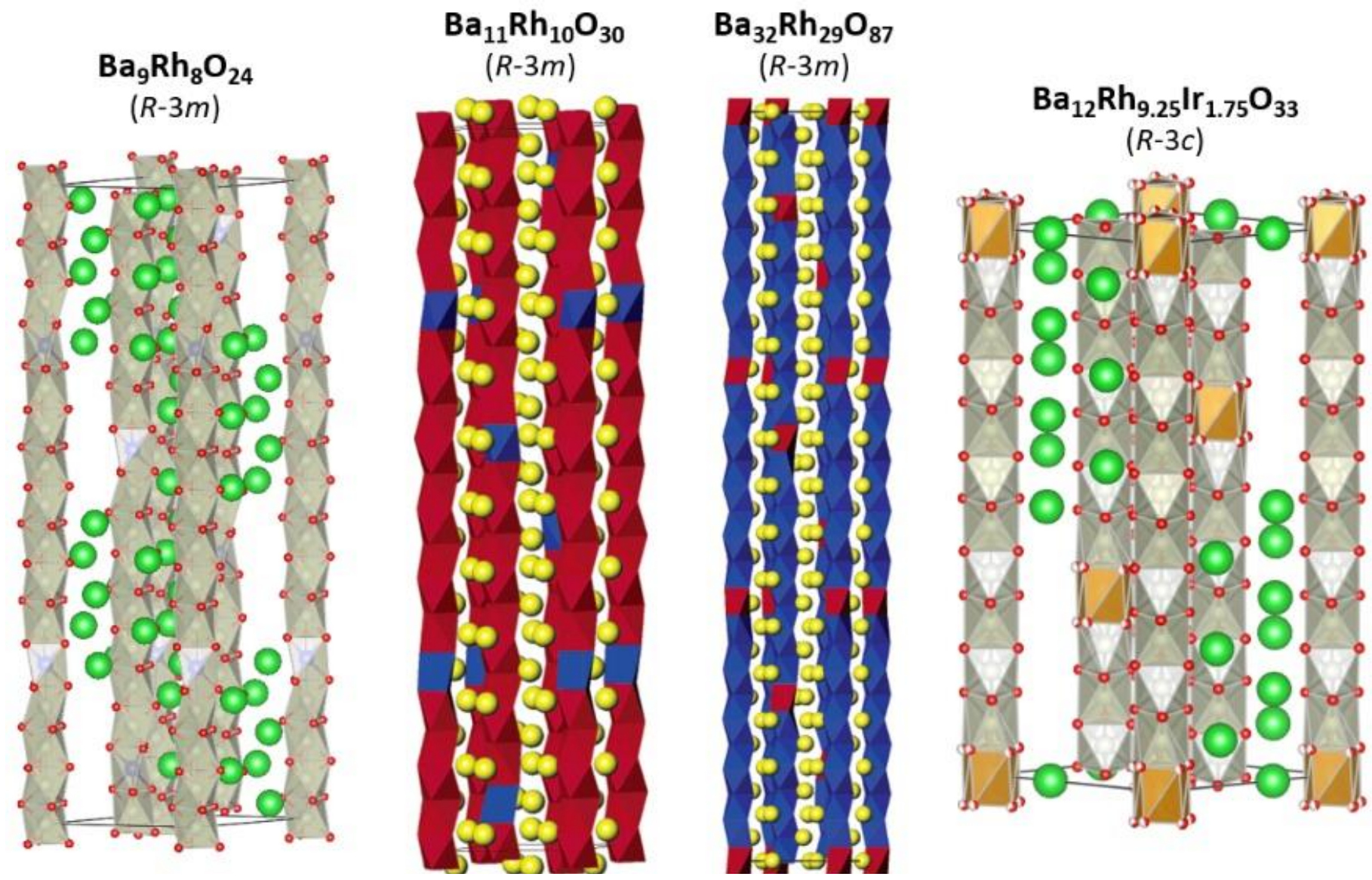

Figure 35. The crystal structures of $Ba_9Rh_8O_{24}$, $Ba_{11}Rh_{10}O_{30}$, $Ba_{32}Rh_{29}O_{87}$ and $Ba_{12}Rh_{9.25}Ir_{1.75}O_{33}$.

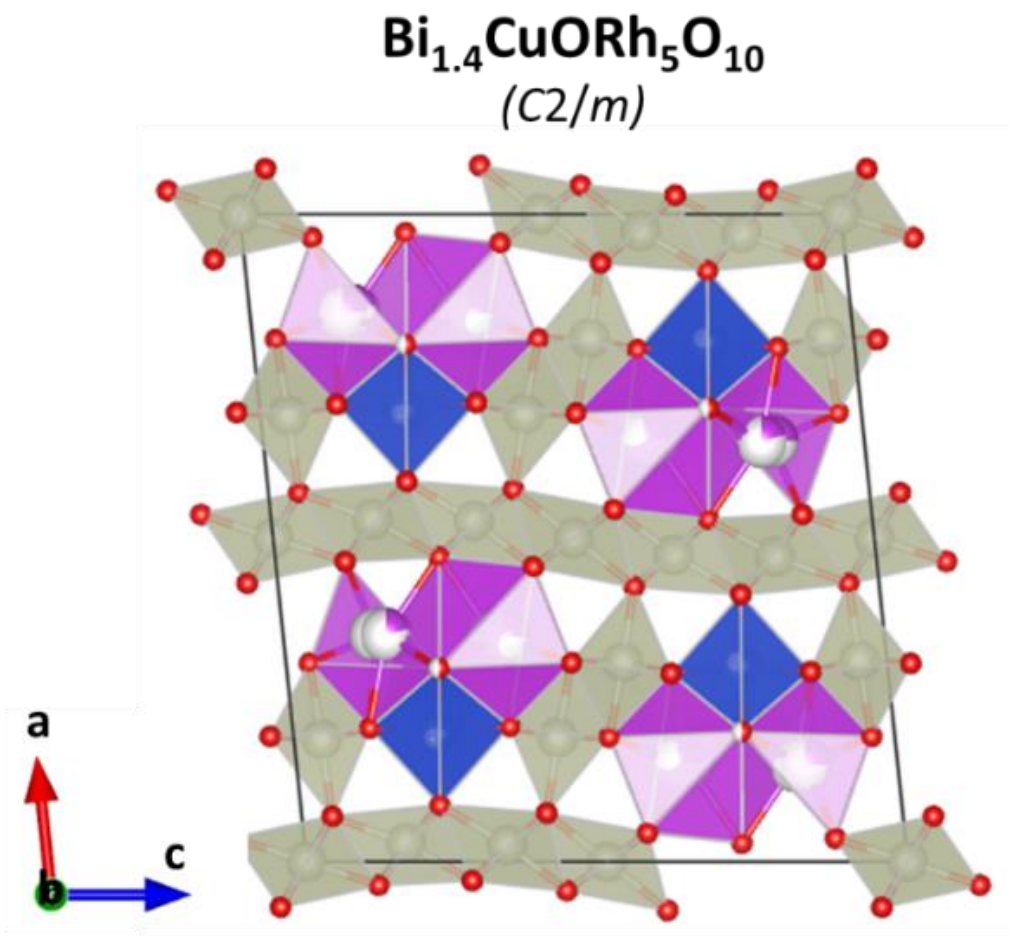

Figure 37. The crystal structure of $Bi_{1.4}CuORh_5O_{10}$.

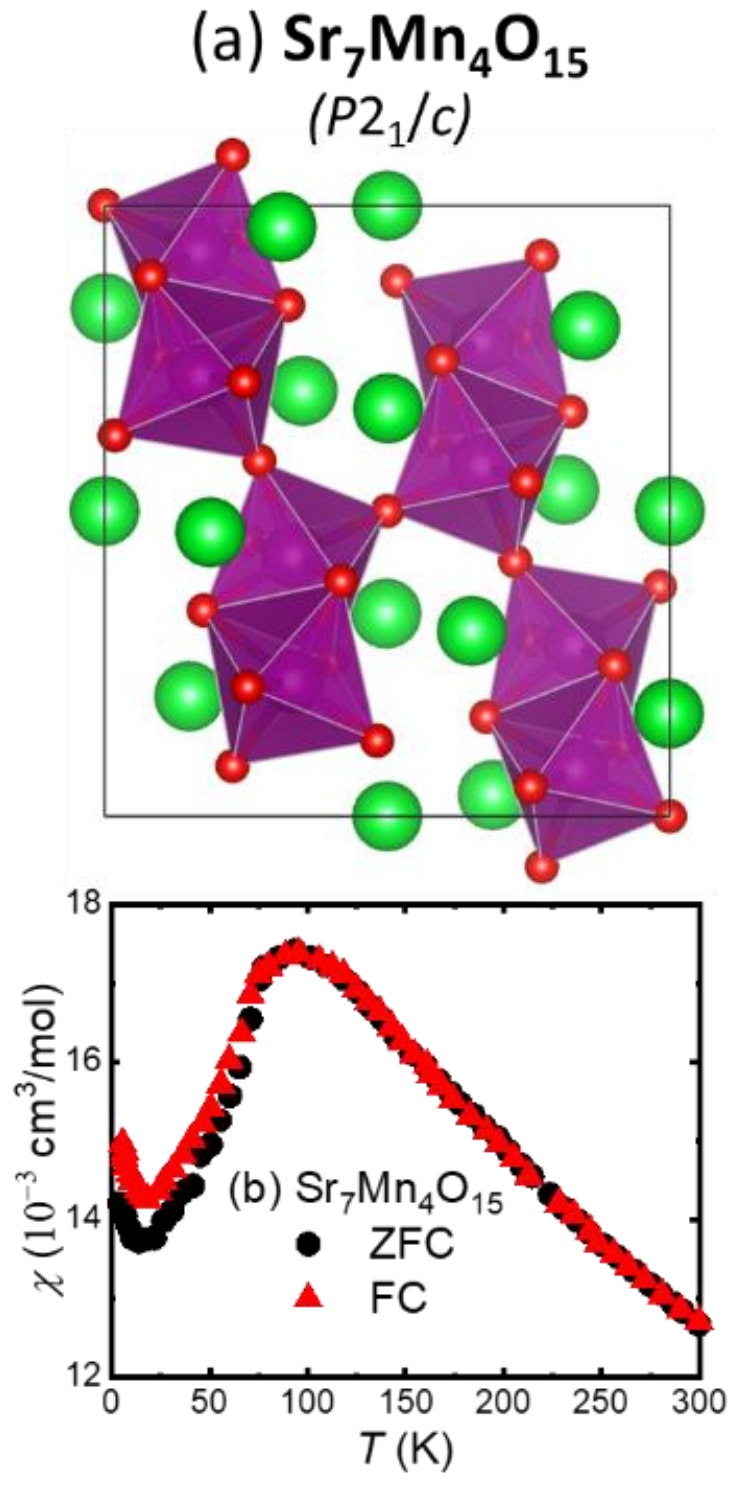

Figure 38. (a) The crystal structure of $Sr_7Mn_4O_{15}$, (b) FC/ZFC magnetic susceptibility of $Sr_7Mn_4O_{15}$ under the applied magnetic field of 100 Oe [221].

$Ba_4Ru_3O_{10}$[224] is a distorted monoclinic material based on the 9R-$BaRuO_3$ hexagonal perovskite. It crystallizes in the space group $P2_1/a$ with the lattice parameters $a$ = 5.776 Å, $b$ = 13.076 Å, $c$ = 7.234 Å, and $\beta$ = 113.53°. Its crystal structure, as shown in **Figure 40**, consists of three face-sharing $RuO_6$ octahedra making $Ru_3O_{12}$ trimers, corner-sharing with each other to form zig-zag layers parallel to the (001) plane. The crystal structure is skewed away from simple hexagonal symmetry by the presence of the 4:3 Ba to Ru ratio and the trimer formation. The intra-atomic Ru-Ru distance within a trimer is shorter than that in Ru metal, 2.57 Å compared to 2.65 Å, which the authors interpret as implying the presence of strong metal-metal bonds, as commonly seen in face-sharing octahedra. The temperature dependent magnetic susceptibility of $Ba_4Ru_3O_{10}$ displays a broad peak at 200 K, taken by the authors to be characteristic of a 2D-antiferromagentic transition. Further, the effective moment was calculated to be 3.0 μB/ f.u, which would yield an S =1 for every $Ru^{4+}$-based trimer in $Ba_4Ru_3O_{10}$. The DFT-based calculations that the authors performed on $Ba_4Ru_3O_{10}$ were interpreted as indicating the presence of a nonmagnetic ground state for one-third of the $Ru^{4+}$ ions and a different charge distribution between different Ru sites in the trimer and molecular orbital formation in the trimer.[225] Readers interested in the question of "where's the charge" in this and other materials in the hexagonal perovskite family are referred to the theoretical contribution in this issue by Streltsov and Khomskii.

Continuing along this line of structural complexity, $Ba_5Ru_3O_{12}$ [226] adopts the orthorhombic space group *Pnma* with lattice parameters $a$ = 10.862 Å, $b$ = 5.893 Å, and $c$ = 19.790 Å. As shown in **Figure 39**, the crystal structure also contains the $Ru_3O_{12}$ trimers, similar to the case of $Ba_4Ru_3O_{10}$. However, for this material, the trimers are isolated from each other and separated by Ba atoms. In $Ba_5Ru_3O_{12}$, there exists a formal mix of $Ru^{4+}$ and $Ru^{5+}$ in a 1 to 2 ratio. The material displays antiferromagnetic ordering at 60 K. Fitting of the high temperature magnetic susceptibility results in a Curie-Weiss temperature of -320 K, indicating a relatively strong antiferromagnetic interaction between Ru ions.

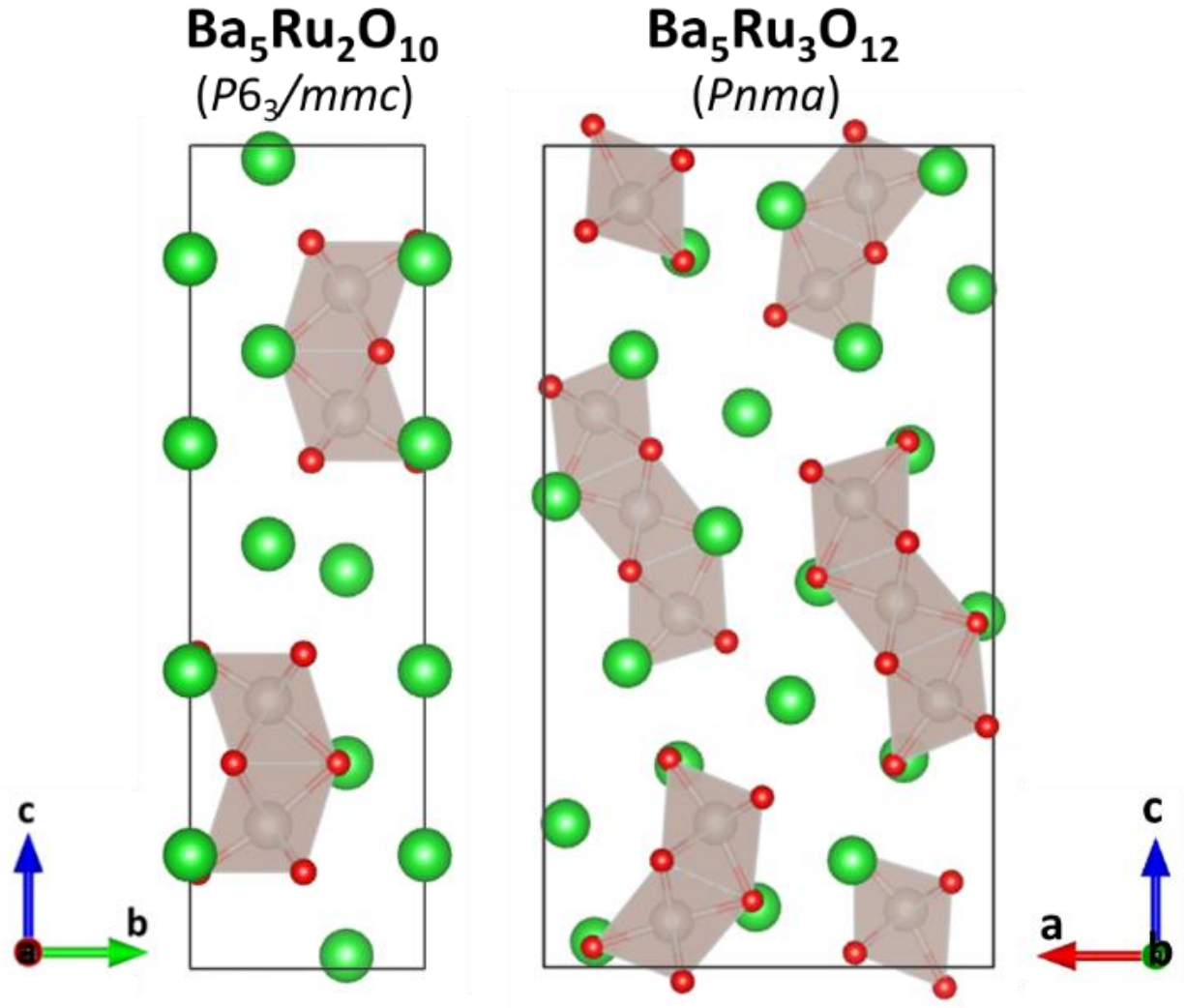

Figure 39. The crystal structures of $Ba_5Ru_2O_{10}$ and $Ba_5Ru_3O_{12}$.

We move on now to barium iridates. Quantum spin liquids are often found in strongly frustrated magnetic systems. Typically, the magnetic ions are arranged in simple triangular, honeycomb, and kagome lattices. $Ba_4Ir_3O_{10}$ is proposed to be a quantum spin liquid candidate, but has the Ir ions arranged in a nominally unfrustrated lattice due to its structural complexity, which is based on zig-zag corner-sharing of $Ru_3O_{12}$ trimers **Figure 40.** The temperature-dependent magnetic susceptibility has been interpreted as yielding a very large negative Curie-Weiss temperature of -766 K, which would indicate the presence of very strong antiferromagnetic interactions in this material. The authors report that no magnetic ordering is observed down to 0.2 K in this material, interpreted as meaning that a quantum spin liquid state exists down to that temperature. The authors further report that only 2% doping of Sr on the Ba sites kills the QSL state and uncovers antiferromagnetic ordering at 130 K. Further, although there is strong intra-trimer magnetic exchange present due to short the Ir-Ir distances present, inter-trimer coupling also plays an important role in fully explaining the low temperature magnetic properties in this quantum magnet.

$Ba_7Ir_6O_{19}$ [227] adopts a distorted monoclinic structure in space group $C2/m$ with the lattice parameters $a$ = 14.913 Å, $b$ = 5.778 Å, $c$ = 10.979 Å and β = 99.58°. As shown in **Figure 40**, the crystal structure consists of three-face-sharing $IrO_6$ octahedra, forming $Ir_3O_{12}$ trimers, corner-shared to each other. Its magnetic properties have not yet been investigated.

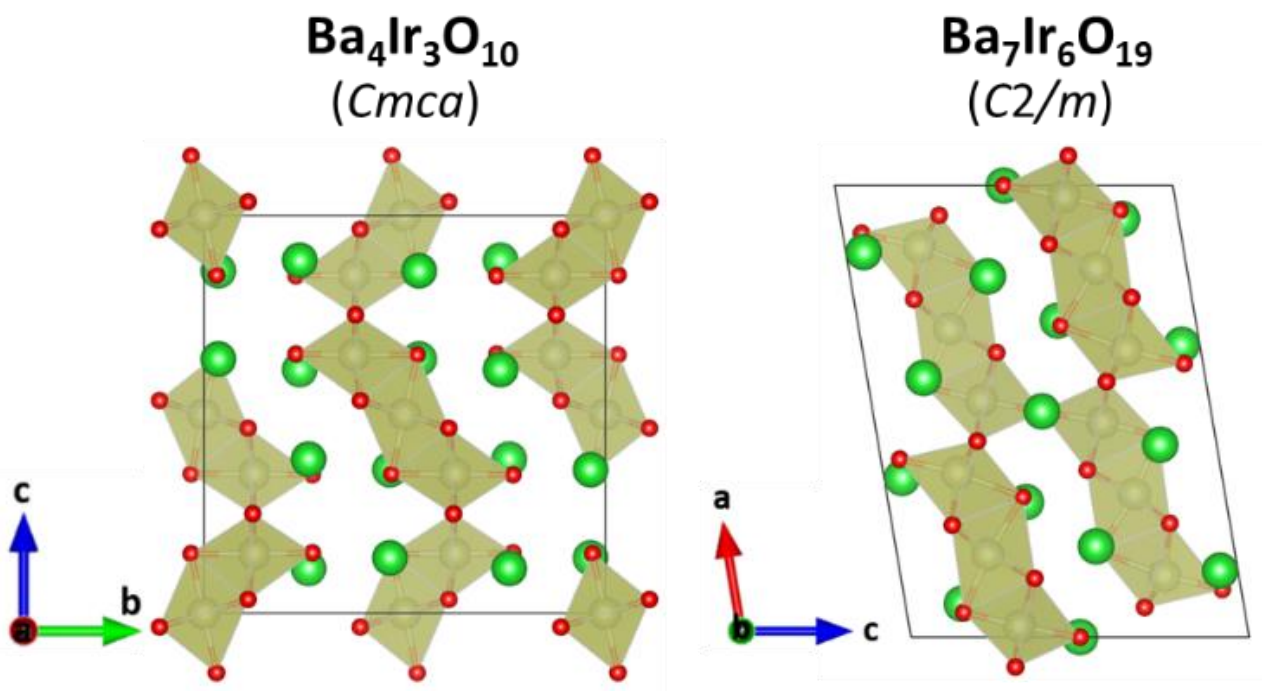

Figure 40. The crystal structures of $Ba_4Ir_3O_{10}$ (left) and $Ba_7Ir_6O_{19}$ (right).

## 11. Conclusions

In a way, the hexagonal oxide perovskites and their distorted derivatives are much richer materials than the cubic perovskites and their derivatives because the crystal structures are more widely varied. Their properties are not studied as much because the hexagonal perovskites are not as jaw-dropping when considered as quantum materials as the classical perovskites are. There are no known superconductors in this family for example. Growing crystals by the optical floating zone method for the simple ones is not often done, presumably due to the volatility of the constituents and relatively little motivation by the physical properties. However, the study of quantum spin liquids is a topic of considerable interest nowadays, especially in the neutron scattering community, and thus single crystals of some of the dimer-based materials have been studied extensively. Quite a few hexagonal-perovskite-based materials have been identified as potential candidates for the quantum spin liquid state at low temperatures (i.e. $Ba_3CuSb_2O_9$, $Ba_3CoSb_2O_9$, $Ba_3InIr_2O_9$, $Ba_3IrTi_2O_9$, $Ba_4NbIr_3O_{12}$, $Ba_4Ir_3O_{10}$ and $Ba_2Y_2Rh_2Ti_2O_{17}$), but so far the field remains "dynamic"; and at the time of this writing there appears to be no community-wide accepted hexagonal perovskite quantum spin liquid, although as described above there are many candidates. Perhaps the problem is that there are too many contenders and different researchers have different favorites among them. Several ferromagnetic hexagonal perovskites have recently been reported, such as $Ba_4NdRu_3O_{12}$ ordering at 11.5 K or $Ba_4NbMn_3O_{12}$ ordering at 42 K. There may be worth further study.

Most of the hexagonal perovskite derivatives are materials in the $Ba_3MM'_2O_9$ family (based on dimers) and next the $Ba_4MM'_3O_{12}$ family (based on trimers). The ferroelectric and dielectric properties of these materials have not generally been reported, though $Ba_6Nd_2Ti_4O_{17}$, a material that includes layers of both tetrahedra and octahedra, is a well-known dielectric. In the near future, in our view, comprehensive studies will reveal more interesting quantum properties in the hexagonal oxide perovskite family. The family does seem to have an excellent chance for hosting the exotic quantum spin liquid state, and that may turn out after several more years of study, to be its primary "claim to fame".

## 12. Acknowledgments

The authors' research on hexagonal oxide perovskites and geometric magnetic frustration has been supported by the Gordon and Betty Moore Foundation, grant GBMF-4412, and by the Basic Energy Sciences Division of the Department of Energy, grant number DE-FG02-08ER46544 to the Institute of Quantum Matter, which continues as an Energy Frontier Research Center funded by the U.S. Department of Energy, Office of Science, Basic Energy Sciences under Award No. DE-SC0019331. The magnetic data presented in this review has been extracted and plotted through use of the program *Origin*. Similarly, the crystal structures presented have been imaged though use of the program *Vesta*, using information available on the International Crystal Structure Database or in the published literature.

## 13. Author contributions

This review article was written through equal contributions of both authors, Loi Nguyen and Robert Cava.

## 14. Notes

The authors declare no competing financial interest.

## 15. Biographies

Loi T. Nguyen graduated from the California State University, Fullerton (CSUF) in 2017, where his solid state chemistry research was on unconventional oxide perovskites. He is currently a third year graduate student in the Department of Chemistry at Princeton University. His research focuses on the synthesis and magnetic properties of hexagonal perovskites. He has discovered several new face-sharing hexagonal perovskites, several of which potentially host a quantum spin liquid state.

Robert J. Cava is the Russell Wellman Moore Professor and former chair of the Chemistry Department at Princeton University. He received his Ph.D. in Ceramics from MIT in 1978, after which he was an NRC Postdoctoral Fellow at the National Bureau of Standards. He then eventually became a Distinguished Member of Technical Staff at Bell Laboratories, where he worked for 17 years. He is a member of the US National Academy of Sciences and a Foreign Member of the Royal Society of London.